\documentclass[nofootinbib,
tightenlines,
superscriptaddress,11pt]{article}
\usepackage[margin=2.4cm]{geometry}
\usepackage{hyperref}
\usepackage{amsmath,amssymb,amsfonts,mathtools}
\usepackage{amsthm}
\usepackage{xcolor}

\usepackage{bbold}
\usepackage{stmaryrd}
\usepackage{authblk} 

\let\oldbibliography\thebibliography
\renewcommand{\thebibliography}[1]{%
  \oldbibliography{#1}%
  \setlength{\itemsep}{3pt}%
  \setlength{\parskip}{0pt}%
}

\newcommand{\nw}[1]{\textcolor{blue}{#1}}
\newcommand{\im}{\mathrm{i}}
\newcommand{\eps}{\epsilon}
\newcommand{\epst}{\tilde\epsilon}
\newcommand{\tr}{\mathrm{Tr}}
\newcommand{\vol}{\mathrm{vol}}

\def\be#1\ee{\begin{align}#1\end{align}}

\def\ba{\begin{eqnarray}}
\def\ea{\end{eqnarray}}
\def\nn{\nonumber}
\def\q{\quad}
\def\R{{\rm Re}}
\def\I{{\rm Im}}

\begin{document}

\title{Area metric gravity \`a la Plebanski:\\ connection, curvature and dynamics}
\author[]{Bianca Dittrich}
\affil[]{\small Perimeter Institute, 31 Caroline Street North, Waterloo, ON, N2L 2Y5, Canada}

\date{}

\maketitle

\begin{abstract}
Area metrics arise as  effective description at intermediate scales in a number of approaches to quantum gravity, including string theory, holography and loop quantum gravity. Recently, it has been shown that area metric actions can be derived from the modified Plebanski framework on the linearized level. Surprisingly these actions avoid extra poles in the propagator for the effective length dynamics, and hence constitute interesting deformations of linearized Einstein gravity. 
To be able to study whether this property persist for the full theory, we construct here non-linear area metric dynamics from modified Plebanski actions. 

We provide the geometric tools for area metrics that do not rely on perturbation
theory and allow a systematic construction and study of general non-linear area metric actions: an algorithm that extracts a length metric from an area metric and a connection on
two-forms that is uniquely determined by the area metric, together with its curvature and curvature invariants. We derive area metric actions from modified Plebanski actions in first and second order versions. We also find that area metric actions can be reformulated as Poincare gauge theories.
 We determine the effective length metric
 action, which on a flat background describes a graviton and a massive spin-two field with
parity-violating couplings set by the Barbero--Immirzi parameter. Additionally we discuss
cosmological backgrounds and the role of the matter coupling, which together open the way to
cosmological applications.
\end{abstract}

\section{Introduction}\label{sec:intro}

Area metric dynamics has been proposed in a number of quantum gravity approaches to describe quantum gravity dynamics at intermediate scales between the deep ultraviolet and the infrared regime. 
These approaches include string theory~\cite{Schuller:2005yt,Punzi:2006hy,Ho:2015cza,Borissova:2024cpx}, AdS/CFT holography \cite{Bhattacharya:2025hag}, and, in particular,  loop quantum gravity and spin foams \cite{Dittrich:2021kzs,Dittrich:2022yoo,Borissova:2022clg,BDK,Dittrich:2023ava}. A first study of the functional renormalisation flow of area metric theories has been undertaken in \cite{Borissova:2025frj}.

Area metrics generalise length metrics: instead of assigning lengths to tangent
vectors they assign areas to tangent bivectors. The (cyclic) area metric tensor is a rank-$4$ tensor with the same symmetries as the Riemann curvature tensor, and therefore has in four spacetime dimensions 20 independent components.~\footnote{A further generalisation to so-called acyclic area metrics has 21 independent components~\cite{Schuller:2005yt}. We will consider cyclic area metrics in the main text and discuss the inclusion of the acyclic area metric degree of freedom in Appendix \ref{app:acyclic}.} A given length metric $g$  induces an area metric $G$ via $
G_{\mu\nu\rho \sigma} = g_{\mu\rho}g_{\nu\sigma} - g_{\mu\sigma}g_{\nu\rho}$, the area metric configuration space is therefore a genuine extension of the length metric configuration space.

\paragraph{Area metrics from loop quantum gravity and spin foams:} Loop quantum gravity \cite{Rovellibook,Thiemannbook} and spin foams \cite{PerezLR,HandbookSpeziale,Hrytseniak:2026vpt} are quantisations of general relativity, which lead to an extension of the (quantum) length configuration space   on the discrete level \cite{Dittrich:2008va,Dittrich:2008ar,Dittrich:2010ey,Dittrich:2012rj,Freidel:2010aq}. A geometric interpretation of this extension in terms of area metrics has been recently provided in \cite{Dittrich:2023ava}. The work \cite{EffSFE} showed that this extension is forced by the discrete and asymptotically equidistant spectrum for area operators \cite{Rovelli:1994ge}. These eigenvalues are summed over in the spin foam path integral \cite{PerezLR,EffSFE,EffSFL}, areas are therefore more fundamental ingredients than lengths. 

The interpretation of these additional degrees of freedom as area metric was reinforced by computations which established the perturbative continuum limit of (effective) spin foams on a regular lattice \cite{Dittrich:2021kzs,Dittrich:2022yoo}: the leading order recovered (linearised) general relativity, the next to leading order arose from the presence of massive area metric degrees of freedom \cite{Dittrich:2022yoo,AsanteToAppear}.

These results led to the question, whether an extension of the configuration space and a corresponding dynamics can be also constructed directly in the continuum.  Spin foam quantisation is based on the Plebanski formulation \cite{Plebanski:1977zz} (with Holst term \cite{Holst:1995pc}) of general relativity. Here one starts with an $SO(3,1)$ $BF$ action\footnote{An $SO(3,1)$ BF action is used in the so-called non-chiral version of the Plebanski action, where one can choose all fields to be real, and which is used for spin foam quantisation. The chiral version works only with the self-dual part of $SO(3,1)$, which necessarily introduces complex fields and require reality conditions.} (with Holst term), which in itself defines a  topological theory \cite{Horowitz:1989ng}. Gravitational dynamics is enforced by adding Lagrange multiplier terms, which constrain the Lie-algebra valued two-forms to be simple, i.e. $B^{IJ}_{\mu\nu}=\pm\star e^I\wedge e^J$ (in the so-called gravitational sector), where $e^I_\mu$ is identified with a tetrad, and $\star$ is the Hodge dual on the Lie-algebra of $SO(3,1)$. These are the so-called simplicity constraints. On the simplicity constraint surface the $BF$ action evaluates to the Palatini action (with Holst term). 

Loop quantum gravity and spin foams quantise instead of the Lie algebra valued connection the corresponding group valued holonomies. In the holonomy representation (where wave functions are functions of the holonomies, i.e. of group elements) the $B$-fields, which are conjugated to the connection, are quantised as left- or right-invariant derivatives. Since 
$SO(3,1)$ is non-Abelian, the operators corresponding to the $B$-fields do not commute.\footnote{There is also a flux representation \cite{Baratin:2010nn}, where the operators corresponding to the $B$-fields act diagonally, but come with a non-commutative product.}  As a consequence, the simplicity constraints on the $B$-fields split into two sets, a commuting part and a part which does not commute with itself and constitutes a second class constraint set \cite{Dittrich:2008ar,Dittrich:2010ey,Dittrich:2012rj,PerezLR}. The latter means that, due to the Heisenberg uncertainty relations, one cannot impose the second set sharply. Consequently, this set is imposed `weakly' \cite{Engle:2007wy,Freidel:2007py, EffSFE,EffSFL}, e.g. via coherent states which are peaked on the constraint hypersurface. This `weak' imposition of part of the simplicity constraints explains the enlargement of the configuration space in spin foams \cite{Dittrich:2008ar,Dittrich:2010ey,Dittrich:2012rj,Freidel:2010aq}. 

To mirror this behaviour in the continuum one would have to also weaken the imposition of the simplicity constraints there. Such a framework has been suggested by Krasnov  in \cite{Krasnov:2006du,Krasnov:2008fm,Krasnov:2009iy}.  Here one replaces the simplicity constraints by a potential, which then imposes the constraints only weakly by making the corresponding degrees of freedom  heavy.  The chiral version of the modified Plebanski theory yielded a surprise --- despite the additional degrees of freedom (a massive spin-2 particle), only the graviton turned out to be propagating \cite{Freidel:2008ku,Krasnov:2009ik}, the massive spin-2 just yielded a contact term in the propagator. Replacing all constraints by a quadratic potential in the non-chiral Plebanski theory (which is used in spin foams) led however to eight propagating degrees of freedom \cite{Alexandrov:2008fs,Speziale:2010cf,Beke:2011mu}. 

But the simplicity constraints on the $B$-fields in spin foams are divided into two sets, one of which is sharply imposed and the other is only weakly imposed. \cite{Borissova:2022clg} therefore kept a sharply imposed subset of simplicity constraints in the continuum, and only introduced a potential for the remaining set of constraints. The split of the continuum constraints was chosen such that the un-constrained and weakly constrained (gauge-invariant) parts of the $B$-fields could be faithfully encoded into area metric perturbations.\footnote{The area metric is constructed as a quadratic form from the $B$-fields, the $B$-fields are square roots of the area metric in a similar way as tetrads are square roots of the length metric.}  The work \cite{Borissova:2022clg} succeeded to construct the (linearised) action for the area metric perturbations. Interestingly, integrating out the 10 non-length metric degrees of freedom resulted also in a propagator which featured only the massless graviton part and a contact term. The Lorentzian version of the linearised theory studied in \cite{BDK,Dittrich26} confirmed this result: there are 20 degrees of freedom, 10 of which constitute the length metric sector (and result in the two physical degree of freedom for the massless graviton), with another 10 degrees encoded into two massive spin-2 fields. The kinetic term of the action has however a left-over shift symmetry from its $BF$-origin. This leads to a second class constraint set, which allows to remove one spin-2 field. The remaining spin-2 field turns out to come with negative kinetic and mass terms. However, its dynamics, at least at linearised level, decouples from the dynamics of the length metric, and this allows for the dynamics to be stable despite the negative energy terms.  \cite{Dittrich26} provides a detailed analysis of the covariant equation of motion, which shows that the dynamics of the massless graviton dynamics is not changed from linearised general relativity: it still satisfies $\Box h_{\mu\nu}=0$ (in TT gauge). The only effect is that non-vanishing massless graviton fields are dressed with excitations from the non-metric sector.\footnote{There is also a Klein-Gordon wave equation for the non-metric degrees of freedom.} If one couples the area metric to the electro-magnetic fields, one obtains  a (extremely small) signal in gravitational wave interferometers due to birefringence for the photons sourced by the non-length degrees of freedom in the area metric \cite{Dittrich26}.

\paragraph{Linearised  area-metric actions.}
The linearised area metric actions constructed in \cite{Borissova:2022clg} are a subset of general linearised area metric actions, which can be constructed by demanding invariance under linearised diffeomorphisms \cite{Alex:2019ari,BDK}. For cyclic area metrics there are two types of additional couplings as compared to (linearised) general relativity: (after a canonical normalisation for the kinematical term of the non-length degrees of freedom) one type is for the term coupling the non-length and length metric degrees of freedom, the other type for the mass of the non-length degrees of freedom. But each type of couplings includes two parameters for the two massive spin-2 fields (corresponding to a decomposition of these degrees of freedom into self-dual and anti-self dual). To obtain a stable dynamics (that is only the graviton pole and a contact term for the propagator in the length metric effective theory) the first type of couplings needs to satisfy a so-called shift-symmetry condition - the remaining parameter is then the Barbero-Immirzi parameter in front of the (parity symmetry violating) Holst term.  Demanding that one does not have an additional pole for the propagator in the effective length metric theory restricts also the two mass parameters to be determined by one parameter. 

\paragraph{Non-linear  area-metric actions.} An important question is to understand the dynamics of non-linear area metric actions, in particular those which can be constructed via the modified Plebanski route.  So far there were two main obstacles: (a) There has not yet been a proposal for a connection that is determined by the area
metric through geometrically natural conditions, from which one can then define a curvature and curvature invariants.\footnote{Several different notions have been proposed \cite{Schuller:2005yt,Schuller:2005ru,Punzi:2006nx}. The earliest notion \cite{Schuller:2005yt} requires a (non-unique) decomposition of the area metric into a sum of length metrics-induced area metrics, the resulting connection does however depend on this non-unique decomposition. Area metric pre-connections were introduced in \cite{Schuller:2005ru}, their symmetric part is fixed by area metric compatibility, but their anti-symmetric part remained undetermined. \cite{Ho:2015cza} showed that compatibility with the area metric and vanishing area torsion
do not determine an area connection uniquely, and fixed the remaining freedom by additional
algebraic conditions. This leads to a three-parameter family of connections, of which the
simplest member is chosen as the `area Levi-Civita connection'. A dynamics was postulated in
the form of a generalised Einstein equation, as the construction does not seem to admit an
action principle. An alternative route is proposed in
\cite{Punzi:2006nx}, which  defines an effective length metric from the Fresnel polynomial of the cyclic part of the area metric: this construction does however need an area metric whose inverse has a non-vanishing totally anti-symmetric part $\phi$; the Fresnel polynomial is then evaluated at $d\phi$. This makes the construction degenerate at the length metric-induced limit. Since the effective length metric depends on $d\phi$ the equations for generic area metrics appear to be
of higher than second order in $\phi$. The non-length modes enter only through derivative terms and are massless. 
Another possibility to construct area metric actions is to demand gravitational closure \cite{Dull:2016eki,Schuller:2020xyv}, that is a consistent canonical evolution for the matter dynamics. This can be done perturbatively.  In the covariant version \cite{Alex:2019ari}, also on the perturbative level, diffeomorphism invariant area metric actions up to order three and four have been constructed, but without a notion of curvature the various interaction terms  lack a geometric interpretation.
}
A main hindrance for constructing a connection is the absence of a soldering form. The
area metric by itself fixes only the part of a connection on two-forms that expresses
compatibility. Another part would be fixed by a torsion condition, which can be formulated
by using a soldering form. Here we will use the $B$-fields as soldering form. We will see that a further condition to obtain uniqueness for the connection is, that it preserves the split of the space of two-forms into self-dual and anti-self-dual forms. This provides then a unique connection. (b) Another issue is to have a natural split of the area metric into length-metric and non-length metric degrees of freedom at the non-linear level.  This will allow to extract an effective length metric dynamics and straightforward comparison to length metric based actions. Here we will provide such a split.

\paragraph{This work.}
Here we construct area-metric actions from the modified Plebanski formalism, which
provides the soldering form: the  two-form fields $B^i_\pm$, which are the main variables in the Plebanski formalism,  identify an internal
three-dimensional bundle with the (anti-)self-dual two-forms. We thus have access to a
connection, its curvature and actions with a direct geometric interpretation, at the
full non-linear level. In particular the torsion-free, area-metric-compatible 
connection (which preserves the split into self- and anti-self dual two-forms) is determined uniquely.  This does not only allow to construct actions resulting from the modified Plebanski framework, but using the full area metric curvature and its invariants also more general actions. We will also show that area metric theories constructed via $B$-fields (which correspond to the tetrads for a length metric) can be brought into a Poincare gauge formulation \cite{Blagojevic:2003cg,Blagojevic:2013xpa}, one can therefore hope that the analysis of Poincare gauge gravitational actions does, at least partially, apply to area metric actions.\footnote{The Poincare gauge reformulation relies on the split of area metrics into a length and non-length metric degree of freedom. As separate entities one could choose different coupling constants for these two parts. But if one treats the area metric as one geometrical object these couplings might not be different.}

The extraction of a length metric from a given area metric is straightforward on the linearised level, where the trace degrees of freedom of the area metric perturbations correspond to a length metric perturbation.
On the non-linear level \cite{Borissova:2022clg} proposed to use 
a parametrisation of the $B$-fields by two tetrads $e_\pm$ and two pairs of determinant one (symmetric, modulo an internal gauge) $3\times 3$ matrices $b_\pm$ introduced in \cite{Freidel:2008ku}. The sharp simplicity constraints identify the two tetrads with each other, $e=e_+=e_-$. The resulting tetrad defines the length metric part. The pair of (symmetric) determinant one $3\times 3$ matrices $b_\pm$ provide the 10 additional degrees of freedom. Here we will use a slight alteration of this parametrisation. \cite{Borissova:2022clg} expressed the area metric in these parameters, but inverting these equations constitutes  a system of 20 polynomial equations, whose solution cannot be explicitly given.\footnote{See \cite{Asante:2024rrd} for techniques to solve a closely related equation system, which determines lengths from areas.}  Here we will present an algorithm, which condenses the algebraic difficulty into determining the spectral decomposition of a $6\times6$ matrix constructed from the area metric. This spectral decomposition determines a split of the space of two-forms into self-dual and anti-selfdual parts, which allows to determine, via Urbantke's formula \cite{Urbantke:1984eb}, a conformal equivalence class of length metrics. The determinant of the length metric and the non-length parameters $b_\pm$ can then be straightforwardly determined. 
Additionally, we will prove that, given a certain non-degeneracy condition, the map between the parameters and the space of area metrics is locally an isomorphism.\footnote{The non-degeneracy conditions is closely related to conditions for the area metric to be physically viable (e.g. for Maxwell's equation to be well-posed) \cite{Schuller:2009hn}.}

Using the modified Plebanski framework (with only a subset of simplicity constraints replaced by a potential) we will derive area metric actions in various forms: $(i)$ The initial action will include apart from the $B$-fields, expressed via the tetrad $e$ and the matrices $b_\pm$, and a connection, also Lagrange multipliers. We will integrate out these Lagrange multipliers, which gives as a first order formulation of area metric gravity. We then proceed to integrate out the connection, which  leads to a $(ii)$ second order version of our action, which will consist of a curvature term, a contorsion term and the potential.  As a final step we can also integrate out the $b_\pm$-fields in order to obtain an $(iii)$ effective action for the length metric.  This action can be computed perturbatively around length-metric induced area metrics.  The result will include the usual Einstein-Hilbert part and a Weyl squared term with form factor as well as terms of higher order in the Weyl curvature.   On the other hand, before integrating out the connection, we can integrate out the $b_\pm$-fields (perturbatively around length metric induced configurations), this gives a Poincare gauge formulation of area metric dynamics resulting from the modified Plebanski framework.

\paragraph{Questions.}
We hope that this formalism can help to understand area metric dynamics more systematically. A non-linear action is needed to understand consequences for black hole physics \cite{Borissova:2025nvj} and cosmology. In cosmology it is important to understand the fate of parity-breaking terms and the possibility of measuring the Barbero--Immirzi parameter \cite{Bianchi:2024mrt,Dittrich26}.  For perturbations on flat space the absence of an additional pole in the effective length-metric action is an astonishing result \cite{Borissova:2022clg,BDK}, but it is not clear, whether it will persist to non-linear order. Likewise it is important to understand the fate of the negative-energy terms that appear in the linearised
Lorentzian action \cite{BDK,Dittrich26}. Another question is to understand the behaviour of area metric dynamics \cite{Borissova:2025frj} and modified Plebanski actions \cite{Krasnov:2006du,Krasnov:2008fm,Krasnov:2009iy} under the renormalisation flow. This is another viability test: the masses of the non-length metric degrees of freedom should remain heavy and the parity violating couplings should be small (at least those couplings which lead to measurable physical effects) in the infra-red to obtain a theory compatible with observations.

\paragraph{Outline.}
Section~\ref{sec:conv} collects conventions. Section~\ref{sec:pleb} introduces the
modified Plebanski action and Section~\ref{sec:param} the parametrisation of the
$B$-fields. Section~\ref{sec:area} discusses the resulting area metrics, and a condition on area metrics, proposed by Reisenberger \cite{Reisenberger:1998fk},
 to be induced by length metrics. 
 Section~\ref{sec:faithful} presents an algorithm to extract a length metric from an area metric and proves the
faithfulness of the map from (semi-constrained) $B$-fields to area metrics. In Section~\ref{sec:eff} we
integrate out the multipliers and in Section~\ref{sec:intcon} the connection, obtaining
the action in terms of the Levi--Civita connection, the contorsion and the potential.  In Section~\ref{sec:areaconn} we define a connection and curvature tensor for area metrics and discuss the conditions the connection has to satisfy to be uniquely determined. 
Section~\ref{Sec:beom} discusses the equations of motion for $b_\pm$, their solution
at second order, and the resulting effective actions for the length metric on flat and
FRW backgrounds. Section~\ref{sec:poincare} rewrites the action as a Poincar\'e gauge theory, and Section~\ref{sec:disc} summarises the results and discusses open questions. The appendices contain various technical derivations and sketch how to include an acyclic component of the area metric, which is in particular interesting for cosmological applications.

\section{Conventions}\label{sec:conv}

Here we collect conventions and definitions of key objects needed throughout the paper.

Spacetime and internal signature are mostly plus, $\eta_{IJ}=\mathrm{diag}(-1,1,1,1)$,
with internal indices $I,J=0,\dots,3$ and spatial indices $i,j,k=1,2,3$.  Spacetime indices are denoted  by greek letters $\mu,\nu,\ldots = 0,\ldots,3$, which are lowered/raised by a spacetime metric $g_{\mu\nu}$ and its inverse $g^{\mu\nu}$. A spacetime metric can be defined from a tetrad $g_{\mu\nu}=\eta_{IJ} e^I_\mu e^J_\nu$.

\paragraph{Levi-Civita tensors and density.} We distinguish three objects:
\begin{itemize}
\item[(i)] the \emph{internal} Levi-Civita tensor $\eps_{IJKL}$, defined by
$\eps^{0123}=+1$, indices moved with $\eta$; hence
$\eps_{0123}=\det(\eta)=-1$;
\item[(ii)] the auxiliary Levi-Civita \emph{densities}
$\epst^{\mu\nu\rho\sigma}$ (weight $+1$) and $\epst_{\mu\nu\rho\sigma}$ (weight $-1$),
both equal to the permutation symbol with $\epst^{0123}=\epst_{0123}=+1$; their indices
are \emph{never} moved with the metric. They are needed where formulas must be
metric independent (e.g. the Urbantke formula). 
\item[(iii)] the \emph{spacetime} Levi-Civita tensor
$\eps^{\mu\nu\rho\sigma}:=\epst^{\mu\nu\rho\sigma}/\sqrt{-g}$ in positively oriented
coordinates, indices moved with $g_{\mu\nu}$. 
Thus
\begin{equation}\label{epsrel}
\eps^{\mu\nu\rho\sigma}=\frac{1}{\sqrt{-g}}\,\epst^{\mu\nu\rho\sigma}\,,\qquad
\eps_{\mu\nu\rho\sigma}=-\sqrt{-g}\;\epst_{\mu\nu\rho\sigma}\,,\qquad
\sqrt{-g}=\det(e)\,.
\end{equation}
\end{itemize}
The internal contraction identities are
\begin{equation}\label{epseps}
\eps_{IJKL}\,\eps^{IJKL}=-24\,,\qquad
\eps_{IJKL}\,\eps^{IJMN}=-2\big(\delta_K^M\delta_L^N-\delta_K^N\delta_L^M\big)\,.
\end{equation}
There are two invariant bilinear forms on the Lie algebra $\mathfrak{so}(3,1)$ given by
\begin{equation}\label{bilin}
\eta_{IJKL}:=\tfrac12\big(\eta_{IK}\eta_{JL}-\eta_{IL}\eta_{JK}\big)\,,
\qquad \tfrac12\,\eps_{IJKL}\,.
\end{equation}

\paragraph{Differential forms and wedge product.} A $p$-form carries the component
convention $\omega=\frac{1}{p!}\,\omega_{\mu_1\dots\mu_p}\,
dx^{\mu_1}\!\wedge\dots\wedge dx^{\mu_p}$, and the wedge product the antisymmetrisation
convention
\begin{equation}\label{wedgeconv}
(\alpha\wedge\beta)_{\mu_1\dots\mu_{p+q}}
=\frac{(p+q)!}{p!\,q!}\,
\alpha_{[\mu_1\dots\mu_p}\beta_{\mu_{p+1}\dots\mu_{p+q}]}\,,
\end{equation}
so that $(e^I\wedge e^J)_{\mu\nu}=e^I_\mu e^J_\nu-e^I_\nu e^J_\mu$. Consequently, for
two two-forms $X,Y$ and with $d^4x=dx^0\wedge dx^1\wedge dx^2\wedge dx^3$,
\begin{equation}\label{wedge22}
X\wedge Y=\tfrac14\,X_{\mu\nu}Y_{\rho\sigma}\,\epst^{\mu\nu\rho\sigma}\,d^4x\,,
\qquad
e^0\wedge e^1\wedge e^2\wedge e^3=\det(e)\,d^4x=\vol\,,
\end{equation}
and in the conventions of item (i) above,
$\eps_{IJKL}\,e^I\wedge e^J\wedge e^K\wedge e^L=-24\,\vol$ \, .

\paragraph{Internal Hodge operator and chiral basis.} The internal Hodge operator
$(\star X)^{IJ}=\tfrac12\,\eps^{IJ}{}_{KL}X^{KL}$ satisfies
\begin{equation}\label{starstar}
\star\star=-\,\mathbb{1}\,, 
\end{equation}
in contrast to $\star\star=+\mathbb{1}$ in Euclidean signature; its eigenvalues are
$\pm\im$ and the chiral split lives in the complexification,
$\mathfrak{so}(3,1)\otimes\mathbb{C}
=\mathfrak{sl}(2,\mathbb{C})_+\oplus\mathfrak{sl}(2,\mathbb{C})_-$, with complex
conjugation exchanging the summands. Rotation and boost generators are defined by
\begin{equation}\label{JK}
(J_i)^{jk}=\eps_{ijk}\,,\qquad (K_i)^{0j}=\delta_i^j=-(K_i)^{j0}\,,
\end{equation}
where $\eps_{ijk}$ is the three-dimensional symbol with $\eps_{123}=+1$ (which in the
present conventions coincides with $\eps^{0ijk}$). All components not listed in
\eqref{JK} vanish. With the defining condition $\eps^{0123}=+1$ one finds
\begin{equation}\label{starJK}
\star K_i=-J_i\,,\qquad \star J_i=+K_i\,,
\end{equation}
and the chiral
basis is
\begin{equation}\label{Pdef}
P^{IJ}_{\pm i}:=\tfrac12\big(J_i\mp\im\,K_i\big)^{IJ}=\tfrac12\,\eps^{0iIJ}\mp\im \eta^{IJ}_{0i} \,,\qquad
\star P_{\pm i}=\pm\im\,P_{\pm i}\,\,,\qquad \;
\overline{P_{+i}}=P_{-i}\,,
\end{equation}
with trace pairings
\begin{equation}\label{gram}
\eta_{IK}\eta_{JL}\,P^{IJ}_{\pm i}P^{KL}_{\pm j}=\delta_{ij}\,,\qquad
\eta_{IK}\eta_{JL}\,P^{IJ}_{+ i}P^{KL}_{- j}=0\,.
\end{equation}
The companion $\eps$-pairings, used repeatedly below, are
\begin{equation}\label{gramEps}
\eps_{IJKL}\,P^{IJ}_{\pm i}P^{KL}_{\pm j}=\pm\,2\im\,\delta_{ij}\,,\qquad
\eps_{IJKL}\,P^{IJ}_{+ i}P^{KL}_{- j}=0\,,
\end{equation}
which follow from (\ref{gram}) using 
$\eps_{IJKL}X^{IJ}Y^{KL}=2\,\eta_{IK}\eta_{JL}X^{IJ}(\star Y)^{KL}$ and (\ref{Pdef}) .

A $SO(3,1)$ valued field $B^{IJ}$ can be split into self-dual and anti-selfdual parts 
using the basis \eqref{Pdef}, that is
\begin{equation}\label{BsplitIntro}
B^{IJ}=P^{IJ}_{+i}\,(B^i_+)+P^{IJ}_{-i}\,(B^i_-)\, .
\end{equation}
Since the basis (\ref{Pdef}) is orthonormal by \eqref{gram}, the decomposition
\eqref{BsplitIntro} is inverted by plain contraction,
\begin{equation}\label{Binv}
B^i_\pm=P_{\pm i\,IJ}\,B^{IJ}\,,\qquad P_{\pm i\,IJ}:=\eta_{IK}\eta_{JL}P^{KL}_{\pm i}\,,
\end{equation}
with no relative factor, and the same rule applies to every
$\mathfrak{so}(3,1)$-valued object, in particular to the connection,
$\omega^i_\pm:=P_{\pm i\,IJ}\,\omega^{IJ}$, with
$\overline{\omega^i_+}=\omega^i_-$ for real $\omega^{IJ}$.
 The absence of a
factor is a Lorentzian accident: the Killing pairings are $J_i\!\cdot\!J_j=+2\delta_{ij}$
but $K_i\!\cdot\!K_j=-2\delta_{ij}$, so
$P_{\pm i}\!\cdot\!P_{\pm j}=\tfrac14\big[J\!\cdot\!J-K\!\cdot\!K\big]_{ij}=\delta_{ij}$;
in Euclidean signature the same $\tfrac12(J\mp K)$ basis has Gram
$\tfrac12\delta_{ij}$ and the inversion carries a factor $2$. For later use we record
the chiral algebra,
\begin{equation}\label{Palg}
[P_{\pm i},P_{\pm j}]=-\,\eps_{ijk}\,P_{\pm k}\,,\qquad [P_{+i},P_{-j}]=0\,,
\end{equation}
so that the curvature $F^{IJ}(\omega)=d\omega^{IJ}+\omega^{I}{}_{K}\wedge\omega^{KJ}$ decomposes as
$F^{IJ}(\omega)=P^{IJ}_{+i}F^i_++P^{IJ}_{-i}F^i_-$ with
\begin{equation}\label{Fchiral}
F^i_\pm=d\omega^i_\pm-\tfrac12\,\eps^i{}_{jk}\,\omega^j_\pm\wedge\omega^k_\pm\,,
\end{equation}
each sector carrying its own $SL(2,\mathbb C)$ connection.


One can also define chiral projectors without singling out the time direction: Introduce the 
$\star$-projectors
\begin{equation}\label{Piproj}
\Pi_{\pm\,IJKL}:=\tfrac12\,\eta_{IJKL}\mp\tfrac{\im}{4}\,\eps_{IJKL}\,,\qquad
(\Pi_\pm B)_{IJ}=\Pi_{\pm\,IJKL}B^{KL}=\tfrac12\big(B\mp\im\star B\big)_{IJ}\,,
\end{equation}
which satisfy $\Pi_\pm^2=\Pi_\pm$, $\Pi_+\Pi_-=0$, $\Pi_++\Pi_-=\eta_{IJKL}$ and
$\Pi_+-\Pi_-=-\tfrac{\im}{2}\eps_{IJKL}$, and project onto the $\star$-eigenvalue
$\pm\im$. Then the chiral basis \eqref{Pdef} is the self-dual part of the
rotation generators,
\begin{equation}\label{PviaPi}
P^{IJ}_{\pm i}=\Pi_\pm{}^{IJ}{}_{KL}\,(J_i)^{KL}=\mp\im\,\Pi_\pm{}^{IJ}{}_{KL}\,(K_i)^{KL}
=\Big[\tfrac12\,\eta^{IJ}{}_{KL}\mp\tfrac{\im}{4}\,\eps^{IJ}{}_{KL}\Big]\eps^{0iKL}
=\tfrac12\,\eps^{0iIJ}\mp\im \eta^{IJ}_{0i}\,.
\end{equation}

The
completeness relation dual to \eqref{gram}--\eqref{gramEps} is
\begin{equation}\label{Pcomplete}
\sum_{i}P^{IJ}_{\pm i}\,P_{\pm i\,KL}=\Pi_\pm{}^{IJ}{}_{KL}\,;
\end{equation}
and the Holst bilinear form appearing in the Plebanski action \eqref{Spleb} below is the weighted sum of the two
projectors,
\begin{equation}\label{HolstPi}
\eta_{IJKL}+\tfrac{1}{2\gamma}\,\eps_{IJKL}=\gamma_+\,\Pi_{+\,IJKL}+\gamma_-\,\Pi_{-\,IJKL}\,,
\qquad \gamma_\pm=1\pm\tfrac{\im}{\gamma}\,.
\end{equation}
Note that we have chosen the convention $\epsilon^{0123}=+1$ and thus $\epsilon_{0123}=-1$.  Flipping the signs would amount to multiplying $\epsilon$ with $-1$ and thus introduce a minus sign into the Hodge operator and therefore exchange the self-dual and anti-self-dual sectors. With such a sign flip (and an appropriate adjustment of signs in the action below) one would then have weights $\gamma^{\rm flip}_\pm=1\mp\im/\gamma$. In the flipped definition the weights for a given sector are actually the same as for the original definition, but what changed is whether the given sector is referred to as self-dual or anti-self-dual.

\paragraph{Plebanski two-forms.}
The Lorentzian Plebanski two-forms are defined as
\begin{equation}\label{Sigdef}
\Sigma^j_{\pm\,\mu\nu}(e):=2\,P^{IJ}_{\pm i}\,\delta^{ij}\,\eta_{IJKL}\,
e^K_\mu e^L_\nu
\;=\;\pm\,\im\,\big(e^0_\mu e^j_\nu-e^0_\nu e^j_\mu\big)
+\eps^{j}{}_{kl}\,e^k_\mu e^l_\nu\, .
\end{equation}
Note that the
$\eta$-lowering inside \eqref{Sigdef} flips the boost coefficient relative to
\eqref{Pdef}. These two-forms satisfy:
\begin{align}
&\text{(a) spacetime duality: }\;
{*}\,\Sigma^j_\pm=\pm\,\im\,\Sigma^j_\pm\,,\qquad
({*}X)_{\mu\nu}:=\tfrac12\,\eps_{\mu\nu\rho\sigma}X^{\rho\sigma}\,,\qquad
{*}{*}=-\mathbb{1}\,,
\label{SigSD}\\[1mm]
&\text{(b) reality: }\;\overline{\Sigma^j_+}=\Sigma^j_-\,,\label{Sigreal}\\[1mm]
&\text{(c) wedge normalisation: }\;
\Sigma^i_\pm\wedge\Sigma^j_\pm=\pm\,2\,\im\;\delta^{ij}\,\vol\,,\qquad
\Sigma^i_+\wedge\Sigma^j_-=0\,,\label{wedge}
\end{align}
where (c) is equivalent to
$\Sigma^i_{\pm\,\mu\nu}\Sigma^j_{\pm\,\rho\sigma}\,\epst^{\mu\nu\rho\sigma}
=\pm8\,\im\,\det(e)\,\delta^{ij}$. 

\vspace{2mm}

Urbantke's formula \cite{Urbantke:1984eb} allows to extract the conformal class of the metric $g_{\mu\nu}= e^I_\mu e^J_\nu\eta_{IJ}$ from the Plebanski two-forms:
\begin{equation}\label{urb}
\eps_{ijk}\,\epst^{\alpha\beta\gamma\delta}\,
\Sigma^i_{\pm\,\mu\alpha}\Sigma^j_{\pm\,\beta\gamma}\Sigma^k_{\pm\,\delta\nu}
=\pm\,12\,\im\,\det(e)\,g_{\mu\nu}\,,\qquad g_{\mu\nu}=\eta_{IJ}e^I_\mu e^J_\nu\, .
\end{equation}
Later-on we will need the following identities for the Plebanski two-forms:
\begin{align}
&\tfrac12\,\delta_{ij}\,\Sigma^i_{\pm\,\mu\nu}\Sigma^j_{\pm\,\rho\sigma}
=\tfrac12\big(g_{\mu\rho}g_{\nu\sigma}-g_{\mu\sigma}g_{\nu\rho}\big)
\mp\tfrac{\im}{2}\,\eps_{\mu\nu\rho\sigma}\,,\label{id74}\\
&\Sigma^i_{\pm\,\mu\nu}\,\Sigma^j_{\pm\,\rho\tau}\,g^{\nu\tau}
=\delta^{ij}g_{\mu\rho}+\eps^{ij}{}_{l}\,\Sigma^l_{\pm\,\mu\rho}\,,\label{id75}\\
&\Sigma^i_{\pm\,\mu\nu}\,\Sigma^{j\,\mu\rho}_{\pm}\,\Sigma^k_{\pm\,\rho\sigma}
=\delta^{ij}\Sigma^k_{\pm\,\nu\sigma}-\delta^{ik}\Sigma^j_{\pm\,\nu\sigma}
+\delta^{jk}\Sigma^i_{\pm\,\nu\sigma}-\eps^{ijk}g_{\nu\sigma}\,,\label{id76}\\
&\epst^{\mu\nu\rho\sigma}\,\Sigma^i_{\pm\,\rho\sigma}
=\pm2\,\im\,\det(e)\,\Sigma^{i\,\mu\nu}_\pm\,.\label{iddual}
\end{align}

\paragraph{Curvature conventions.} We use
$R^\rho{}_{\sigma\mu\nu}=\partial_\mu\Gamma^\rho{}_{\nu\sigma}
-\partial_\nu\Gamma^\rho{}_{\mu\sigma}
+\Gamma^\rho{}_{\mu\lambda}\Gamma^\lambda{}_{\nu\sigma}
-\Gamma^\rho{}_{\nu\lambda}\Gamma^\lambda{}_{\mu\sigma}$,
$R_{\sigma\nu}=R^\mu{}_{\sigma\mu\nu}$, $R=g^{\sigma\nu}R_{\sigma\nu}$, so that the
curvature of the Levi--Civita connection \eqref{LCspin} is
\begin{equation}\label{FGamma}
F^{IJ}(\Gamma)=d\Gamma^{IJ}+\Gamma^I{}_K\wedge\Gamma^{KJ}
=\tfrac12\,e^I{}_\rho\,e^J{}_\sigma\,R^{\rho\sigma}{}_{\mu\nu}\,dx^\mu\wedge dx^\nu \, .
\end{equation}
The curvature $F^{IJ}(\omega)=d\omega^{IJ}+\omega^{I}{}_{K}\wedge\omega^{KJ}$ of a $SO(3,1)$ connection $\omega$ can be decomposed into chiral components as follows
$F^{IJ}(\omega)=P^{IJ}_{+i}F^i_++P^{IJ}_{-i}F^i_-$.  

For $\omega=\Gamma$ we can use the 
$\Sigma_\pm$-basis to extract various curvature tensors. Define $X$ and $Y$ via
\begin{equation}\label{Fdecomp}
F^i_\pm(\Gamma)=\tfrac12\,X^{ij}_\pm\,\Sigma^j_\pm+\tfrac12\,Y^{ij}_\pm\,\Sigma^j_\mp\,,
\qquad
X_-=\overline{X_+}\,,\quad Y_-=\overline{Y_+}\,,
\end{equation}
where $X$ and $Y$ can be  extracted from $F^i_\pm$ via $\Sigma^j_\pm\wedge F^i_\pm=\pm\im\,X^{ij}_\pm\vol$ and
$\Sigma^j_\mp\wedge F^i_\pm=\mp\im\,Y^{ij}_\pm\vol$. The Ricci scalar $R$ , the traceless Ricci tensor $\mathring R_{\mu\nu}$ and Weyl curvature tensor $W$ are encoded in $X$ and $Y$ as follows
\ba\label{XYids}
&& X_\pm=X_\pm^T\,,\qquad
\tr X_\pm=\tfrac{R}{2}\,,\qquad
\mathring X^{ij}_\pm=\tfrac14\,(\Sigma^i_\pm)^{\mu\nu}\,W_{\mu\nu\rho\sigma}\,
(\Sigma^j_\pm)^{\rho\sigma}\,,\nn\\
&& Y^{ij}_\pm=-\tfrac12\,(\Sigma^i_\pm)^{\mu}{}_{\lambda}\,(\Sigma^j_\mp)^{\lambda\nu}\,
\mathring R_{\mu\nu}\, .
\ea
 Thus the $X$-blocks carry the scalar curvature and the chiral Weyl curvature and the $Y$-block
the traceless Ricci curvature.

\section{The (modified)  Plebanski action}\label{sec:pleb}

In this section we define the Plebanski action \cite{Plebanski:1977zz} with Holst term \cite{Holst:1995pc} and explain the modified Plebanski framework as introduced by \cite{Krasnov:2006du,Krasnov:2008fm,Krasnov:2009iy}.

We consider as variables an $\mathfrak{so}(3,1)$-valued two-form
$B^{IJ}_{\mu\nu}$, an $SO(3,1)$ connection $\omega^{IJ}_\mu$ with curvature
$F^{IJ}(\omega)=d\omega^{IJ}+\omega^{I}{}_{K}\wedge\omega^{KJ}$, and Lagrange multipliers $\phi_{IJKL}$ with the symmetries
$\phi_{IJKL}=-\phi_{JIKL}=-\phi_{IJLK}=\phi_{KLIJ}$, $\phi_{IJKL}\eps^{IJKL}=0$, which give 
20 independent components. The Lorentzian Plebanski action with Holst term and cosmological
constant reads
\begin{equation}\label{Spleb}
S_{\rm Pleb}=-\int \Big[\eta_{IJKL}\,+\,\frac{1}{2\gamma}\,\eps_{IJKL}\Big]B^{IJ}\wedge F^{KL}
\;+\;\frac12\int \Big[\phi_{IJKL}\,-\,\frac{\Lambda}{6}\,\eps_{IJKL}\Big]B^{IJ}\wedge B^{KL}\,,
\end{equation}
with $\gamma$ the Barbero--Immirzi parameter. Varying $\phi$ gives the simplicity
constraints
\begin{equation}\label{simpl}
B^{IJ}\wedge B^{KL}=c\;\eps^{IJKL}\,,\qquad
c=-\frac{1}{24}\,\eps_{IJKL}\,B^{IJ}\wedge B^{KL}\,.
\end{equation}
There are four sectors of solutions, which can be parametrised by a tetrad $e^I_\mu$ as
\begin{equation}\label{sectors}
B^{IJ}=\pm\star(e\wedge e)^{IJ}\quad\text{(gravitational)}\,,\qquad
B^{IJ}=\pm(e\wedge e)^{IJ}\quad\text{(topological)}\,,
\end{equation}
with $(e\wedge e)^{IJ}:=e^I\wedge e^J$. On the gravitational sectors one finds
\begin{equation}\label{cval}
\eps_{IJKL}\,B^{IJ}\wedge B^{KL}=+24\,\vol\,,\qquad
c=-\,\vol\,,\qquad \vol:=\det(e)\,d^4x\,,
\end{equation}
so that the $\Lambda$-term in \eqref{Spleb},
$-\frac{\Lambda}{12}\eps_{IJKL}B\wedge B$, evaluates to $-2\Lambda\!\int\!\vol$.  The first term, on the sector $B=+\star e\wedge e$, 
 evaluates to the Palatini--Holst form; altogether
\begin{equation}\label{onshell}
S\big|_{B=\star(e\wedge e)}=-\int\Big[\tfrac12\,\eps_{IJKL}\,e^Ie^JF^{KL}
-\tfrac1\gamma\,e_Ie_JF^{IJ}\Big]-2\Lambda\int\vol\,,
\end{equation}
i.e.\ general relativity with cosmological constant $\Lambda$ in units
$2\kappa=1$: 
in our
conventions the first term equals $+\int\det(e)\,R$  (with $\eps_{0123}=-1$ and mostly-plus signature we have\\
$\tfrac12\eps_{IJKL}e^Ie^JF^{KL}(\Gamma(e))=-\det(e)\,R$) and
$+\tfrac1\gamma\int e_Ie_JF^{IJ}$ is the Holst term, which vanishes when the connection is torsion free.  Thus, with an on-shell connection and $\det(e)>0$, the total is $+\int\sqrt{-g}\,(R-2\Lambda)$, whose field equations
are $G_{\mu\nu}+\Lambda g_{\mu\nu}=0$.  With the choice of signs in (\ref{Spleb}) the chiral weights are
$\gamma_\pm=1\pm\im/\gamma$,  see the remark below (\ref{HolstPi}) about the dependence on conventions.

\paragraph{Modification.} In the modified Plebanski framework \cite{Krasnov:2006du,Krasnov:2008fm,Krasnov:2009iy}, one changes the constraint term by a potential $V(\phi)$ term
\begin{equation}\label{modV}
\frac12\,\phi_{IJKL}\,B^{IJ}\wedge B^{KL}\;\longrightarrow\;
\frac12\Big[\phi_{IJKL}+\frac16 V(\phi)\,\eps_{IJKL}\Big]B^{IJ}\wedge B^{KL}\,,
\end{equation}
where we keep the cosmological constant term separate from the potential. 

The
$\phi$ equation of motion are thus altered to
\begin{equation}\label{eomphi}
  B^{IJ}\wedge B^{KL} -c \, \eps^{IJKL}   = 4 c\,\frac{\partial V(\phi)}
{\partial\phi_{IJKL}}\,,\qquad \text{with}\;\; 
c=-\frac{1}{24}\,\eps_{IJKL}\,B^{IJ}\wedge B^{KL}\, ,
\end{equation}
Note that the left hand side of the first equation in (\ref{eomphi}) is $\mathbb{P}_{\epsilon}(B\wedge B)$  where $\mathbb{P}_{\epsilon}$ is the projector which projects out the totally anti-symmetric part from  a tensor in ${\rm Sym}^2\big(\Lambda^2\mathbb R^{4}\big)$, as  $\phi_{IJKL}\eps^{IJKL}=0$ by definition. The derivative on the right hand side has therefore to satisfy $\mathbb{P}_{\epsilon}(\partial V(\phi)/
\partial\phi_{IJKL})=0$ and the simplicity constraints (\ref{simpl}) can  be expressed as $\mathbb{P}_{\epsilon}(B\wedge B)=0$.

Consider a quadratic potential
\begin{equation}\label{Vquad}
V(\phi)=\tfrac12\,\phi\cdot M\cdot\phi \q .
\end{equation}
parametrised by a symmetric matrix on $\mathbb{P}_{\epsilon}({\rm Sym}^2\big(\Lambda^2\mathbb R^{1,3}\big))$.
If $M$ is invertible on this space, one can solve \eqref{eomphi} for
$\phi$ and reinsert into the right hand side of (\ref{modV}), which gives
\ba
\frac12\Big[\phi+\frac16 V(\phi)\,\eps\Big]\cdot B\wedge B\;\Big|_{\rm sol} \; \; =\;\;
\frac{1}{16 c}  \mathbb{P}_\epsilon(B\wedge B) \cdot M^{-1} \cdot  \mathbb{P}_\epsilon(B\wedge B) \, , 
\ea
that is a potential which suppresses  all the non-simple directions
of $B$. This corresponds to a ``weak" imposition of the simplicity constraints, with masses set by $M$.

Every null eigenvectors $v$ of $M$ lead to sharply imposed constraint, this follows from contracting the $\phi$ equation of motion (\ref{eomphi}) with $v$, which leads to $v\cdot \mathbb{P}_\epsilon(B\wedge B)=0$.


\section{Parametrisation of the $B$-fields by $(e,b)$-data}\label{sec:param}

Here we will introduce a convenient parametrisation of the $B$-fields by a pair of tetrads and a pair of $3\times 3$ matrices. This parametrisation will facilitate the split of the simplicity constraints into two sets, one of which we will impose sharply and the other ``weakly" via the potential introduced above.  After imposing the sharp constraints, the $B$-field degrees of freedom can be packaged into an area metric. The parametrisation we use has the advantage to provide a natural split of the area metric degrees of freedom into length metric and non-length metric degrees of freedom. 

 To define this parametrisation of the $B$-fields we split them into self-dual and anti-selfdual parts 
using the basis \eqref{Pdef}, that is
\begin{equation}\label{Bsplit}
B^{IJ}_{\mu\nu}=P^{IJ}_{+i}\,(B^i_+)_{\mu\nu}+P^{IJ}_{-i}\,(B^i_-)_{\mu\nu}\, .
\end{equation}
As the $P_\pm$ are complex, we have to allow for complex two-forms $B^i_\pm$. The reality property (\ref{Pdef}) for the $P_\pm$ leads to a  reality condition for the $B^i_\pm$
\begin{equation}\label{Brel}
\overline{B^i_+}=B^i_-\,.
\end{equation}
Similarly, the gauge group acting on \eqref{Bsplit} is  a single $SL(2,\mathbb C)$,
whose action on the $-$-sector is determined by that on the $+$-sector.

Following \cite{Freidel:2008ku, Krasnov:2009ik, Speziale:2010cf,Borissova:2022clg}, each sector is parametrised by a (now complex) tetrad and a complex
invertible $3\times3$ matrix,
\begin{equation}\label{param}
B^i_{\pm\,\mu\nu}(b_\pm,e_\pm)=\sigma_\pm\,(b_\pm)^i{}_a\,
\Sigma^a_{\pm\,\mu\nu}(e_\pm)\,
\end{equation}
with reality conditions $\overline{b_+}=b_-$, $\overline{e_+}=e_-$ and sector factors
$\sigma_\pm$ discussed below. 

In the works \cite{Freidel:2008ku, Krasnov:2009ik, Speziale:2010cf,Borissova:2022clg} the two conditions 
$ \det b_\pm=1$ are imposed.  We will {\it not} use these exact conditions, instead we will use the two conditions
\ba\label{phaseCondnew}
 \det(b_+)\det(b_-)=1 \q \text{and} \q {\rm arg} (\det e_+)=0 \q .
\ea
This choice is more in tune with the split of the 20 simplicity constraints into 10  sharply and 10 weakly imposed constraints. With the conditions \eqref{phaseCondnew} the 10 sharp constraints will impose that $g^+=e_+^I e_+^J \eta_{IJ}$ is real, up to the discrete transformations $e_+\to\im^k\,e_+$ left unfixed by \eqref{phaseCondnew}, see below. Note that \eqref{phaseCondnew} implies $\arg\det g^+=2\arg\det e_+=0$ identically, so that the reality of $g^+$ amounts to nine independent conditions; the tenth sharp constraint will be the trace condition ${\rm Im}\,\tr q^+=0$.


\paragraph{Recovering the gauge-invariant content.}  Next we show that we can extract the $SL(2,\mathbb C)$ invariant part of the parametrisation (\ref{param}) from the $B$-fields.

Urbantke's formula (\ref{urb}) applied  to the parametrisation \eqref{param}
(using $\eps_{ijk}b^i{}_lb^j{}_mb^k{}_n=\det(b)\,\eps_{lmn}$) gives
\begin{equation}\label{urbB}
\sigma_\pm^3\,\det(e_\pm)\,\det(b_\pm)\;g^\pm_{\mu\nu}
=\mp\,\frac{\im}{12}\;
\eps_{ijk}\,\epst^{\alpha\beta\gamma\delta}\,
B^i_{\pm\,\mu\alpha}B^j_{\pm\,\beta\gamma}B^k_{\pm\,\delta\nu}\, .
\end{equation}
The determinants $\det(e_\pm)$ and $m_\pm^{ij}:=(b_\pm)^i{}_a(b_\pm)^j{}_b\,\delta^{ab}$ (with $\det m_\pm=(\det b_\pm)^2$, a pure phase by \eqref{phaseCondnew}) can be computed from
\begin{equation}\label{qext}
\sigma_\pm^{2}\, \det(e_\pm)  \;m_\pm^{ij}
=\mp\,\frac{\im}{8}\;B^i_{\pm\,\mu\nu}B^j_{\pm\,\rho\sigma}\,
\tilde \eps^{\mu\nu\rho\sigma}\, .
\end{equation}
We will have to distinguish
 two internal metrics built from $b$: the wedge extraction \eqref{qext} yields
$m^{ij}=(bb^{T})^{ij}$ with gauge ($i$) indices, while the area metric below  uses
\begin{equation}\label{qab}
q^\pm_{ab}:=(b_\pm)^i{}_a(b_\pm)^j{}_b\,\delta_{ij}=(b^{T}b)_{ab}\,,\,\,
\end{equation}
with auxiliary ($a$) indices. 
The two matrices are conjugate, $m=b\,q\,b^{-1}$
(as $bb^{T}=b(b^{T}b)b^{-1}$), hence share all invariants
($\tr\,m^{n}=\tr\,q^{n}$, $\det m=\det q=(\det b)^{2}$, of unit modulus by \eqref{phaseCondnew}) and coincide to first order around
$q=\delta$.

\paragraph{Sectors.} For a gravitational sector solution $B=+ \star e\wedge e$ one finds the chiral decomposition $B^i_\pm=\pm \im \Sigma^i_\pm (e)$ and for a topological sector solution $B=+ e\wedge e$ one has $B^i_\pm =\Sigma^i_\pm (e)$.  We thus have for the $\sigma_\pm$-factors
\begin{equation}\label{sigmavals}
B=\pm\star(e\wedge e):\;\;\sigma_+=\pm\im\,,\;\sigma_-=\mp\im\,;
\qquad
B=\pm(e\wedge e):\;\;\sigma_+=\sigma_-=\pm1\,.
\end{equation}
From now on we will fix $\sigma_+=+\im$ and $\sigma_-=-\im$, and thus $\sigma_+/\sigma_-=-1$.  With $\sigma_\pm$ frozen, the scaling ambiguity of the parametrisation \eqref{param} acts as $(b_+,e_+)\to(\lambda^{-2}b_+,\lambda\,e_+)$, $\lambda\in\mathbb C^\times$ (and by $\bar\lambda$ in the anti-selfdual sector), since $\Sigma(\lambda e)=\lambda^2\Sigma(e)$. The two real conditions \eqref{phaseCondnew} fix it up to $\lambda\in\{1,\im,-1,-\im\}$: $|\det b_+|$ fixes $|\lambda|=1$ and $\arg\det e_+$ fixes $\arg\lambda^4=0$. 

The gravitational sector solution $B=+ \star e\wedge e$ is reached if the parametrisation data satisfy
\begin{equation}\label{simplsol}
(e_+)^I_\mu=(e_-)^I_\mu=e^I_\mu\;\;(\text{real})\,,\qquad
b^i_{+j}=b^i_{-j}=\delta^i_j\,,\qquad \sigma_\pm=\pm\im\,.
\end{equation}
$B$ is left invariant by the scaling ambiguity parametrised by $\lambda$ above. The allowed re-scalings $\lambda\in\{1,\im,-1,-\im\}$ lead therefore to the same gravitational sector solution $B=+ \star e\wedge e$. But we will later demand that $e_+=e_-$ are real (as a sharp simplicity constraint), which excludes $\lambda\in\{\im,-\im\}$.

Note that the first condition in (\ref{simplsol}) now has a double role: it identifies the two sectors
\emph{and}, because $e_-=\overline{e_+}$, it is the \emph{reality condition} for the
tetrad. 

\paragraph{Counting.} $B$ has $36$ real components; $SO(3,1)$ gauge invariance leaves
$30$. In the parametrisation \eqref{param} the data are, per chirality, the complex
tetrad $e_+$ ($16$ complex) and $b_+\in GL(3,\mathbb C)$ ($9$ complex), subject to the two real conditions \eqref{phaseCondnew}. We 
fixed the sector factors to $\sigma_\pm =\pm \im$.  The conjugation
relations $\overline{e_+}=e_-$, $\overline{b_+}=b_-$ determine the $-$ sector. Three
redundancies act at fixed $B^i_+$:

\emph{(i) The stabiliser of $\Sigma_\pm$.} Under an infinitesimal complexified
internal rotation $\delta e^I_\mu=\omega^I{}_J\,e^J_\mu$, with
$\omega\in\mathfrak{so}(3,1)\otimes\mathbb C$ ($6$ complex parameters), the triple
$\Sigma^j_+(e)$ of \eqref{Sigdef} is invariant exactly on the \emph{anti-self-dual}
subalgebra: solving $\delta\Sigma^j_+=0$ for all $j$ gives 
$K_i=-\im J_i$, i.e.\ $\omega$ with vanishing $P_+$-component in the basis
\eqref{Pdef}. Hence
${\rm Stab}\big(\{\Sigma^j_+\}\big)=\mathfrak{sl}(2,\mathbb C)_-$ with complex dimension 3.
 Symmetrically $\mathfrak{sl}(2,\mathbb C)_+$ stabilises $\{\Sigma^j_-\}$. The two
intersect trivially, and each intersects the \emph{real} Lorentz algebra trivially
(a real $\omega$ with $P_+\!\cdot\omega=0$ has $P_-\!\cdot\omega=\overline{P_+\!\cdot\omega}=0$,
hence $\omega=0$), so no physical gauge is lost to the stabiliser.

\emph{(ii) Rotation of the contracted index,} $e_+\to\Lambda_+(R)\,e_+$,
$b_+\to b_+R^{-1}$ with $R=\exp(\theta^a\tau_a) \in SO(3,\mathbb C)$ where $(\tau_a)^b{}_c=\eps_{abc}$. Furthermore $\Lambda_+(R):=\exp\big(\theta^a\,\mathcal T_{+a}\big)$ with $(\mathcal T_{+a})^I{}_J:=P^{IK}_{+a}\,\eta_{KJ}$ satisfies 
$\Sigma^b_+\big(\Lambda_+(R)e\big)=R^b{}_c\,\Sigma^c_+(e)$ and $\Sigma^b_-\big(\Lambda_+(R)e\big)=\Sigma^b_-(e)$.  This gives $3$ complex parameters. The rotation $R^{-1}$ acts on the index
$a$ that is summed in \eqref{param}, i.e.\ on $b$ from the \emph{right}; it must be
distinguished from the residual gauge below, which acts on the free index $i$, i.e.\
on $b$ from the \emph{left}.

\emph{(iii) Scaling,} $(b_+,e_+)\to(\lambda^{-2}b_+,\lambda\,e_+)$ with
$\lambda\in\mathbb C^\times$: $1$ complex, since $\Sigma(\lambda e)=\lambda^2\Sigma(e)$. This scaling redundancy is fixed 
by the two real conditions \eqref{phaseCondnew}, up to
$\lambda^4=1$.

Therefore we have $(16+9)-(3+3+1)=18$ complex parameters, exactly the number of complex components
of $B^i_{+\,\mu\nu}$. The
remaining gauge freedom is the one acting on $B$: this is $SL(2,\mathbb C)$, acting in the $+$ sector as
$SO(3,\mathbb C)$ on the free index, $b_+\to R\,b_+$ (equivalently
$B_+\to R B_+$), with $3$ complex parameters. We thus have 15 gauge invariant complex parameters, that is 30 real parameters.


\section{Area metrics from semi-simple $B$-fields}\label{sec:area}

An area metric on a four-dimensional manifold is a rank-four covariant tensor with the
symmetries $G_{\mu\nu\rho\sigma}=-G_{\nu\mu\rho\sigma}=G_{\rho\sigma\mu\nu}$, i.e.\ a
metric on the space of bivectors.  An area metric assigns areas to infinitesimal parallelograms and
dihedral angles between them, as a length metric does for lengths of and angles between
vectors. Here we restrict to cyclic area metrics, that is we impose the condition $G_{\mu[\nu\rho\sigma]}=0$. This allows to reconstruct the area metric from measuring areas only, as shown in \cite{Borissova:2022clg}.  With this requirement the area metric tensors
carry the same algebraic symmetries as the Riemann curvature tensor and have thus the same number of independent components, namely $20$.  

The signature of an area metric defined on a four-dimensional Lorentzian manifold is $(3,3)$: the space of bi-vectors is six-dimensional and describes planes in tangent space. A basis for this space is given by $e^i \wedge e^j$, $1\leq i< j\leq3$, which provide three spacelike planes,  and $e^0\wedge e^j$, $1\leq j\leq 3$, which provide three timelike planes. 

As the $B$-fields are bi-vectors, it is natural to construct an area metric out of these fields. The $B$-fields have however 30 gauge invariant degrees of freedom, whereas the area metric has only 20. We have therefore to impose 10 constraints on the $B$-fields in order to match the 20 degrees of freedom of the area metric.\footnote{An alternative route, not pursued in this work, is to consider a $SO(3,3)$ gauge symmetry for the $B$-fields, i.e. treat the index pair $(I,J)$ as one super-index (and thus ignore simplicity considerations). $SO(3,3)$ is 15-dimensional, quotienting the $B$-fields by this gauge group would lead to 21 degrees of freedom, which corresponds to area metrics without the cyclicity requirement.}  These will constitute the 10 simplicity constraints, which we will impose sharply on the $B$-fields.

For each sector define the (complex) chiral area
metrics, cyclic by construction\footnote{The sector normalisation $\sigma_\pm^{-2}$ is included in the definition so that the area metric is built from the sector-stripped fields
$\sigma_\pm^{-1}B^i_\pm=b^i{}_a\Sigma^a_\pm$; without it, the extra internal stars in
$B=\star(e\wedge e)$ would produce an overall sign, and the length-metric-induced area
metric would come out as $-(gg-gg)$ on the Lorentzian gravitational sector.}:
\begin{equation}\label{Adef}
A^\pm_{\mu\nu\rho\sigma}=\frac{1}{2\,\sigma_\pm^{2}}\Big[
B^i_{\pm\,\mu\nu}B^j_{\pm\,\rho\sigma}\,\delta_{ij}
-\frac{1}{4!}\,B^i_{\pm\,\mu'\nu'}B^j_{\pm\,\rho'\sigma'}\delta_{ij}\,
\epst^{\mu'\nu'\rho'\sigma'}\,\epst_{\mu\nu\rho\sigma}\Big]\,,
\end{equation}
which also satisfy the reality relation
\begin{equation}\label{Areal}
\overline{A^+_{\mu\nu\rho\sigma}}=A^-_{\mu\nu\rho\sigma}\,,
\end{equation}
and thus $A=A^+ +A^-=2 \mathrm{Re}\,A^+$ is real.

In the parametrisation \eqref{param}:
\begin{equation}\label{A59}
A^\pm_{\mu\nu\rho\sigma}
=\frac{1}{2}\Big[
q^\pm_{ab}\,\Sigma^a_{\pm\,\mu\nu}(e_\pm)\,\Sigma^b_{\pm\,\rho\sigma}(e_\pm)
\;\mp\;\frac{\im}{3}\,\tr(q^\pm)\,\det(e_\pm)\tilde \eps_{\mu\nu\rho\sigma}\Big]\, .
\end{equation}

We split $q^\pm$ into its trace and traceless parts,
\begin{equation}\label{qsplit}
q^\pm_{ab}=\tfrac13\,q^\pm_\tr\,\delta_{ab}+\mathring q^\pm_{ab}\,,
\qquad \delta^{ab}\mathring q^\pm_{ab}=0\,,
\end{equation}
and introduce the length metric induced area metrics
\ba
G^\pm_{\mu\nu\rho\sigma}:= g^\pm_{\mu\rho}g^\pm_{\nu\sigma}
-g^\pm_{\mu\sigma}g^\pm_{\nu\rho} \, .
\ea

Using \eqref{id74} in the form
$\delta_{ab}\Sigma^a_\pm\Sigma^b_\pm=G^\pm\mp\im\,\eps$ we can express $A^\pm$ as 
\begin{equation}\label{Aexplicit}
A^\pm_{\mu\nu\rho\sigma}
=\frac{q^\pm_\tr}{6}G^\pm_{\mu\nu\rho\sigma}
+\frac12\,\mathring q^\pm_{ab}\,
\Sigma^a_{\pm\,\mu\nu}(e_\pm)\,\Sigma^b_{\pm\,\rho\sigma}(e_\pm)\; , 
\end{equation}
where the explicit $\epsilon$ term in (\ref{A59}) cancelled with the $\epsilon$ term resulting from using (\ref{id74}).

To re-write the last term in (\ref{Aexplicit}) into terms which are real for real tetrads, we split $\sigma$ into boost and rotation part
\begin{equation}\label{EFdef}
{\cal E}^a_{\pm\,\mu\nu}:=e^0_{\pm\mu}e^a_{\pm\nu}-e^0_{\pm\nu}e^a_{\pm\mu}\,,\qquad
{\cal B}^a_{\pm\,\mu\nu}:=\eps^a{}_{bc}\,e^b_{\pm\mu}e^c_{\pm\nu}\,,
\qquad \Sigma^a_\pm=\pm\im\,{\cal E}^a_\pm+{\cal B}^a_\pm\,. 
\end{equation}

With this notation the $\delta$ contraction of $\Sigma_\pm \Sigma_\pm$ amounts to
\begin{equation}\label{EFcov}
\delta_{ab}\big({\cal B}^a_\pm {\cal B}^b_\pm-{\cal E}^a_\pm {\cal E}^b_\pm\big)
=(g_{\mu\rho}g_{\nu\sigma}
-g_{\mu\sigma}g_{\nu\rho})\, , 
\qquad
\delta_{ab}\big({\cal E}^a_\pm {\cal B}^b_\pm+{\cal B}^a_\pm {\cal E}^b_\pm\big)
=-\,\eps_{\mu\nu\rho\sigma}\,,
\end{equation}
which together reassemble \eqref{id74}. 

For the last term in \eqref{Aexplicit} we contract $\Sigma_\pm \Sigma_\pm$ with a traceless $\mathring q^\pm_{ab}$:
\begin{equation}\label{Aexplicit2}
\tfrac12\mathring q^\pm_{ab}\Sigma^a_\pm\Sigma^b_\pm
=\tfrac12\,\mathring q^\pm_{ab}\Big[
{\cal B}^a_{\pm\mu\nu}{\cal B}^b_{\pm \rho\sigma}-{\cal E}^a_{\pm \mu\nu}{\cal E}^b_{\pm\rho\sigma}
\;\pm\;\im\big({\cal E}^a_{\pm \mu\nu}{\cal B}^b_{\pm\rho\sigma}+{\cal B}^a_{\pm\mu\nu}{\cal E}^b_{\pm \rho\sigma}\big)\Big]\, .
\end{equation}

This allows us to express $\R(A^+)$ and $\I(A^+)$ as
\ba \label{ReAp}
2\,\R(A^+) &=&
\;\;\;\,\, \tfrac{1}{3}\,\R(q^+_\tr) \R(G^+) \,\,-\,\, \tfrac{1}{3}\,\I(q^+_\tr) \I(G^+) \nn\\  
&&  \,+\,  \R(\mathring q^+_{ab}) \R\!\left( {\cal B}^a_+{\cal B}^b_+-{\cal E}^a_+{\cal E}^b_+\right) \,-\,   \R(\mathring q^+_{ab})\I\!\left(  {\cal B}^a_+{\cal E}^b_+ + {\cal E}^a_+ {\cal B}^b_+        \right) \nn\\  
&& \,-\,  \I(\mathring q^+_{ab})   \I\!\left(     {\cal B}^a_+{\cal B}^b_+-{\cal E}^a_+{\cal E}^b_+     \right) \,   -\,     \I(\mathring q^+_{ab}) \R\!\left(  {\cal B}^a_+{\cal E}^b_+ + {\cal E}^a_+ {\cal B}^b_+      \right)   
\ea
and 
\ba \label{ImAp}
2\,\I(A^+) &=&
\;\;\;\,\, \tfrac{1}{3}\,\R(q^+_\tr) \I(G^+) \,\,+\,\, \tfrac{1}{3}\,\I(q^+_\tr) \R(G^+) \nn\\  
&&  \,+\,   \R(\mathring q^+_{ab}) \I\!\left( {\cal B}^a_+{\cal B}^b_+-{\cal E}^a_+{\cal E}^b_+\right) \,+\,   \R(\mathring q^+_{ab})\R\!\left(  {\cal B}^a_+{\cal E}^b_+ + {\cal E}^a_+ {\cal B}^b_+        \right) \nn\\  
&& \,+\,  \I(\mathring q^+_{ab})   \R\!\left(     {\cal B}^a_+{\cal B}^b_+-{\cal E}^a_+{\cal E}^b_+     \right) \,   -\,  \I(\mathring q^+_{ab}) \I\!\left(  {\cal B}^a_+{\cal E}^b_+ + {\cal E}^a_+ {\cal B}^b_+      \right)   \, .
\ea


\paragraph{Three properties of $A$.}  Let us point out three properties for the area metric $A$. To this end we remind the reader of identity (\ref{id75})
\ba\label{id75copy}
\Sigma^i_{\pm\,\mu\nu}\,\Sigma^j_{\pm\,\rho\tau}\,g^{\nu\tau}
=\delta^{ij}g_{\mu\rho}+\eps^{ij}{}_{l}\,\Sigma^l_{\pm\,\mu\rho}
\ea
 Here we assume that $e_+=e_-=e$ is real, greek indices are raised and lowered with respect to the associated real metric $g$.
 
\emph{(a) Chiral blocks.} Starting with $A=A^+ + A^-$ in the form (\ref{Aexplicit}) we contract $A$ with $\Sigma_{\pm}^{a\,\mu\nu}\Sigma_{\pm}^{b\,\rho\sigma}$.  Contracting \eqref{id75copy} with $g^{\mu\rho}$ shows orthogonality
$\Sigma^a_{\pm\mu\nu}\Sigma_\pm^{b\,\mu\nu}=4\delta^{ab}$,  furthermore we have that
$\Sigma^a_{+\mu\nu}\Sigma_-^{b\,\mu\nu}=0$. For the contraction of (\ref{Aexplicit}) we therefore obtain
\begin{equation}\label{Ablocks}
A_{\mu\nu\rho\sigma}\,\Sigma_{\pm}^{a\,\mu\nu}\Sigma_{\pm}^{b\,\rho\sigma}
=\frac83\,\R(q^\pm_\tr)\,\delta^{ab}+8\mathring q^{\pm ab}\,,
\qquad
A_{\mu\nu\rho\sigma}\,\Sigma_+^{a\,\mu\nu}\Sigma_-^{b\,\rho\sigma}=0\,.
\end{equation}

Thus, in the basis of bi-vectors given by $\{\Sigma_+^{a\,\mu\nu}, \Sigma_-^{b\,\mu\nu}\}_{a,b=1,2,3}$,  $A$ is block diagonal.  The (anti-) self-dual
block \eqref{Ablocks} has real trace $\R(q^+_\tr)=\R(q^-_\tr)$.

\emph{(b) Ricci trace.}  The right-hand side of \eqref{id75} is annihilated by a
symmetric traceless $\mathring q^\pm_{ab}$, so the second term for $A^\pm$ in \eqref{Aexplicit} is
Ricci traceless.  With $g^{\mu\rho}G_{\mu\nu\rho\sigma}=3g_{\nu\sigma}$,
\begin{equation}\label{Atrace}
g^{\mu\rho}A_{\mu\nu\rho\sigma}=\R(q^+_\tr)\,g_{\nu\sigma}\,.
\end{equation}
Thus $W:=A-\tfrac13\R(q^+_\tr)G$ is the Ricci-traceless (Weyl type) part of $A$.
 
\emph{(c) Homogeneity.} At fixed $q^\pm$, both $G$ and $\Sigma^a\Sigma^b$ are
homogeneous of degree four in the tetrad, hence so is $A$.

\paragraph{Reisenberger's condition.} In Euclidean signature, Reisenberger \cite{Reisenberger:1998fk} showed that the simplicity conditions are (locally, on the gravitational sector) equivalent to $A^+=A^-$.  By \eqref{Areal} the Lorentzian statement acquires a
reformulation as reality constraint
\begin{equation}
A^+_{\mu\nu\rho\sigma}=A^-_{\mu\nu\rho\sigma}
\quad\Longleftrightarrow\quad
\mathrm{Im}\,A^+_{\mu\nu\rho\sigma}=0\,:
\end{equation}
which suggests that in Lorentzian signature, the simplicity constraints are equivalent to the reality of the self-dual area
metric.

Indeed lets consider a perturbation $g^+=\mathbb{g}^+ +h^+$, with $\mathbb{g}^+=\mathbb{g}$ real and $h^+$ purely imaginary (to describe a violation of the simplicity constraints) and $q^+=\mathbb{I}+\chi^+$ with $\chi^+_{ab}$ symmetric. The conditions  \eqref{phaseCondnew} on the parametrisation included $|\det q^+|=1$.

With $\det(\mathbb{I}+\chi^\pm)=1+\tr(\chi^\pm)+O(\chi^2)$ the condition $\det(\mathbb{I}+\chi^+)\det(\mathbb{I}+\chi^-)$ leads to  ${\rm Re}\tr(\chi^+)=O(\chi^2)$. 
But the imaginary part of the trace  ${\rm Im}\tr(\chi^+)=:{\rm Im}\chi^+_\tr$ is a first order variable. 

Furthermore the normalisation conditions in \eqref{phaseCondnew} demanded that ${\rm Im}\det g^+=0$. We have thus  $\mathbb{g}^{\mu\nu}  h^+_{\nu\mu}=0$ modulo second order terms  (indices are lowered (and raised) with the (inverse) background metric). 

 We have to first order in perturbation variables 
\ba \label{ImAppert}
2\,\delta \I(A^+)_{\mu\nu\rho\sigma} &=&
   \delta \I(G^+)_{\mu\nu\rho\sigma} \,+\, 
    \tfrac{1}{3}{\rm Im}\chi^+_\tr \mathbb{G}_{\mu\nu\rho\sigma}+ \I\big[\mathring\chi^+_{ab}\,\mathbb{\Sigma}^a_+\mathbb{\Sigma}^b_+\big] \ , 
\ea
where $\mathbb{\Sigma}^a_\pm,{\mathbb G}$ are $\Sigma_\pm,G$ evaluated on the background tetrad and therefore real; and $\mathring\chi^+$ is the traceless part of $\chi^+$.
 

The vanishing of (\ref{ImAppert}) gives 20 conditions for 10 variables $h^+_{\mu\nu}$, which are subject to one condition $\mathbb{g}^{\mu\nu}  h^+_{\nu\mu}=0$: 5 variables $ \R(\mathring \chi^+_{ab})$,  5 variables $\I(\mathring\chi^+)$ and one variable ${\rm Im}\chi^+_\tr$. That is we have 20+1 conditions for 21 variables.
 
Now the $\mathring\chi$-terms in (\ref{ImAppert}) vanish when contracted with ${\mathbb g}_+^{\mu\rho}$ or ${\mathbb g}_+^{\nu\sigma}$. This follows from (\ref{id75}), which states $g^{\mu\rho}\,\Sigma^a_{\pm\,\mu\nu}\Sigma^b_{\pm\,\rho\sigma}
=\delta^{ab}g_{\nu\sigma}+\eps^{abc}\,\Sigma^c_{\pm\,\nu\sigma}$.  The right hand side of this equation vanishes if we contract with a symmetric traceless $\mathring\chi^+_{ab}$: the $\eps$-term drops by symmetry and the $\delta$-term due to  $\mathring\chi^+_{ab}$ being traceless.  

As $h^+$ is assumed to be purely imaginary, we have for the remaining terms in (\ref{ImAppert})
\ba\label{eq54a}
\delta \I(G^+)_{\mu\nu\rho\sigma}\,+\, 
    \tfrac13 {\rm Im}\chi^+_\tr \mathbb{G}_{\mu\nu\rho\sigma} \,\,=\,\,\I(h^+_{\mu\rho}{\mathbb g}_{\nu\sigma} + {\mathbb g}_{\mu\rho}{h}^+_{\nu\sigma} -h^+_{\mu\sigma}{\mathbb g}_{\nu\rho} -{\mathbb g}_{\mu\sigma}{h}^+_{\nu\rho})\,+\, 
    \tfrac13 {\rm Im}\chi^+_\tr \mathbb{G}_{\mu\nu\rho\sigma} \; ,
\ea
whose contraction with ${\mathbb g}^{\mu\rho}$ gives (using $h^+_{\mu\rho}{\mathbb g}^{\mu\rho}=0$)
\ba\label{eq54b}
\left(\delta \I(G^+)_{\mu\nu\rho\sigma}\,+\, 
    \tfrac13 {\rm Im}\chi^+_\tr \mathbb{G}_{\mu\nu\rho\sigma}\right){\mathbb g}^{\mu\rho}  \,\,=\,\,\,2 h_{\nu\sigma} + {\rm Im}\chi^+_\tr \,{\mathbb g}_{\nu\sigma} \; .
\ea
Contracting additionally with ${\mathbb g}_+^{\nu\sigma}$ we can conclude that ${\rm Im}\chi^+_\tr=0$, using this in (\ref{eq54b}) shows that also $h^+_{\nu\sigma}=0$.

The  trace-free part of $\delta \I(A^+)$ involves only $\R(\mathring\chi^+)$ and $ \I(\mathring\chi^+)$, and its vanishing (for a non-degenerate $\mathbb{g}^+$) implies $\mathring\chi^+=0$. 
%
To see this we contract the trace-free combination in \eqref{ImAppert}
%
\begin{eqnarray}\label{chiSD}
\I\big[\mathring\chi^+_{ab}\,\mathbb{\Sigma}^a_+\mathbb{\Sigma}^b_+\big]
&=& \tfrac{1}{2\im} \left[ \mathring\chi^+_{ab}\,\mathbb{\Sigma}^a_+\mathbb{\Sigma}^b_+ -\mathring\chi^-_{ab}\,\mathbb{\Sigma}^a_-\mathbb{\Sigma}^b_- \right] \, , 
\end{eqnarray}
 with $\mathbb{\Sigma}_+^{c\,\mu\nu}\mathbb{\Sigma}_+^{d\,\rho\sigma}$. We use that 
$\mathbb{\Sigma}^a_{+\mu\nu}\mathbb{\Sigma}_+^{c\,\mu\nu}=4\delta^{ac}$ and $\mathbb{\Sigma}^a_{-\mu\nu}\mathbb{\Sigma}_+^{c\,\mu\nu}=0$.
Hence  the conjugate term $\mathring\chi^-_{ab}\Sigma^a_-\Sigma^b_-$ drops out of the contraction and the remaining term leads to $-8\,\im\,\mathring\chi^{+\,cd}$. Thus, demanding that (\ref{ImAppert}) vanishes we can conclude that $\mathring\chi^+=0$.  Together with ${\rm Im}\chi^+_\tr=0$ from the trace sector, $\chi^+=0$ at first order, and also $\chi^-=\overline{\chi^+}=0$.

Thus reality of $A^+$ is, to first order around the gravitational
sector, equivalent to the simplicity
constraints.

We note that a configuration $e^+=e^{\im\alpha}\hat e$ with real tetrad $\hat e$ and $q^+=\delta$ leads to
${\rm Im}\,A^+(e^{\im\alpha}\hat e,\delta)\propto\sin(4\alpha)$, as one can see directly from (\ref{ImAp}). So $\I(A^+)=0$ admits
the discrete solutions $\alpha=n\pi/4$. These do describe the four Plebanski sectors,
$\alpha=0,\tfrac\pi2$ giving $B=\pm\star(\hat e\wedge \hat e)$ and $\alpha=\tfrac\pi4,
\tfrac{3\pi}4$ the topological sector $B=\pm\,\hat e\wedge \hat e$.

\paragraph{The semi-constrained theory.}  The $B$-fields have 30 gauge invariant parameters, whereas cyclic area metrics have 20 independent components.  We look for 10 conditions on the parameters\\ $(e_+,e_-, b_+,b_-)$, so that the area metric $A=A^++A^-$ provides at least locally a truthful encoding of the remaining 20 gauge invariant parameters.

As in \cite{Borissova:2022clg}, we will ensure that  the
ten conditions
\begin{equation}\label{gg}
g^+_{\mu\nu}=g^-_{\mu\nu}
\qquad\Longleftrightarrow\qquad
g_{\mu\nu}:=g^+_{\mu\nu}\;\;\text{real}\,
\end{equation}
hold, which in Lorentzian signature amount to the reality condition for the length metric
(the tetrads may then be taken real by a complexified gauge transformation). Under
the normalisation \eqref{phaseCondnew} one has $\arg\det g^+=0$ identically, so
\eqref{gg} contains only \emph{nine} new conditions.  

As tenth
condition we will impose ${\rm Im}\,\tr q^+=0$.  This is the constraint obtained by varying the Lagrange multiplier $\varphi_0$ (the spin 0 or trace part of the Lagrange multiplier field $\phi_{IJKL}$) in 
Section~\ref{sec:eff}.  
This choice also appears naturally from the discussion of Reisenberger's condition above:  
the trace parts of $\delta {\rm Im}A^+=0$ imposed that $h^+$ had to be real (or vanishing with the assumption that $h^+$ is imaginary) {\it and} that $t:={\rm Im}\tr(\chi^+)=0$. 
The 10 trace parts of the $\delta {\rm Im}A^+=0$ equations included $t=0$ because the length metric perturbations already satisfied $\tr({ \mathbb g}^{-1} h^+)=0$ because of our normalisation choice \eqref{phaseCondnew}.

The remaining $20$ real degrees of freedom are the real
length metric ($10$) and the complex symmetric matrix $q^+$ with $|\det q^+|=1$ and
${\rm Im}\,\tr q^+=0$. This gives $12-2=10$ real parameters, with $q^-$ determined by $q^+$ as $q^-=\overline{q^+}$.

For a real tetrad the area metric $A=2\R(A^+)$ simplifies to
\ba \label{ReApReg}
A &=&
\tfrac13\,\R(q^+_\tr) G
  \,+\,  \R(\mathring q^+_{ab}) \left( {\cal B}^a{\cal B}^b-{\cal E}^a{\cal E}^b\right)  
   -\,     \I(\mathring q^+_{ab})\left(  {\cal B}^a{\cal E}^b + {\cal E}^a {\cal B}^b      \right)   \, \nn\\
  &=&  \tfrac16\,(q^+_\tr+ q^-_\tr) G +     \tfrac12\mathring q^+_{ab}\Sigma^a_+(e)\Sigma^b_+(e) + \tfrac12\mathring q^-_{ab}\Sigma^a_-(e)\Sigma^b_-(e) \nn\\
  &=&  \tfrac13 \R(q^+_\tr)G\q\,\,\,+\,\,\q W \q .
\ea
For 
$q^\pm_{ab}=\delta_{ab}$ the area metric is the one induced by the length metric:
\begin{equation}\label{induced}
A_{\mu\nu\rho\sigma}(e,\delta,\delta)
=g_{\mu\rho}g_{\nu\sigma}-g_{\mu\sigma}g_{\nu\rho}\, .
\end{equation}
Since we obtain such a length metric induced area metric from the area metric definition (\ref{Adef}) with simple $B$-fields. we will refer to such area metrics as simple area metrics.

If a certain non-degeneracy condition is satisfied, the area metric defines a faithful map for perturbations $(h,\chi^\pm)$ around $(\mathbb{g},\mathbb{q}^\pm)$ with real length metric $\mathbb{g}_{\mu\nu} $ and ${\rm Im}\,\tr \mathbb{q}^+=0$. This will be proven in Section~\ref{sec:faithful}.

\section{Reconstruction of $(e,b)$-parameters from the area metric and faithfulness of the area metric map}\label{sec:faithful}

It will be more convenient to define area metric actions via their parametrisation with the tetrad $e$ and the $3\times3$ matrices $b_\pm$, rather than as functions of the area metric directly. In order to do this we have to show that at least locally, the map between these parameters (with our normalisation choices and satisfying the sharp constraints specified above and modulo redundancies in the parametrisation discussed in Section \ref{sec:param}) and the area metric is an isomorphism. We will first present an algorithm to construct $(e,b_\pm)$ parameters from an area metric. We will then prove that the area metric defines, locally, a faithful map for perturbations $(h,\chi^\pm)$ around $(\mathbb{g},\mathbb{q}^\pm)$ with a real  length metric $\mathbb{g}_{\mu\nu} $ and ${\rm Im}\,\tr \mathbb{q}^+=0$.

\paragraph{Reconstruction of $(e,b_\pm)$ from the area metric and naturalness of the
 split into length metric and non-length metric degrees of freedom.} 
Constructing area metrics out of $B$ fields, which are parametrised by a (real) tetrad $e$ and complex $3\times 3$ matrices $b_\pm$ gives area metrics in this parametrisation. This parametrisation includes a tetrad and hence length metric, and therefore suggest a split of an area metric in a length metric induced part (or simple part) and a remainder. Indeed, (\ref{ReApReg}) provides such a split.

In the following we will show how this split can be constructed from the area metric itself.  Thus $(e,b)$ provide a natural split of area metric degrees of freedom into simple (length-metric induced) and non-simple parts.   One procedure is to solve equations (\ref{Aexplicit}), however these constitute 20 non-linear equations for 20 variables (ignoring gauge choices), and thus cannot (in general) be solved algebraically.  The following construction relies on an eigenvalue decomposition of a $6\times 6$ matrix, and thus isolates the algebraic unsolvability to finding the roots of the characteristic polynomial of degree 6.

Given a cyclic area metric $A$ consider the following 
endomorphism of the six-dimensional space 
$\Lambda^2$ of two-forms\footnote{A similar endomorphism $\epst^{\mu\nu\kappa\lambda}A_{\kappa\lambda\rho\sigma}$ on the space of bi-vectors underlies the algebraic Segre--Jordan classification of area metrics in \cite{Schuller:2009hn}.}
\begin{equation}\label{endo}
(\hat A\alpha)_{\mu\nu}:=\tfrac12\,
A_{\mu\nu\kappa\lambda}\,\epst^{\kappa\lambda\rho\sigma}\alpha_{\rho\sigma}\,,
\end{equation}
We remind the reader of our conventions for the Levi-Civita symbol: $\epst^{0123}=\epst_{0123}=1$,\\ so
$\epst^{\mu\nu\rho\sigma}\epst_{\mu\nu\alpha\beta}=2(\delta^\rho_\alpha\delta^\sigma_\beta
-\delta^\rho_\beta\delta^\sigma_\alpha)$ with no signature sign.

The map $\hat A$ is
self-adjoint for the wedge pairing of two-forms, $\langle\alpha,\beta\rangle:=\tfrac12
\epst^{\mu\nu\rho\sigma}\alpha_{\mu\nu}\beta_{\rho\sigma}$, 
since $\langle \hat A \alpha,\beta\rangle=\langle  \alpha,\hat A\beta\rangle$ is the symmetric pairing of $A$ with the bi-vectors $\tfrac12 \epst \alpha$ and $\tfrac12 \epst \beta$.

Consider an area metric $A$ induced by the length metric $g$. Then $\hat A=2\sqrt{-g}{*}_g$, and as ${*}_g{*}_g=-\mathbb{1}$ the square
\begin{equation}
\hat A^2=4\,\det(g)\cdot\mathbb{1}\;<\;0\, 
\end{equation}
is strictly negative and a multiple of the identity operator. 
Thus, up to normalisation, $\hat A$ is a complex structure on $\Lambda^2$, whose
$\pm\im$-eigenspaces are the self-dual and anti-self-dual spaces of $g$. Since the
Hodge operator on two-forms determines the conformal class of the metric, $A(g)$
determines $g$ up to scale --- and the scale is fixed by $\hat A^2$.

We will apply the same strategy to a non-simple area metric $A$. As $\hat A$ is self-adjoint with respect to the wedge pairing, the spectrum of $\hat A$ comes in complex conjugated pairs. We have to split the eigenvalues into two sets, which will then define the split of $\Lambda^2$ into a self-dual space $\Lambda^2_+$ and anti-selfdual space $\Lambda^2_-$. 
Here we have a discrete ambiguity in how to make that choice.  For an area metric constructed from $B_\pm$-fields satisfying the sharp simplicity constraints, the spectrum of $\hat A$ restricted to the space spanned by $B_+^j$ is given by 
$\lambda_i=2\imath \det(e) \mu_i$, where $\mu_i$ are the eigenvalues of 
$m_+=b_+b_+^T$. Thus for $b$ sufficiently near the identity, that is for area metrics sufficiently near a simple area metric, and for a positively oriented tetrad, the eigenvalues with positive imaginary part pick out $\Lambda_+$ as spanned by the $B_+^j$ basis. 

We therefore restrict our considerations to area metric which are sufficiently near a simple one, more specifically to area metrics for which $\hat A$ has three eigenvalues with strictly positive imaginary part (and hence also three eigenvalues with strictly negative imaginary part).\footnote{\cite{Schuller:2009hn}  discusses the types of area metrics arising from an algebraic Segre--Jordan classification, which includes as a criterium the degeneracy and  reality of eigenvalues. They point out that not all area metrics are physically viable and that one might want to exclude certain types of area metrics. One possibility is to restrict to area metrics whose $\hat A$ eigenvalues come in two complex-conjugated triples and connect smoothly to a length metric induced area metric without crossing the case where one or more of the eigenvalues are real. An alternative way to proceed is to try all possible choices of subdividing the eigenvalues into two sets of three eigenvalues, reconstruct the $B$-fields from each choice, then compute the area metrics and compare with the original one.}
  We then define
$\Lambda^2_+$ ($\Lambda^2_-$) as the union of eigenspaces of eigenvalues with a positive (negative) imaginary part. These eigenspaces are automatically orthogonal with respect to $\langle \cdot, \cdot\rangle$ and hence $A$-orthogonal, so we can split $A=A^+\oplus A^-$. 

Given any basis $\beta^i_\pm$ of $\Lambda^2_\pm$ we can construct the Urbantke metric $g^{U\pm}_{\mu\nu}\propto\eps_{ijk}
\epst^{\alpha\beta\gamma\delta}\beta^i_{\mu\alpha}\beta^j_{\beta\gamma}\beta^k_{\delta\nu}$, which determines the conformal class of the length metrics, with respect to which $\Lambda^2_\pm$ provides a splitting into self-dual and anti-selfdual forms.  For an area metric whose spectrum splits into two complex-conjugate triples, the Urbantke metric is real up to a phase.\footnote{A three-dimensional subspace
$\Lambda\subset\Lambda^2_{\mathbb C}$ with $\langle\Lambda,\bar\Lambda\rangle=0$ and
$\langle\cdot,\cdot\rangle$ nondegenerate on it is the self-dual space of a
real conformal Lorentzian metric: such subspaces form a nine-dimensional family, a single orbit of $SO(3,3)$ with stabiliser $SO(3,\mathbb C)\cong
SL(2,\mathbb C)$, and the self-dual spaces of real metrics form the same orbit. Since
the eigenspace $\Lambda^2_+$ of $\hat A$ is $\langle\cdot,\cdot\rangle$-orthogonal
to its conjugate, its Urbantke metric is therefore real up to a phase.}

Since $\hat A$ is self-adjoint with respect to $\langle \cdot, \cdot \rangle$ the eigenvectors for different eigenvalues will be orthogonal. (If eigenvalues agree we will choose an orthogonal subbasis.) We will additionally normalise these eigenvectors (with respect to $\langle \cdot, \cdot \rangle$), and denote them by $\beta^i_\pm$, with $\beta^i_-=\overline{\beta^i_+}$. We can then construct the area metric
\ba
A(\beta) = \sum_{i} \left(\lambda_i \beta^i_+ \beta^i_+ +  \overline{\lambda}_i \beta^i_- \beta^i_- \right)\, , 
\ea
where $\lambda_i, i=1,2,3$ is the triple of eigenvalues of ${\cal A}$ with positive imaginary part. By construction $\hat A(\beta)=\hat A$ and hence $A(\beta)=A$.

Let us check that $A(\beta)$ has indeed no acyclic part:
\ba
\tilde \epsilon \cdot A(\beta) &=&2 \sum_{i} \left(\lambda_i \langle\beta^i_+ \beta^i_+\rangle +  \overline{\lambda}_i \langle\beta^i_- \beta^i_-\rangle \right)
\,\,=\,\, 2 \sum_{i}\left( \lambda_i +  \overline{\lambda}_i \right) \,\,=\,\,2 \tr \hat A\, .
\ea
Now the trace of $\hat A$ is proportional to the acyclic part of the area metric $A$: $2 \tr \hat A = \tilde \epsilon^{\mu\rho\sigma\kappa} A_{\mu\rho\sigma\kappa}$. This vanishes as we assumed $A$ to be cyclic. 

We can therefore define $B^i_+=\sqrt{-2 \lambda_i} \beta_+^i$ and $B^i_-=\overline{\sqrt{-2\lambda_i}} \beta_-^i$, which satisfy
\ba
B^i_{+\mu\nu} B^j_{+\rho\sigma} \epst^{\mu\nu\rho\sigma} \,=\, -4\lambda_i \delta^{ij} \q .
\ea
Thus, according to (\ref{qext}) (with $\sigma^2_\pm=-1$) we have 
\ba\label{lambdam}
\det(e_+) m_+^{ij} =- \tfrac{\im}{2} \lambda_i \delta^{ij}\, .
\ea
   We have therefore ${b_+^i}_a$ in an internal gauge, in which $m_+=b_+b_+^T$ is diagonal; this does not mean that $q_+=b_+^T b_+$ is diagonal.

The scaling normalisation (\ref{phaseCondnew}) demands $|\det(m_+)|=1$, from $\det(e_+) m_+^{ij} = \tfrac{\im}{4} \lambda_i \delta^{ij}$ we can therefore determine 
\ba
|\det e| \,=\, \tfrac12 \bigg(\prod_{i=1,2,3} |\lambda_i|\bigg)^{\frac{1}{3}} \, .
\ea
The phase of $e$ is also subject to the normalisation condition  (\ref{phaseCondnew}):  ${\rm arg} (\det e_+)=0$, which fixes the phase modulo $\pi/2$ factors.

This also provides a straightforward criterium for when an area metric $A$ is simple. Above we have seen, that for a simple $A$ the map $\hat A^2$ has a completely degenerate spectrum with one strictly negative eigenvalue, and thus $\hat A$ has two  conjugated, purely imaginary and thrice degenerate eigenvalues.  This criterium ($\hat A^2$ has a completely degenerate spectrum with one strictly negative eigenvalue)  is also sufficient: it follows from (\ref{lambdam}) that $m_\pm$ is proportional to the identity matrix, and hence one can choose $b_\pm$ to be equal to the identity matrix also.

We note that this extraction of a length metric from an area metric differs from the one discussed in \cite{Punzi:2006hy,Punzi:2006nx}, which is based on the principal Fresnel polynomial defined by the area metric.  The construction in \cite{Punzi:2006hy,Punzi:2006nx} requires the inverse to the area metric to have an acyclic part. The differential of this acyclic part is then chosen as a canonical cotangent section. The construction of \cite{Punzi:2006hy,Punzi:2006nx} is not applicable to simple area metrics and introduces a (mild) non-locality, whereas the construction in this paper is completely local and does not require a non-vanishing acyclic part of the inverse area metric.

\paragraph{Faithfulness of the area metric map.}  Here we show that for parameters with $g^+= g^-=g$ and $\I \tr q^+=0$  the map $(g,q^\pm)\mapsto A$ is faithful at the
linearised level: its differential has vanishing kernel on perturbations whose
length-metric part is \emph{real}. Since a real cyclic area metric has $20$ components
and the perturbations number $10+10=20$, the linearised map is then an isomorphism. These considerations do not rely on finding a diagonalisation of the area metric, and are therefore more direct. They provide a non-degeneracy criterion, which is closely related to the condition on the eigenvalues $\hat A$ discussed above.
 
\paragraph{Set-up.} The background is a real tetrad with real length metric
$\mathbb{g}_{\mu\nu}$, together with $\mathbb{q}^\pm$ complex symmetric,
$\overline{\mathbb{q}^+}=\mathbb{q}^-$, normalised as in \eqref{phaseCondnew} by
$|\det\mathbb{q}^\pm|=1$, with the additional condition ${\rm Im}\,\tr\mathbb{q}^\pm=0$. Furthermore ${\mathbb \Sigma}^a_\pm$
denotes the Plebanski two-forms of the \emph{background} tetrad, and $\mathbb{A}, \mathbb{W}, \mathbb{G}$
denote $A,W$, $G$ evaluated on the background (with $W=A-\tfrac13\R(q^+_\tr)G$ the Weyl part of $A$). We perturb
\begin{equation}\label{pert}
g_{\mu\nu}=\mathbb{g}_{\mu\nu}+h_{\mu\nu}\,,\qquad
q^\pm_{ab}=\mathbb{q}^\pm_{ab}+\chi^\pm_{ab}\,,
\end{equation}
with $h$ real symmetric ($10$ real), realised by a real tetrad perturbation, and
$\overline{\chi^+}=\chi^-$. The two normalisation conditions $|\det q^\pm|=1$ and
${\rm Im}\,\tr q^\pm=0$ linearise to
\begin{equation}\label{unimod}
{\rm Re}\,\tr\!\big[(\mathbb{q}^+)^{-1}\chi^+\big]=0\,,\qquad
{\rm Im}\,\tr\chi^+=0\,,
\end{equation}
leaving $\chi^+$ with $12-2=10$ real parameters.
 
\paragraph{Non-degeneracy assumptions.} Consider ${\mathbb W}_{\mu\rho}^{\alpha \beta}={\mathbb W}_{\mu\nu\rho\sigma}{\mathbb g}^{\alpha\nu}{\mathbb g}^{\beta\sigma}$ as a map from the space of
traceless symmetric tensors onto itself. 
We
assume
\begin{equation}\label{nondeg}
\text{(*)}\quad\quad -\tfrac23\,\R(\mathbb{q}^+_\tr)\ \text{is not an eigenvalue of }
{\mathbb W} \,.
\end{equation}
In terms of the map $\hat A$ introduced in (\ref{endo}), this condition is equivalent to demanding that the eigenvalues of $\hat A$ on $\Lambda^2_+$ and  $\Lambda^2_-$ are pairwise distinct: $\lambda_i \neq \overline{\lambda}_j$ for all $i,j=1,2,3$. This is satisfied for the area metrics considered above, where we demanded that $\hat A$ has three eigenvalues with strictly positive imaginary part, which then defined the subspace $\Lambda^2_+$.

Under the normalisation conditions \eqref{phaseCondnew} this is the only assumption needed;
with the normalisation $\det q^\pm=1$ of \cite{Freidel:2008ku, Krasnov:2009ik, Speziale:2010cf,Borissova:2022clg} one would
in addition need (**) $\R\,\tr[(\mathbb{q}^+)^{-1}]\neq0$, see the Remark below.
Assumption (*) holds at $\mathbb{q}^\pm=\delta$, where ${\mathbb W}=0$ while
$\tfrac23\R(\mathbb{q}^+_\tr)=2$. Being an open
condition, it persists in a neighbourhood. 
 
\paragraph{Proposition.} \emph{With the assumption\nw{}\ \eqref{nondeg}, if $\delta A=0$ for a
perturbation \eqref{pert} with $h$ real, then $h=0$ and $\chi^\pm=0$.}
 
 ~\\
\emph{Proof.} 
\emph{ Step 1 ($h$ is conformal).} We perturb \eqref{Atrace}, that is $g^{\mu\rho}A_{\mu\nu\rho\sigma}=\R(q^+_\tr)\,g_{\nu\sigma}$.
With $\delta g^{\mu\rho}=-h^{\mu\rho}$ and $\delta A=0$ we obtain
\begin{equation}\label{step1}
-\,h^{\mu\rho}\,\mathbb{A}_{\mu\nu\rho\sigma}
=\R(\chi^+_\tr)\,\mathbb{g}_{\nu\sigma}+\R(\mathbb{q}^+_\tr)\,h_{\nu\sigma}\,.
\end{equation}
We project both sides of (\ref{step1}) onto their traceless parts. To this end, note that
\ba
h^{\mu\rho}\,\mathbb{A}_{\mu\nu\rho\sigma} \left( {\mathbb g}^\nu_\alpha {\mathbb g}^\sigma_\beta -\tfrac14{\mathbb g}^{\nu\sigma} {\mathbb g}_{\alpha \beta}\right)
&=&h^{\mu\rho} \left( \mathbb{A}_{\mu\alpha\rho\beta} -\tfrac{1}{4} \R(\mathbb{q}^+_\tr){\mathbb g}_{\alpha\beta} {\mathbb g}_{\mu\rho}\right) \nn\\
&=& \mathring h^{\mu\rho} \mathbb{A}_{\mu\alpha\rho\beta} \, ,
\ea
where $\mathring h$ is the traceless part of $h$ defined via $h=\mathring h+\tfrac14(\tr h)\mathbb{g}$. We now replace ${\mathbb A}$ with ${\mathbb A}={\mathbb W}+ \tfrac13\R(\mathbb{q}^+_\tr)\mathbb G$ and use $\mathring h^{\mu\rho}\mathbb{G}_{\mu\nu\rho\sigma}=-\mathring h_{\nu\sigma}$ to arrive at 
\begin{equation}\label{step1b}
\tfrac13\R(\mathbb{q}^+_\tr)\,\mathring h-\mathbb{W}(\mathring h)
=\R(\mathbb{q}^+_\tr)\,\mathring h
\qquad\Longleftrightarrow\qquad
\mathbb{W}(\mathring h)=-\tfrac23\R(\mathbb{q}^+_\tr)\,\mathring h\,.
\end{equation}
By assumption (*) we can conclude that $\mathring h=0$, so $h=2\omega\,\mathbb{g}$ with $\omega$ a single real
function.
\vspace{4mm}

{}\q\q \;\;\emph{Step 2 (self-dual block).}  For a conformal perturbation $h=2\omega\,\mathbb{g}$ the degree four homogeneity of $A$ in the tetrad $e$ implies $\delta_h A=4\,\omega\,\mathbb{A}$. 

We therefore obtain for a perturbation of $A$  restricted to a conformal $h$
\begin{equation}\label{step2a}
\delta A=4\omega\,\mathbb{A}
+\tfrac13\R(\chi^+_\tr)\,\mathbb{G}
+\R\big[\mathring\chi^+_{ab}\, \mathbb{\Sigma}^a_+\mathbb{\Sigma}^b_+\big]\,.
\end{equation}
We contract with $\mathbb{\Sigma}_+^{a}\mathbb{\Sigma}_+^{b}$ and use \eqref{Ablocks}; the conjugate term
$\mathring\chi^-\mathbb{\Sigma}_-\mathbb{\Sigma}_-$ drops because
$\mathbb{\Sigma}^a_{+\mu\nu}\mathbb{\Sigma}_-^{b\,\mu\nu}=0$. 
Setting $\delta A=0$ in (\ref{step2a}) leads then to
\begin{equation}\label{step2}
4\omega\Big[\tfrac13\R(\mathbb{q}^+_\tr)\delta^{ab}+\mathring{\mathbb{q}}^{+ab}\Big]
+\tfrac13\R(\chi^+_\tr)\,\delta^{ab}+\mathring\chi^{+ab}=0\,,
\end{equation}
whose trace and traceless parts give $\R(\chi^+_\tr)=-4\omega\,\R(\mathbb{q}^+_\tr)$
and $\mathring\chi^+=-4\omega\,\mathring{\mathbb{q}}^+$. Reassembling,
\begin{equation}\label{chiform}
\chi^+=-4\omega\,\mathbb{q}^++\im\lambda\,\delta\,,\qquad
\lambda:=\tfrac13\big[4\omega\,\I(\mathbb{q}^+_\tr)+\I(\chi^+_\tr)\big]\in\mathbb{R}\,.
\end{equation}
Only $\R(\chi^+_\tr)$ and $\mathring\chi^+$ are fixed at this stage: $\I(\chi^+_\tr)$
is invisible to $\delta A$ and has been absorbed into $\lambda$.
 
{}\q\q\;\emph{Step 3 (normalisation conditions).} Insert \eqref{chiform} into the two conditions
\eqref{unimod}. Taking the trace of \eqref{chiform} gives
$\tr\chi^+=-4\omega\,\tr\mathbb{q}^+_{\phantom\tr}+3\im\lambda$, and since
${\rm Im}\,\tr\mathbb{q}^+=0$ on the slice \eqref{phaseCondnew}, its imaginary part is
\begin{equation}\label{step3}
{\rm Im}\,\tr\chi^+=3\,\lambda=0\,,
\end{equation}
so $\lambda=0$ directly. With $\lambda=0$ and $\tr[(\mathbb{q}^+)^{-1}\mathbb{q}^+]=3$,
the first condition of \eqref{unimod} then reads
${\rm Re}\,\tr[(\mathbb{q}^+)^{-1}\chi^+]=-12\,\omega=0$. Hence $h=0$, and
\eqref{chiform} gives $\chi^+=0$, $\chi^-=\overline{\chi^+}=0$. $\square$

\paragraph{Remark.}   Under the normalisation $\det q^\pm=1$ of
\cite{Freidel:2008ku, Krasnov:2009ik, Speziale:2010cf,Borissova:2022clg}, Step 3 instead inserts \eqref{chiform} into
$\tr[(\mathbb{q}^+)^{-1}\chi^+]=0$ and yields
$0=-12\omega+\im\lambda\,\tr[(\mathbb{q}^+)^{-1}]$, which fixes
$(\omega,\lambda)=0$ only if additionally (**)
$\R\,\tr[(\mathbb{q}^+)^{-1}]\neq0$ holds. That condition is violated for instance at
$\mathbb{q}^+=\mathrm{diag}(z,z,z^{-2})$, symmetric and unimodular, where
$\tr[(\mathbb{q}^+)^{-1}]=2z^{-1}+z^{2}$, whose real part vanishes at
$z=2^{1/3}e^{\im\pi/3}$; the linearised map has then a one-dimensional kernel, along
the direction $(\omega,\lambda)$ with $12\omega=-\lambda \I (\tr[(\mathbb{q}^+)^{-1}])$.

This degeneracy is an artifact of that slice: rotating the same configuration onto
the slice \eqref{phaseCondnew}, i.e.\ replacing $\mathbb{q}^+$ by
$e^{4\im\alpha}\mathbb{q}^+$ with $\alpha$ chosen so that
${\rm Im}\,\tr(e^{4\im\alpha}\mathbb{q}^+)=0$, the kernel is trivial, with smallest
singular value $\approx0.45$. 

There is also a strikingly simple example violating assumption (*).
For $\mathbb{g}=\mathrm{diag}(-1,1,1,1)$ and real and diagonal $\mathbb{q}$ such that
$\mathring{\mathbb{q}}=\mathrm{diag}(\lambda_1,\lambda_2,\lambda_3)$ with
$\sum_a\lambda_a=0$, the spectrum of $\mathbb{W}$ can be computed to be
\begin{equation}\label{Wspec}
\mathrm{spec}\,\mathbb{W}
=\big\{-\lambda_1,-\lambda_2,-\lambda_3\ (\text{each twice})\big\}
\,\cup\,\big\{2\lambda_1,2\lambda_2,2\lambda_3\big\}\,.
\end{equation}
Take as example   
\begin{equation}\label{qflip}
\mathbb{q}^\pm=\mathrm{diag}(1,-1,-1)\,,
\end{equation}
a rotation by $\pi$ around the $x$-axis. There
$\tr[(\mathbb{q}^+)^{-1}]=-1\neq0$, so (**) holds.  We have $-\tfrac23\tr\mathbb{q}=\tfrac23$ and $\mathring{\mathbb{q}}=\mathrm{diag}(\tfrac43,-\tfrac23,-\tfrac23)$, so the eigenvalue $\tfrac23$ appears with multiplicity four.
The
linearised area metric map at \eqref{qflip} has indeed a kernel of dimension exactly four,
consisting of \emph{pure metric} perturbations ($\chi^\pm=0$) spanned by the traceless
directions $h_{12},h_{13},h_{02},h_{03}$ --- precisely the eigenvalue-$\tfrac23$
eigenspace of $\mathbb{W}$.

\section{Integrating out  the multipliers: the  first order
action }\label{sec:eff}

Here we first choose  the potential for the modified Plebanski action such that the constraints enforcing
$e_+=e_-$ and ${\rm Im}\,\tr q^\pm=0$ are kept \emph{sharp}, while the remaining simplicity constraints are
imposed weakly, that is via the potential. We then integrate out the multipliers and  obtain
an action for the real tetrad $e$,  the complex-conjugate pair
$q^\pm_{ab}$ of \eqref{qab} and the connection. 

\paragraph{Choice of potential: sharp reality constraints, weak $q$-constraints.}
Decompose the multiplier in the chiral basis \eqref{Pdef},
\ba\label{phisplit}
\phi_{IJKL}&=&\Big[\varphi^+_{ij}+\tfrac13\varphi_0\,\delta_{ij}\Big]P_{+i\,IJ}P_{+j\,KL}
\,\,+\,\,\Big[\varphi^-_{ij}+\tfrac13\varphi_0\,\delta_{ij}\Big]P_{-i\,IJ}P_{-j\,KL}
\nn\\
&&\,\,+\,\,\psi_{ij}\big(P_{+i\,IJ}P_{-j\,KL}+P_{-j\,IJ}P_{+i\,KL}\big)\,,
\ea
where
$\varphi^\pm$ are complex symmetric $3\times3$ matrices that are 
traceless (corresponding to representation (2,0) and (0,2)), $\varphi_0$ is a single real scalar (corresponding to representation (0,0)) and $\psi$ is hermitian ($9$ real components, corresponding to representation (1,1)):
\begin{equation}\label{phitl}
\tr\varphi^+=\tr\varphi^-=0\,,\qquad
\overline{\varphi^+}=\varphi^-\,,\qquad
\overline{\varphi_0}=\varphi_0\,,\qquad \psi^\dagger=\psi\, .
\end{equation}
With the Gram matrix \eqref{gram} and
$\eps_{IJKL}P^{IJ}_{\pm i}P^{KL}_{\pm j}=\pm2\im\,\delta_{ij}$ one gets
$\eps^{IJKL}\phi_{IJKL}=2\im\big(\tr\varphi^+-\tr\varphi^-\big)$, so in this
parametrisation the $\eps$-trace condition on $\phi$ is satisfied, and the reality of $\phi_{IJKL}$  imposes \eqref{phitl}.
 The count is $10$ (traceless $\varphi^\pm$) $+\,9$ ($\psi$)
$+\,1$ ($\varphi_0$) $=20$.

The Lagrange multiplier term in the Plebanski action (\ref{Spleb})
 reads in the same parametrisation as 
\begin{equation}\label{potchiral}
\frac12\,\phi_{IJKL}B^{IJ}\wedge B^{KL}
=\frac12\Big[\varphi^+_{ij}\,B^i_+\!\wedge B^j_+
+\varphi^-_{ij}\,B^i_-\!\wedge B^j_-
+2\,\psi_{ij}\,B^i_+\!\wedge B^j_-
+\tfrac13\varphi_0\,\delta_{ij}\big(B^i_+\!\wedge B^j_++B^i_-\!\wedge B^j_-\big)
\Big]\,.
\end{equation}

Varying (\ref{potchiral}) with respect to $\psi$ and $\phi_0$ leads to the constraint equations
\begin{equation}\label{sharp}
\psi:\;\;B^{i}_+\wedge B^{j}_-=0\,,\qquad
\varphi_0:\;\;\delta_{ij}\big(B^i_+\wedge B^j_++B^i_-\wedge B^j_-\big)=0\,.
\end{equation}
Inserting the parametrisation $B^i_\pm=\sigma_\pm\,(b_\pm)^i{}_a\Sigma^a_\pm$ into the $\psi$-equation we see that it implies $\Sigma_+(e_+)\wedge\Sigma_-(e_-)=0$. This will be satisfied if and only if the conformal classes of
$g_+$ and $g_-$ agree, and since $g_-=\overline{g_+}$  the nine
$\psi$-constraints force $g_+=e^{2\im\alpha}\hat g$ with $\hat g$ a \emph{real}
Lorentzian metric and $\alpha$ a phase. Under the normalisation \eqref{phaseCondnew} the phase is not free:
$\arg\det e_+=0$ forces $e^{4\im\alpha}$ to be real, i.e.\
$\alpha\in\{0,\tfrac\pi4,\tfrac\pi2,\tfrac{3\pi}4\}$ modulo $\pi$ --- the tetrad is
real up to the discrete companions $e_+\to\im^k e_+$, which correspond to the four
Plebanski sectors, see the remark at the end of the subsection discussing Reisenberger's condition in Section \ref{sec:area}. To fix the gravitational $+$ branch we set $\alpha=0$.

For the $\varphi_0$-equation we use the identity 
 \eqref{qext}, that is
$B^i_{\pm\,\mu\nu}B^j_{\pm\,\rho\sigma}\,
\tilde \eps^{\mu\nu\rho\sigma} = \pm 8\im \sigma_\pm^{2}\, \det(e_\pm)  \;m_\pm^{ij}
$.  Remembering that  $\tr\,m^\pm=\tr\,q^\pm$ and we fixed $\sigma_\pm^2=-1$
 we obtain
\begin{equation}\label{trconstr}
{\rm Im}\big(\sigma_+^2\;e^{4\im\alpha}\,\tr q_+\,\det\hat e\big)=0
\q\q \Rightarrow\q\q {\rm Im}\,\tr q_+ =0\, ,
\end{equation}
where in the second step we used that under \eqref{phaseCondnew} the prefactor
$e^{4\im\alpha}\det\hat e$ is real and nonvanishing. The trace constraint  thus
fixes ${\rm Im}\,\tr q_+$ to be vanishing.   In contrast, would we have chosen  the normalisation
$\det b_\pm=1$ instead of (\ref{phaseCondnew}), the trace constraint would have resulted in $4\alpha+\arg(\tr q_+)=k\pi$, $k\in\mathbb Z$, which 
fixes the tetrad phase only in combination with the dynamical phase of $\tr q_+$. 

The
count for the degrees of freedom after imposing the ten sharp constraints is $30-10=20$: a real tetrad $\hat e$ ($10$ degrees)
and the complex $q_+$ with $|\det q_+|=1$ ($11$ degrees), minus the single relation
${\rm Im}\,\tr q_+=0$. Together, the two conditions on $q_+$ lead to the
linearised version \eqref{unimod} used in Section~\ref{sec:faithful}.

We therefore choose the potential $V$ to be independent of $\varphi_0$ and $\psi$ and quadratic in $\varphi^\pm$:
\begin{equation}\label{Vchoice}
V(\phi)=-\frac{3}{2}\bigg[\frac{1}{\mu}\,\tr\big[(\varphi^+)^{2}\big]
+\frac{1}{\bar{\mu}}\,\tr\big[(\varphi^-)^{2}\big]\bigg]\,,
\end{equation}

with a single \emph{complex} coupling $\mu$. The Hessian $M$ of \eqref{Vquad} is
thus nondegenerate exactly on the ten traceless
chiral directions.

Lorentzian invariance forbids the
mixed contraction $\tr(\varphi^+\varphi^-)$, which is $SU(2)$- but not
$SL(2,\mathbb C)$-invariant.  
The overall sign is fixed so that
${\rm Re}\,\mu>0$ is the non-tachyonic branch.

\paragraph{Integrating out the chiral multipliers: the potential for $q^\pm$.}
The object we integrate out the multipliers from is the non-derivative part
of the modified action \eqref{Spleb}, \eqref{modV}: writing
$S=S_{BF}+S_{\rm pot}$ with $S_{BF}=-\int[\eta+\tfrac1{2\gamma}\eps]\,B\wedge F$, we
define
\begin{equation}\label{Spotdef}
S_{\rm pot}[B,\phi]\;:=\;+\,\frac12\int
\Big[\phi_{IJKL}+\Big(\tfrac{1}{6}V(\phi)\,-\,\tfrac{\Lambda}{6}\Big)\eps_{IJKL}\Big]\,
B^{IJ}\wedge B^{KL}\,.
\end{equation}
The connection term $S_{BF}$ contains no multipliers and is a spectator throughout;
$\phi$ appears in \eqref{Spotdef} without derivatives, so integrating it out is
algebraic. 

We remind the reader that $\phi_{IJKL}$ can be split into the following (SO(3,1)-irreducible) components 
\ba\label{phisplitb}
\phi_{IJKL}&=&\Big[\varphi^+_{ij}+\tfrac13\varphi_0\,\delta_{ij}\Big]P_{+i\,IJ}P_{+j\,KL}
\,\,+\,\,\Big[\varphi^-_{ij}+\tfrac13\varphi_0\,\delta_{ij}\Big]P_{-i\,IJ}P_{-j\,KL}
\nn\\
&&\,\,+\,\,\psi_{ij}\big(P_{+i\,IJ}P_{-j\,KL}+P_{-j\,IJ}P_{+i\,KL}\big)\, ,
\ea
and that the potential $V$ is independent of $\varphi_0$ and $\psi_{ij}$, but quadratic in $\varphi^+_{ij}$ and $\varphi^-_{ij}$.

We will apply the following strategy:  (1) solve the \emph{sharp} constraints, which eliminates
$\psi$ and $\varphi_0$ together with their conjugate constraint directions of
$B\wedge B$; (2) evaluate the surviving wedge products on the constraint surface,
which turns \eqref{Spotdef} into a function of $(e,q^\pm)$ with a linear and quadratic dependence in the remaining multipliers $\varphi^\pm$; (3) solve the
$\varphi^\pm$ equations of motion and reinsert.

\emph{Step 1 (sharp surface).} By the choice of potential above, the sharp constraints set
$e_\pm=e$ real (gravitational branch of \eqref{phaseCondnew}) and
${\rm Im} q^+_\tr=0$. On this surface the wedge products evaluate, by \eqref{wedge}
and \eqref{qext}, to
\begin{equation}\label{wedgeeval}
B^i_\pm\wedge B^j_\pm=\pm2\im\,\sigma_\pm^2\,m_\pm^{ij}\,\vol
=\mp2\im\,m_\pm^{ij}\,\vol\,,
\qquad
B^i_+\wedge B^j_-=0\,,
\end{equation}
using $\sigma_\pm^2=-1$. Consequently the $\psi$-term, which is the contraction with  $B^i_+\wedge B^j_-$ in \eqref{potchiral}, vanishes.
So does the $\varphi_0$-term, as
$\delta_{ij}(B^i_+\wedge B^j_++B^i_-\wedge B^j_-)=4\,{\rm Im}(\tr q_+)\vol=0$.  Thus the
sharp multipliers drop out of \eqref{Spotdef} identically, as they must. 

For the
$\eps$-trace one finds from \eqref{wedgeeval}, with
$\eps_{IJKL}P^{IJ}_{\pm i}P^{KL}_{\pm j}=\pm2\im\,\delta_{ij}$,
\begin{equation}\label{chat0}
\eps_{IJKL}B^{IJ}\wedge B^{KL}=4\big(\tr m_++\tr m_-\big)\vol \, .
\end{equation}
Note, that we use $\tr m_\pm=\tr q_\pm$ throughout.
We introduced $c$ in (\ref{simpl}) as $c=-\tfrac{1}{24}\,\eps \cdot B\wedge B$, we have thus (with $\I q^+_\tr=0$)
\begin{equation}\label{chat}
c=-\tfrac13 q^+_\tr \,\vol\,\,,
\end{equation}
which reduces to $ c=-\vol$ at
$q=\delta$, consistent with \eqref{cval}.

\emph{Step 2 (the Gaussian action for $\varphi^\pm$).} Inserting
\eqref{wedgeeval}--\eqref{chat} into \eqref{Spotdef}: the multiplier term contributes
linearly, $+\tfrac12\varphi^+_{ij}B^i_+\wedge B^j_+=-\im\,\tr(\varphi^+m_+)\vol$ and
$+\tfrac12\varphi^-_{ij}B^i_-\wedge B^j_-=+\im\,\tr(\varphi^-m_-)\vol$; the potential
and cosmological terms contribute through the $\eps$-trace,
$+\tfrac1{12}\big[V(\phi)-\Lambda\big]\eps_{IJKL}B\wedge B
=\tfrac23 q^+_\tr\,\big[V(\phi)-\Lambda\big]\vol$. With the choice \eqref{Vchoice} for $V$ we obtain
\begin{equation}\label{Sgauss}
S_{\rm pot}[e,q^\pm,\varphi^\pm]=\int\vol\,\Big[
-\im\,\tr\!\big(\varphi^+\mathring m_+\big)+\im\,\tr\!\big(\varphi^-\mathring m_-\big)
-\frac{q^+_\tr}{\mu}\,\tr\!\big[(\varphi^+)^2\big]
-\frac{q^+_\tr}{\bar{\mu}}\,\tr\!\big[(\varphi^-)^2\big]
-\tfrac23 \Lambda\,q^+_\tr\Big]\,,
\end{equation}
where only the traceless parts $\mathring m_\pm=m_\pm-\tfrac13(\tr m_\pm)\delta$
couple, since $\varphi^\pm$ are traceless. \eqref{Sgauss} is real by conjugation of
the two sectors, and manifestly Gaussian in $\varphi^\pm$.

\emph{Step 3 (solve and reinsert).} The $\varphi^+$ equation of motion of
\eqref{Sgauss} and its solution are
\begin{equation}\label{phisol}
-\im\,\mathring m_+-\frac{2q^+_\tr}{\mu}\,\varphi^+=0
\qquad\Longrightarrow\qquad
\varphi^+=-\frac{\im\,\mu}{2q^+_\tr}\,\mathring m_+\,,
\end{equation}
and its conjugate for $\varphi^-$. Reinserting, and using that 
$\tr(\mathring m_\pm^2)=\tr(\mathring q_\pm^2)$ (due to $m=b\,q\,b^{-1}$),
yields, on the gravitational branch,
\begin{equation}\label{Spot}
S_{\rm pot}[e,q^\pm]=-\int\vol\;\bigg[
\frac{\mu}{4q^+_\tr}\,\tr\big(\mathring q_+^{\,2}\big)
+\frac{\bar{\mu}}{4q^+_\tr}\,\tr\big(\mathring q_-^{\,2}\big)
+\tfrac23 \Lambda\,q^+_\tr\bigg]\, .
\end{equation}
$S_{\rm pot}$ is manifestly real by conjugation of the two sectors. 

As a consistency check, at $q^\pm=\delta$ the potential part of the action evaluates to $S_{\rm pot}=-2\Lambda\!\int\!\vol$,
 reproducing \eqref{onshell}.

\paragraph{Expansion around simple $B$-fields.}
We expand around $b_\pm=\delta$ as $b_\pm=\delta+\tfrac12\mathring\chi^\pm+\tfrac16\chi^\pm_\tr\,\delta$. We are using the $SO(3,\mathbb{C})_\pm$ symmetry acting on $B$ to fix $b_\pm=\sqrt{q}$ to be symmetric.\footnote{Apply a complex-orthogonal polar decomposition to $b$: $b=Rs$ where $s=\pm\sqrt{b^T b}$ is complex-symmetric and $R=bs^{-1}$ is in $SO(3,\mathbb{C})$.  For $R$ we then have $R^TR=s^{-1} (b^Tb) s^{-1}=\mathbb{1}$, that is $R$ is orthogonal, and we choose the sign for $s$ such that $\det R=1$.}
 Thus we can assume
 $\mathring\chi^\pm$ to be symmetric and traceless.
The trace is fixed by the conditions $|\det b_\pm|=1$ and ${\rm Im}\,q^\pm_\tr=0$. To see this we use
 the expansion of the determinant of a matrix $M$
\ba\label{expdet}
 \det(\delta +M)&=& 1+ \tr M + \tfrac{1}{2} \left( (\tr M)^2- \tr(M^2)     \right) +O(M^3) \q ,
 \ea
and  compute
\ba
1&\stackrel{!}{=}&\det(\delta+\tfrac12\mathring\chi^++\tfrac16\chi^+_\tr\delta)\det(\delta+\tfrac12\mathring\chi^-+\tfrac16\chi^-_\tr\delta)\nn\\
&=& 1 + \tfrac12 (\chi^+_\tr  + \chi^-_\tr ) + \tfrac{1}{12}\left(  (\chi^+_\tr)^2  + (\chi^-_\tr)^2   \right) 
+\tfrac{1}{4} \chi^+_\tr \chi^-_\tr
-  \tfrac{1}{8} \tr\left( \mathring\chi^+ \mathring\chi^++ \mathring\chi^- \mathring\chi^-\right) \,\,+\,\, O(\mathring \chi^3)\,.\nn\\
\ea
On the other hand ${\rm Im}\,q^\pm_\tr=0$ with $q_+=b_+^T b_+$ implies 
\ba
0&\stackrel{!}{=}&\tr q^+-\tr q^- 
\nn\\
&=& \chi^+_\tr  - \chi^-_\tr
+ \tfrac{1}{12}\left(  (\chi^+_\tr)^2  -(\chi^-_\tr)^2   \right)
\,+\,  \tfrac{1}{4} \tr\left( \mathring\chi^+ \mathring\chi^+- \mathring\chi^- \mathring\chi^-\right)
\,\,+\,\, O(\mathring \chi^3)\, \, .
\ea
Thus $\chi^+_\tr$ and $\chi^-_\tr$ are of second order, and we can neglect the $(\chi^\pm_\tr)^2$ terms in the two  equations above. These two equations lead then to 
\begin{equation}\label{tauFix}
\chi^\pm_\tr=\tfrac14\tr(\mathring\chi^\mp\mathring\chi^\mp)+O(\mathring\chi^3)\,.
\end{equation}
The trace is thus fixed by the traceless part of the opposite chirality. For $q^+_\tr$ we obtain
\ba
q^+_\tr &=& 3+ \tfrac14 \tr(\mathring\chi^+\mathring\chi^+)+ \tfrac14 \tr(\mathring\chi^-\mathring\chi^-)+O(\mathring\chi^3)\, ,
\ea
which inserted into (\ref{Spot}) gives
\begin{equation}\label{Spotquad}
S^{(2)}_{\rm pot}=-\int\vol\;\bigg[
\Big(\frac{\mu}{12}+\frac{\Lambda}{6}\Big)\tr\big[(\mathring\chi^+)^2\big]
+\Big(\frac{\bar{\mu}}{12}+\frac{\Lambda}{6}\Big)\tr\big[(\mathring\chi^-)^2\big]
\bigg]-2\Lambda\!\int\!\vol+\, O(\mathring\chi^3) \, .
\end{equation}
This reproduces the mass structure of the linearised Lorentzian area-metric
action of \cite{BDK,Dittrich26}\footnote{ 
\cite{BDK} use a canonical normalisation of the $\chi$-kinetic terms, which can be obtained by rescaling $\chi$ with $\gamma_\pm$ appropriately. Demanding that within this canonical normalisation the masses are equal (which avoids the appearance of additional poles in the propagator of the linearised theory) fixes $\arg\mu$ as a function of $\gamma$, whereas $|\mathcal \mu|$ would determine the mass.}
, with the cosmological constant shifting the $\chi$-masses on an
(A)dS background. There will be a further contribution to the $\chi$-masses from the curvature term, see (\ref{Ahat}).

 The $\mu\to\infty$ limit restores the sharply
constrained theory, which is general relativity.  We note that no term
\emph{linear} in $\chi^\pm$ survives in $S_{\rm pot}$: the $\mu$-terms are manifestly quadratic in the traceless parts, and the only candidate
--- the $\Lambda q^+_\tr$-term --- turns out to be quadratic in $\chi$, due to the $|\det(b_+)|=1$ normalisation condition.

After solving for the connection, linear couplings of $\chi^\pm$
will reside solely in the curvature term \eqref{SGamma}, through the chiral
Weyl block. For this reason $q^\pm=\delta$ will solve the
theory exactly on conformally flat backgrounds.

\vspace{5mm}

\paragraph{Higher-order potentials.}\label{higherorderpot}
The potential $V$ can also include higher order polynomials in $\varphi_\pm$. These have to be invariant under Lorentz transformations, which act on the two chiral sectors by mutually conjugate complex
rotations, $\varphi^+\to R\varphi^+R^T$, $\varphi^-\to\bar R\varphi^-\bar R^T$ with
$R\in SO(3,\mathbb C)$.
 For a symmetric traceless $3\times3$ matrix the invariants under
$\varphi\to R\varphi R^T$ are generated by $I=\tr\varphi^2$ and $J=\tr\varphi^3$ (all
higher traces are polynomials in these), so the Lorentz-invariant potentials
compatible with the reality condition $\varphi^-=\overline{\varphi^+}$ are the real
functions $F(I_+,J_+,\bar I_+,\bar J_+)$. 
Mixed traces such as $\tr(\varphi^+\varphi^-)$
are excluded by this alone: they transform with $R^T\bar R=R^{-1}\bar R$, which is
the identity only for pure rotations, so boosts break them. Invariance does \emph{not}, however, exclude non-holomorphic products
of invariants such as $|I_+|^2=\tr(\varphi^+)^2\tr(\varphi^-)^2$, which start at quartic order.  

The most general invariant and real
polynomial potential is
\ba\label{Vgen}
V(\phi)&=&-\frac32\sum_\pm\Big[\frac{1}{\mu_\pm}\,I_\pm+\frac{1}{\mu^{(3)}_\pm} J_\pm
+\frac{1}{\mu^{(4)}_\pm} I_\pm^2+\frac{1}{\mu^{(5)}_\pm} I_\pm J_\pm+\dots\Big] \,\,+\,\, 
\nn\\
&& \q\q -\frac{1}{\nu^{(4)}}  I_+I_- -  \left(\frac{1}{\nu^{(5)}}  I_+J_-+\frac{1}{\bar\nu^{(5)}}  I_-J_+\right) + \cdots  \,,
\ea
where $\mu^{(n)}_-=\bar\mu^{(n)}_+$ and $\nu^{(4)}=\bar \nu^{(4)}$ and so on.
There is one new complex coupling at cubic order. At quartic  order there is one holomorphic complex coupling and a non-holomorphic real coupling, and quintic order there are two complex couplings.

Integrating out $\varphi$ is a Legendre transform\footnote{If we assume the potential to be convex.} to $\mathring m_\pm$ as $\varphi^\pm$ enters linearly into $S_{\rm pot}$
through $-\im\,\tr\!\big(\varphi^+\mathring m_+\big)+\im\,\tr\!\big(\varphi^-\mathring m_-\big)$, see (\ref{Sgauss}). This leads to a higher order effective potential for $\mathring q^\pm/q^+_\tr$, which at second order coincides (for convex potential) with (\ref{Spot}).  

Clearly, the higher order potential terms are relevant for higher order perturbation theory.  We will see that integrating out the $b$ fields relates $\mathring q^\pm/q^+_\tr$ to the chiral Weyl curvature parts. A non-vanishing $J_\pm$-term in the potential would therefore lead to a parity-odd cubic Weyl invariant. A non-convex $V$  could lead to non-simple area metric solutions with zero curvature.

\paragraph{The complete first order action with quadratic potential.} To end this section, we state the modified Plebanski action with (degenerate) quadratic potential and with multipliers integrated out:
\ba\label{MPlebMout}
S_{\rm MPleb}\,=\, -\sum_{\pm} \gamma_\pm \int B^i_\pm \wedge F^i_\pm \,\,-\,\, \int\vol\;\bigg[
\frac{\mu}{4q^+_\tr}\,\tr\big(\mathring q_+^{\,2}\big)
+\frac{\bar{\mu}}{4 q^+_\tr}\,\tr\big(\mathring q_-^{\,2}\big)
+\tfrac{2}{3}\Lambda\,q^+_\tr\bigg]
\ea
where $B^i_\pm=\pm \im {(b_\pm)^i}_a\Sigma^a_\pm(e)$.  The $b_\pm$ have to satisfy the conditions $|\det(b_\pm)|=1$ and\\ ${\rm Im}(\tr(q_\pm))=0$ and the tetrad $e$ has to be real.

\section{Integrating out the connection}\label{sec:intcon}

We now proceed to integrate out the connection. This is also done in \cite{Freidel:2008ku} and in \cite{Speziale:2010cf} for Euclidean signature, to be self-contained, we provide a derivation here. The connection appears only in the $BF$ part of the action, which splits chirally into 
\begin{equation}\label{SBFchiral}
S_{BF}=-\int\Big[\eta_{IJKL}+\tfrac{1}{2\gamma}\eps_{IJKL}\Big]B^{IJ}\wedge F^{KL}
=-\sum_\pm\gamma_\pm\int B^i_\pm\wedge F^i_\pm\,,\qquad \gamma_\pm=1\pm\tfrac{\im}{\gamma}\,,
\end{equation}
where we defined $F^i_\pm=d\omega^i_\pm-\tfrac12\,\eps^i{}_{jk}\,\omega^j_\pm\wedge\omega^k_\pm$ in (\ref{Fchiral}) with the sector connections $\omega^i_\pm:=P_{\pm i\,IJ}\,\omega^{IJ}$.

The equation of motion for the sector connections is the torsion-freeness condition
\begin{equation}\label{torsion}
d_{\omega_\pm}B^i_\pm
=\Big(dB^i_\pm-\eps^i{}_{jk}\,\omega^j_\pm\wedge B^k_\pm\Big)=0\, ,
\end{equation}
which determines $\omega^j_\pm$ uniquely (for non-degenerate $B^k_\pm$). With the parametrisation $B^i_\pm=\sigma_\pm b_\pm\Sigma_\pm(e)$ it will be convenient to further split $\omega_\pm$ into
\begin{equation}\label{omsplit}
\omega^i_\pm=\Gamma^i_\pm(e)+\rho^i_\pm\,,\qquad
\Gamma^i_\pm(e):=P_{\pm i\,IJ}\,\Gamma^{IJ}(e)\,,\qquad
\overline{\rho^i_+}=\rho^i_-\,,
\end{equation}
with $\Gamma^{IJ}(e)$ the Levi--Civita (torsion-free, metric) spin connection of the
real tetrad, i.e.\ the unique solution of
$de^I+\Gamma^I{}_J\wedge e^J=0$ with $\Gamma^{IJ}=-\Gamma^{JI}$, explicitly 
\begin{equation}\label{LCspin}    
\Gamma_\mu{}^{IJ}
=e^{I\nu}\partial_{[\mu}e^J{}_{\nu]}-e^{J\nu}\partial_{[\mu}e^I{}_{\nu]}
-e^{I\rho}\,e^{J\sigma}\,e^K{}_{\mu}\,\partial_{[\rho}e_{K\,\sigma]}\,,
\end{equation}
where $e_I{}^\mu$ is the inverse tetrad.
Projecting $d(e^I\wedge e^J)$ with $P_{\pm\,a}$  and then using the Cartan equation for ${\Gamma^I}_J$ (see Appendix (\ref{app:Gamma})) leads to Cartan equations for $\Gamma^i_\pm$, which are given by     
\begin{equation}\label{cartanchi}
d_{\Gamma_\pm}\Sigma^a_\pm(e)
:=d\Sigma^a_\pm-\eps^a{}_{bc}\,\Gamma^b_\pm\wedge\Sigma^c_\pm=0\, ,
\end{equation}
and which can be solved for $\Gamma_\pm$ in closed form. To express the solution, we introduce the dual of
$d\Sigma_\pm$,
\begin{equation}\label{Gdual}
G^a_{\pm\,\mu}:=g_{\mu\sigma}\,\frac{1}{3!\det e}\,\epst^{\sigma\alpha\beta\gamma}
(d\Sigma^a_\pm)_{\alpha\beta\gamma}\,,
\end{equation}
and introduce the operator ${\mathbb S}_\pm$ acting on $t^k{}_\nu$ as
\begin{equation}\label{Sop}
(\mathbb{S}_\pm t)^p{}_\mu:=\eps^{pkl}\,(\Sigma^l_\pm)^\nu{}_\mu\,t^k{}_\nu\, .
\end{equation}
The operator satisfies $\mathbb S_\pm^2+\mathbb S_\pm-2=0$, hence $\mathbb S_\pm^{-1}=\tfrac12(\mathbb1+\mathbb S_\pm)$.
As shown in Appendix~\ref{app:Gamma} one then finds
\begin{equation}\label{Gammasol}
\Gamma^a_{\pm\,\mu}=\pm\frac{\im}{2}\big[(\mathbb1+\mathbb S_\pm)G_\pm\big]^a{}_\mu\, .
\end{equation}

The difference $\rho^i_\pm$ is an adjoint-valued
\emph{tensor}, not a connection. Inserting \eqref{omsplit} into \eqref{torsion} and
using $d_{\Gamma_\pm}\Sigma^a_\pm=0$, the derivative lands entirely on $b_\pm$, and the
$\rho$-equation is \emph{algebraic}:
\begin{equation}\label{rhoeq}
\eps^i{}_{jk}\,\rho^j_\pm\wedge (b_\pm)^k{}_a\,\Sigma^a_\pm
=(D_{\Gamma_\pm}b_\pm)^i{}_a\wedge\Sigma^a_\pm\,,
\qquad
(D_{\Gamma_\pm}b)^i{}_a:=db^i{}_a-\eps^i{}_{jk}\Gamma^j_\pm\,b^k{}_a
-\,\eps_{ab}{}^{c}\,\Gamma^b_\pm\,b^i{}_c\, .
\end{equation}
The system (\ref{rhoeq}) provides twelve complex linear equations per sector for the twelve complex $\rho^i_{\pm\mu}$.

\paragraph{Solution for $\rho$.} 
The equations for $\rho_\pm$ can be solved in the same manner as the equations for $\Gamma_\pm$. We introduce the dual of the source term  
\ba\label{Tdual}
\hat T^{\,l\,\sigma}_\pm&:=&\frac{1}{3!\,\det(e)\,\det b_\pm}\;
\epst^{\sigma\mu\nu\rho}\,(b^T_\pm)^{l}{}_{i}
\Big[(D_{\Gamma_\pm}b_\pm)^i{}_a\wedge\Sigma^a_\pm\Big]_{\mu\nu\rho}
\nn\\
&=& \pm\frac{\im}{\det b_\pm}\,(b^T_\pm)^{l}{}_{i}\,
\big(\Sigma^a_\pm\big)^{\sigma\nu}\,(D_{\Gamma_\pm}b_\pm)^i{}_{a\,\nu}\,,
\ea
where spacetime indices are moved with $g(e)$.  The solution can then be expressed as 
\ba\label{rhoexplicit}
\rho^i_{\pm\,\mu}&=&\pm\,\frac{\im}{2}\,\big[b_\pm(\mathbb1+\mathbb{S}_\pm)\hat T_\pm\big]^i{}_\mu\, \nn\\
&=&
-\frac{1}{2\det b_\pm}\,(b_\pm)^i{}_p\Big[\delta^{pk}\delta^\nu_\mu
+\eps^{pkl}\big(\Sigma^l_\pm\big)^\nu{}_\mu\Big](b^T_\pm)^k{}_j\,(\Sigma^a_\pm)_\nu{}^\lambda\,
(D_{\Gamma_\pm}b_\pm)^j{}_{a\lambda} \, .
\ea
Clearly,  $\rho$ is linear in
$D_\Gamma b$, and $\rho=0\Leftrightarrow D_\Gamma b=0$.

\paragraph{The action in $(\Gamma,\rho)$ variables.}  It is instructive to express the $BF$ part of the action in $(\Gamma,\rho)$ variables. Inserting
$\omega_\pm=\Gamma_\pm+\rho_\pm$ into $F^i_\pm(\omega)$ we obtain
\ba
F^i_\pm(\omega)=F^i_\pm(\Gamma)+d_{\Gamma_\pm}\rho^i_\pm
-\tfrac12\eps^i{}_{jk}\rho^j_\pm\wedge\rho^k_\pm \q\q  \text{with} \q\q d_{\Gamma_\pm}\rho^i_\pm=d\rho^i_\pm -{\epsilon^i}_{jk}\Gamma^j\wedge \rho^k \, .
\ea

We insert this into the $BF$-action (\ref{SBFchiral}) and integrate the middle term by parts.  Because of the chiral Cartan equations (\ref{cartanchi}) the $d_{\Gamma_\pm}$-derivatives descends from $B_\pm$ to $b$
\ba
d_{\Gamma_\pm}B^i_\pm=\sigma_\pm(D_{\Gamma_\pm}b)^i{}_a\wedge\Sigma^a_\pm
\ea
with $D_{\Gamma_\pm}b$ defined in (\ref{rhoeq}).
Modulo a boundary term we therefore obtain
\begin{equation}\label{SBFrho}
S_{BF}=-\sum_\pm\gamma_\pm\int\Big[
B^i_\pm\wedge F^i_\pm(\Gamma)
-\sigma_\pm\,(D_{\Gamma_\pm}b_\pm)^i{}_a\wedge\Sigma^a_\pm\wedge\rho^i_\pm
-\tfrac12\,\eps_{ijk}\,B^i_\pm\wedge\rho^j_\pm\wedge\rho^k_\pm\Big]\,.
\end{equation}
The $\rho_\pm$ equation of motion from \eqref{SBFrho} is precisely \eqref{rhoeq}.  Since the action \eqref{SBFrho} is only linear and quadratic (that is Gaussian) in $\rho_\pm$, its
on-shell value is the curvature term minus the quadratic term evaluated on
the solution $\rho_*$. (Write \eqref{SBFrho} as $S_0+L[\rho]+Q[\rho]$
with $L$ linear and $Q$ quadratic, the stationarity condition gives
$L[\rho_*]=-2Q[\rho_*]$, hence $S\big|_{\rho_*}=S_0-Q[\rho_*]$.). We thus have 
\begin{equation}\label{SBFrhostar}
S_{BF}(\rho_*)=-\sum_\pm\gamma_\pm\int\Big[
B^i_\pm\wedge F^i_\pm(\Gamma)
+\tfrac12\,\eps_{ijk}\,B^i_\pm\wedge\rho^j_{\pm*}\wedge\rho^k_{\pm*}\Big]\,.
\end{equation}
We thus obtain the usual Levi-Civita curvature term, altered by the presence of a potentially non-trivial ${b^i_\pm}_a$ in $B^i_\pm$, and a term quadratic in $\rho_*$. As we will explain below $\rho_*$ is the chiral contorsion of $\omega$ with respect to the tetrad and, at the same time, determines the torsion of the Levi-Civita connection $\Gamma$ with respect to $B$.
The expression 
$\eps_{ijk}B^i\wedge\rho^j\wedge\rho^k$ is the area-metric-weighted
Einstein--Cartan invariant  quadratic in the
contorsion $\rho$. It encodes the dynamics of the area metric deformation beyond the curvature coupling.

\paragraph{Torsion and contorsion.}  We defined the split $\omega=\Gamma(e)+\rho$ with
$\Gamma(e)$  the Levi--Civita connection of the tetrad. $\rho$ acts through
$\eps$, hence preserves $\delta$. Thus $\omega$ is a metric connection with torsion
with respect to the length metric, and $\rho$ is its contorsion: the torsion
of $e$ under $\omega$ is, chirally,
\begin{equation}\label{torsionomega}
T^a_\pm:=d_{\omega_\pm}\Sigma^a_\pm=-\,\eps^a{}_{bc}\,\rho^b_\pm\wedge\Sigma^c_\pm\, .
\end{equation}
In terms of the real Lorentz connection,
$\rho^i_\pm=P_{\pm i\,IJ}\,K^{IJ}$ is the chiral projection of the contorsion
one-form $K^{IJ}=\omega^{IJ}-\Gamma^{IJ}(e)$, and the $12$ complex components of
$\rho_+$ are the $24$ real components of $K$, equivalently of
the torsion $T^I=d_\omega e^I$.  

Whereas $\omega$ has torsion with respect to $\Sigma^a$, the connection $\Gamma$ has torsion with respect to $B^i$.
The torsion of $\omega$ relative to $\Sigma$,
$d_\omega\Sigma^a=-\eps^a{}_{bc}\rho^b\wedge\Sigma^c$, determines also
 the torsion of $\Gamma$ relative to $B$, $d_\Gamma B^i=\eps^i{}_{jk}\rho^j\wedge
B^k$ (from $d_\omega B=0$).

\paragraph{Evaluating the $BF$-action.} 
Equation (\ref{SBFrhostar}) reduces the $BF$-action to two terms, the Levi-Civita curvature term and the term quadratic in the contorsion. For the Levi-Civita curvature term we remind the reader of the curvature decomposition (\ref{Fdecomp}), which is given by
\ba
F^i_\pm(\Gamma)=\tfrac12\,X^{ij}_\pm\,\Sigma^j_\pm+\tfrac12\,Y^{ij}_\pm\,\Sigma^j_\mp\, ,
\ea
this is wedged with $B^i_\pm =\sigma_\pm {b^i_\pm}_a \Sigma^a_\pm$, where $\sigma_\pm=\pm\im$.

With $e$ real, the wedge identities
$\Sigma^a_\pm\wedge\Sigma^b_\pm=\pm2\im\,\delta^{ab}\vol$ and
$\Sigma^a_+\wedge\Sigma^b_-=0$ give for the curvature term
\begin{equation}\label{SGamma}
B^i_\pm\wedge F^i_\pm(\Gamma)\,=\,-\,\tr\!\big(X_\pm\,b_\pm^{\,T}\big)\,\vol\,
\,=\, -\left(\tfrac{R}{6}\,\tr b_\pm+\tr(\mathring X_\pm\!\cdot\!\mathring b^T_\pm)\right) \,\vol\,
\end{equation}
where $\mathring X^{ij}_\pm=\tfrac14\,(\Sigma^i_\pm)^{\mu\nu}\,W_{\mu\nu\rho\sigma}\,
(\Sigma^j_\pm)^{\rho\sigma}$ encodes the (anti-) selfdual Weyl curvatures.

Thus we get the Ricci scalar $R$ coupled to $\tr b_\pm$ and the traceless part of $b_\pm$ coupled to the (anti-) selfdual Weyl curvature, whereas the traceless Ricci tensor (encoded in the $Y$-block) does not appear.

For the quadratic contorsion
term in (\ref{SBFrhostar}) one finds (see Appendix~\ref{app:contorsion})
\begin{equation}\label{cubicrho}
\tfrac12\,\eps_{ijk}\,B^i_\pm\wedge\rho^j_\pm\wedge\rho^k_\pm
\,=\,-\frac{\det b_\pm}{4}\,\hat T_\pm(\mathbb1+\mathbb{S}_\pm)\hat T_\pm\;\vol
\end{equation}

Together, (\ref{SGamma}) and (\ref{cubicrho}) lead to the connection-on-shell $BF$-action
\begin{equation}\label{SBFeff}
S^{\rm eff}_{BF}[e,q^\pm]=+\sum_\pm\gamma_\pm\int\vol\;\Big[
\tr\!\big(X_\pm\,b_\pm^{\,T}\big)
+\frac{\det b_\pm}{4}\;
g^{\mu\nu}\,\hat T^{p}_{\pm\,\mu}\,
\big[(\mathbb{1}+\mathbb{S}_\pm)\,\hat T_\pm\big]^{p}{}_{\nu}\Big]\,.
\end{equation}

To obtain the full modified Plebanski action we have to add the potential term \eqref{Spot},
with $q_\pm=b_\pm^Tb_\pm$; see also \eqref{MPlebMout}.

\paragraph{Expansion in $b=\delta +\tfrac12 \chi$.}
We expand again around $b_\pm=\delta$ with $b_\pm=\delta+\tfrac12\mathring\chi^\pm+\tfrac16\chi^\pm_\tr
\,\delta$, with $\mathring\chi^\pm$ symmetric and traceless, 
and
$
\chi^\pm_\tr
=\tfrac14\,\tr\mathring\chi_\mp^{\,2}+O(\mathring\chi^3)$, see (\ref{tauFix}).

The $\Gamma$-covariant derivative of $\mathring\chi$ is given by
\begin{equation}\label{DGammachi}
(D_\mu\mathring\chi)^{ja}=\partial_\mu\mathring\chi^{ja}-\eps^j{}_{nk}\Gamma^n_{\pm\mu}
\mathring\chi^{ka}-\eps^a{}_{nk}\Gamma^n_{\pm\mu}\mathring\chi^{jk}\,,
\end{equation}
so that $D_\Gamma b=\tfrac12
D\mathring\chi+O(\mathring\chi^2)$. The three parts of the action contribute as follows.
\begin{itemize}
\item The curvature term \eqref{SGamma} gives $\tr(X_\pm b_\pm^T)=\tfrac R2+\tfrac12
\tr(\mathring X_\pm\mathring\chi^\pm)+\tfrac R12\chi^\pm_\tr$.
\item The contorsion term, which enters \eqref{SBFeff} as $-\tfrac12\eps_{ijk}B^i\wedge
\rho^j\wedge\rho^k$ per unit $\gamma_\pm$, becomes
\begin{equation}\label{Qdef}
-\tfrac18\Big[D^\mu\mathring\chi\cdot D_\mu\mathring\chi+\eps^{abc}\Sigma^{c\,\mu\nu}_\pm
D_\mu\mathring\chi^{ja}D_\nu\mathring\chi^{jb}\Big]+O(\mathring\chi^3)\,.
\end{equation}
The $\Sigma$-twisted term is a total derivative only on flat space. On a curved
background, integrating by parts produces the commutator $[D_\mu,D_\nu]$, i.e.\ the
chiral curvature, and the term equals
\begin{equation}\label{twistcurv}
-\tfrac{R}{16}\,\mathring\chi\cdot\mathring\chi+\tfrac38\,\tr(\mathring X_\pm\mathring\chi^2)
\end{equation}
up to a total derivative (Appendix~\ref{app:expansion}).
\item The potential contributes \eqref{Spotquad}.
\end{itemize}
Altogether, on an arbitrary background and up to total derivatives,
\begin{align}
S_{\rm MPleb}[e,\mathring\chi^\pm]\simeq\int\sqrt{-g}\,(R-2\Lambda)+\sum_\pm\int\vol\Big\{&
\frac{\gamma_\pm}{2}\,\mathring X_\pm\cdot\mathring\chi^\pm-\frac{\hat\mu_\pm}{12}\,
\mathring\chi^\pm\cdot\mathring\chi^\pm\nonumber\\
&-\frac{\gamma_\pm}{8}D^\mu\mathring\chi^\pm\cdot D_\mu\mathring\chi^\pm
+\frac{3\gamma_\pm}{8}\tr(\mathring X_\pm\mathring\chi^{\pm2})\Big\}+O(\mathring\chi^3)\,,
\label{Seffchi}
\end{align}
where $X\cdot Y$ denotes the trace over $(ab)$ indices and (see Appendix \ref{app:expansion})
\begin{equation}\label{Ahat}
\hat{\mu}_\pm:=\mu_\pm+2\Lambda
-\tfrac14 \gamma_\mp R +\tfrac34 \gamma_\pm R\,,\qquad
\hat{\mu}_\pm=\tilde{\mu}_\pm+\tfrac34\gamma_\pm R\,,\qquad
\tilde{\mu}_\pm:=\mu_\pm+2\Lambda-\tfrac14\gamma_\mp R\,,
\end{equation}
with $\mu_+=\mu$ and $\mu_-=\bar\mu$. $\tilde{\mu}$ collects the non-derivative (trace-sector) contributions:
\begin{itemize}
\item the $2\Lambda$ comes from $\tfrac23 \Lambda q^+_\tr$ in \eqref{Spot} with $q^+_\tr=3+\tfrac12{\rm Re}\,\tr\mathring
\chi_+^2$;
\item the $-\tfrac14\gamma_\mp R$ is the cross-chiral trace coupling $\tfrac{\gamma_\mp R}{12}
\chi^\mp_\tr$ of the curvature term, evaluated with \eqref{tauFix}.
\end{itemize}
The remaining $\tfrac34\gamma_\pm R$ in $\hat{\mu}_\pm$ comes from the kinetic curvature
term \eqref{twistcurv}. The last term of \eqref{Seffchi} is a traceless, Weyl-dependent
mass matrix. It contributes to neither $\tilde{\mu}_\pm$ nor $\hat{\mu}_\pm$, and
once $\mathring\chi\propto\mathring X$ it is cubic in the Weyl tensor.

\section{A connection for area metrics}\label{sec:areaconn}

Here we point out that the $B$-forms allow a natural definition of a connection on two-forms compatible with area metrics.

\paragraph{A covariant derivative for two-forms.} The two-forms $B^i_\pm$ can be understood as soldering forms between $\Lambda^2_\pm(T^*{\cal M})$ to internal three-dimensional bundles, and we can use these forms to decompose any two-form with respect to $B^i_\pm$, that is $\alpha=\sum_\pm\alpha^\pm_iB^i_\pm$. 

Given $\mathfrak{so}(3,\mathbb C)_\pm$-connections $\omega_\pm$ we can define a covariant derivative on the internal three-dimensional bundles, and thus on $\Lambda^2_\pm(T^*{\cal M})$ (and its tensor products). 
\begin{equation}\label{DomegaTwoform}
D^{\omega}_\mu\alpha=\sum_\pm\Big(\partial_\mu\alpha^\pm_l+\alpha^\pm_j\,\eps^j{}_{kl}\,
\omega^k_{\pm\mu}\Big)B^l_\pm\, .
\end{equation}
Torsion freeness of $\omega$ with respect to $B$, that is $d_\omega B=0$, fixes uniquely the $\mathfrak{so}(3,\mathbb C)_\pm$-connections, and ties the exterior derivative of two-forms to the covariant derivative
\ba
dx^\mu\wedge D^\omega_\mu\alpha-d\alpha\,=\,-\alpha_l\,(d_\omega B)^l\,=\,0 .
\ea
The $\mathfrak{so}(3,\mathbb C)_\pm$ connections preserve automatically $\delta_{ij}$ on the internal bundles, and thus the area metric. By construction the covariant derivative (\ref{DomegaTwoform}) preserves also the split of $\Lambda^2$ into selfdual and anti-selfdual sector.

We can express the same covariant derivative using $\Sigma^a_\pm$ as soldering forms.  With a decomposition $\alpha=\sum_\pm\alpha^\pm_a\,\Sigma^a_\pm$ (so that $\alpha^\pm_a=\sigma_\pm\alpha^\pm_ib^i_{\pm a}$) we can express $D^{\omega}=D^{\cal A}$ as 
\begin{equation}\label{DAtwoform}
D^{\cal A}_\mu\alpha=\sum_\pm\Big(\partial_\mu\alpha^\pm_a-\alpha^\pm_b\,{\cal A}^b_{\pm\,a\,\mu}
\Big)\,\Sigma^a_\pm\,  ,
\end{equation}
where ${\cal A}^b_{\pm\,a\,\mu}$ arises from redefining torsion-freeness with respect to $B^i_\pm$ as torsion-freeness with respect to $\Sigma^a_\pm$:
\begin{equation}\label{Aind}
d_{\omega_\pm}B^i_\pm=0
\;\Longleftrightarrow\;
d\Sigma^a_\pm+\mathcal{A}^a_{\pm\,b}\wedge\Sigma^b_\pm=0\,,\qquad
\mathcal{A}^a_{\pm\,b}:=(b_\pm^{-1})^a{}_i\,
\big[\,db^i_{\pm\,b}-\eps^i{}_{jk}\,\omega^j_\pm\,b^k_{\pm\,b}\big]\,,
\end{equation}
We thus need to be careful with distinguishing the internal $i,j,k,\ldots$ indices from the internal $a,b,c,\ldots$ indices.

$\mathcal{A}_\pm$ is metric with respect to $q^\pm$ but
not with respect to $\delta$:
\begin{equation}\label{qmetric}
D^{\mathcal A}_\mu\,q_{ab}\,=\, \partial_\mu q_{ab}-\mathcal{A}^c{}_a\,q_{cb}-\mathcal{A}^c{}_b\,q_{ac}=0\,,
\qquad
D^{\mathcal{A}}_\mu\,\delta_{ab}=-2\,\mathcal{A}_{(ab)}\neq0\,,
\end{equation}
whereas $\omega_\pm$ is metric with respect to $\delta_{ij}$. 

We now have the two equations
\ba
d\Sigma^a_\pm-\eps^a{}_{bc}\,\Gamma^b_\pm\wedge\Sigma^c_\pm=0 \q \text{and}\q  d\Sigma^a_\pm+\mathcal{A}^a_{\pm\,b}\wedge\Sigma^b_\pm=0 \q .
\ea
Thus one may ask whether
$\mathcal{A}^a_{\pm\,b}=\eps^a{}_{bc}\Gamma^c_\pm$.  But this is, in general, not the case. ${\cal A}$ is a $GL(3,\mathbb C)$ connection, whereas $\Gamma$ is a $SO(3,\mathbb C)$ connection (which are metric with respect to $\delta$).

The equation $d\Sigma_\pm^a+{X^a}_b\wedge \Sigma^b_{\pm}$ 
 determines $X$ only up to the $24$-dimensional kernel of
$X\mapsto X^a{}_b\wedge\Sigma^b_\pm$ on $\mathfrak{gl}(3,\mathbb C)$-valued
one-forms, and only within $\mathfrak{so}(3,\mathbb C)$ (where the kernel is trivial) is the
solution unique, namely $\Gamma_\pm(\Sigma)$ of \eqref{Gammasol}. 

Splitting $\omega=\Gamma+\rho$ in \eqref{Aind} and using the definition of $D_{\Gamma}b$ in
\eqref{rhoeq} gives the  decomposition
\begin{equation}\label{AGammaR}
\mathcal{A}^a_{\pm\,b}=\eps^a{}_{bc}\,\Gamma^c_\pm+\tau^a_{\pm\,b}\,,\qquad
\tau^a_{\pm\,b}:=(b_\pm^{-1})^a{}_i\Big[(D_{\Gamma_\pm}b_\pm)^i{}_b
-\eps^i{}_{jk}\,\rho^j_\pm\,(b_\pm)^k{}_b\Big]\,,
\end{equation}
and $\tau^a_{\pm\,b}\wedge\Sigma^b_\pm=0$ is precisely the $\rho$-equation
\eqref{rhoeq}.

Now the covariant derivative $D^\omega$ (or equivalently $D^{\cal A}$) does \emph{not} preserve simplicity of two-forms.  That is a two-form $\alpha\wedge \beta$ will in general be transported to a non-simple two-form.   For a connection to preserve simplicity, it has to preserve the wedge pairing (up to a multiplicative factor) $\langle\alpha,\beta\rangle:=\tfrac12
\epst^{\mu\nu\rho\sigma}\alpha_{\mu\nu}\beta_{\rho\sigma}$: if $\alpha$ is simple we have $\langle\alpha,\alpha\rangle=0$. On $\Lambda_\pm$ the wedge pairing in the $\Sigma^a_\pm$ basis is $\propto\delta$, which is \emph{not} preserved by ${\cal A}$, but is preserved by $\epsilon\Gamma$. In the $B^i_\pm$ basis the wedge pairing is, in general, not proportional to $\delta$  (which is preserved by $\omega$).

\paragraph{Characterisation of the covariant derivative on two-forms.}

Define a covariant derivative on two-forms by
\ba
D_\mu\alpha_{\nu\rho}=\partial_\mu\alpha_{\nu\rho}-C_{\mu\,\nu\rho}{}^{\kappa\lambda}\alpha_{\kappa\lambda}
\ea
for a general connection $C$ on $\Lambda^2$.   $D^\omega$ as defined in (\ref{DomegaTwoform}) is the unique connection on two-forms which satisfies the following three conditions:
\begin{itemize}
\item[(i)]  It preserves the area metric. In terms of the $C$-coefficients this condition is
\ba
\partial_\mu A_{\nu\rho\sigma\tau}={C_{\mu\nu\rho}}^{\kappa\lambda}A_{\kappa\lambda\sigma\tau} + 
{C_{\mu\sigma\tau}}^{\kappa\lambda}A_{\nu\rho\kappa\lambda} \, .
\ea
\item[(ii)] 
It preserves a complex structure, which determines  the chiral split.  This complex structure $\mathcal J$ can be defined from the area metric (and an
orientation encoded in $\tilde \epsilon^{\mu\nu\rho\sigma}$) as a matrix function of the endomorphism $\hat A$ of \eqref{endo},
\ba\label{Jfromhat}
\mathcal J=\hat A\,\big(-\hat A^{2}\big)^{-1/2}\, ,
\ea
with the principal square root. On an eigenvector of $\hat A$ with eigenvalue $\lambda$ it acts
as $\lambda/\sqrt{-\lambda^2}=\pm\im$ for ${\rm Im}\,\lambda\gtrless0$. Hence $\mathcal J$ is real, satisfies $\mathcal J^2=-1$, and equals $+\im$
on the span of the eigenvectors with ${\rm Im}\,\lambda>0$, which is $\Lambda^2_+$ for configurations
considered in Section \ref{sec:faithful}, that is where $\hat A$ has three eigenvalues with strictly positive imaginary part. For a metric-induced area metric, $\hat A=2\sqrt{-g}\,\star$
and \eqref{Jfromhat} reduces to the Hodge star.  Reversing
the orientation changes the sign of $\mathcal J$, but not the connection preserving it.

In terms of the $C$-coefficients the preservation of ${\cal J}$ can be expressed as 
\ba
\partial_\mu {{\cal J}_{\nu\rho}}^{\kappa\lambda} ={C_{\mu\nu\rho}}^{\sigma\tau}{{\cal J}_{\sigma\tau}}^{\kappa\lambda}-{{\cal J}_{\nu\rho}}^{\sigma\tau}{C_{\mu\sigma\tau}}^{\kappa\lambda}\, .
\ea
\item[(iii)] It is torsion-free on two-forms,
\begin{equation}
dx^\mu\wedge D^\omega_\mu\alpha-d\alpha=-\alpha_l\,(d_\omega B)^l=0
\quad\Longleftrightarrow\quad C_{[\mu\,\nu\rho]}{}^{\kappa\lambda}=0\,,
\end{equation}
i.e.\ $d_\omega B=0$.
\end{itemize}

Uniqueness follows by counting. Conditions (i) and (ii)
leave an $\mathfrak{so}(3,\mathbb C)$-valued one-form per sector ($12$ complex components). Condition (iii)
imposes $3\times4=12$ complex conditions, and their solution $\omega=\Gamma+\rho_*$ is unique because
$\mathbb S$ (defined in (\ref{Sop})) is invertible. Without (ii), a general area metric-compatible connection has $60$ real
components against $24$ torsion conditions \cite{Ho:2015cza}, thus (ii) is needed to uniquely fix an area metric connection.


\paragraph{Curvature of the connection on two-forms.}
Since $D_\mu$ acts only on the two-form indices, its curvature is defined by the commutator
of covariant derivatives along the coordinate directions,
\ba
[D_\mu,D_\nu]\,\alpha_{\rho\sigma}=-\,\mathcal R_{\mu\nu\,\rho\sigma}{}^{\kappa\lambda}\,\alpha_{\kappa\lambda}\, ,
\ea
where $D_\mu(D_\nu\alpha)_{\rho\sigma}=\partial_\mu(D_\nu\alpha)_{\rho\sigma}-C_{\mu\,\rho\sigma}{}^{\kappa\lambda}(D_\nu\alpha)_{\kappa\lambda}$. The terms with derivatives of
$\alpha$ cancel in the commutator, and
\ba\label{Rtwoform}
\mathcal R_{\mu\nu\,\rho\sigma}{}^{\kappa\lambda}=\partial_\mu C_{\nu\,\rho\sigma}{}^{\kappa\lambda}-\partial_\nu C_{\mu\,\rho\sigma}{}^{\kappa\lambda}
-C_{\mu\,\rho\sigma}{}^{\alpha\beta}\,C_{\nu\,\alpha\beta}{}^{\kappa\lambda}+C_{\nu\,\rho\sigma}{}^{\alpha\beta}\,C_{\mu\,\alpha\beta}{}^{\kappa\lambda}\,,
\ea
with all index pairs summed over. In matrix notation on $\Lambda^2$, $\mathcal R_{\mu\nu}=\partial_\mu C_\nu-\partial_\nu C_\mu-[C_\mu,C_\nu]$.
Here $D_\mu$ acts only on the two-form indices: for fixed $\nu$, $D_\nu\alpha$ is again treated as a two-form,
and $\mathcal R_{\mu\nu}$ is the curvature of the bundle connection evaluated on the coordinate vector
fields $\partial_\mu,\partial_\nu$. \footnote{For $\nabla$ a connection on $TM$ the second derivative also acts on the index $\nu$ of the first derivative $\nabla_\nu v_\rho$. The commutator thus includes also a torsion term
$
[\nabla_\mu,\nabla_\nu]\,v_\rho=-R_{\mu\nu\rho}{}^{\kappa}\,v_\kappa-T^\lambda{}_{\mu\nu}\,\nabla_\lambda v_\rho\, .$
}

 By (i) and (ii), $\mathcal R_{\mu\nu}$ preserves the area metric and
the chiral split, i.e.\ it takes values in $\mathfrak{so}(3,\mathbb C)_+\oplus\mathfrak{so}(3,\mathbb C)_-$.
For $D^\omega$ the curvature is the adjoint action of $F(\omega)$ in the $B$-frame:
\ba
[D^\omega_\mu,D^\omega_\nu]\big(\alpha_lB^l\big)=\alpha_j\,\eps^j{}_{kl}\,F^k_{\mu\nu}(\omega)\,B^l\, .
\ea
With the dual frame $\alpha^\pm_l=(\mathbb M_\pm^{-1})_{ln}\langle B^n_\pm,\alpha\rangle$, $\mathbb M^{ln}_\pm=\langle B^l_\pm,B^n_\pm\rangle$, this gives
\ba\label{Rtwoformomega}
\mathcal R_{\mu\nu\,\rho\sigma}{}^{\kappa\lambda}=-\tfrac12\sum_\pm\eps^l{}_{km}\,F^k_{\pm\,\mu\nu}\,B^m_{\pm\,\rho\sigma}\,(\mathbb M_\pm^{-1})_{ln}\,
\epst^{\alpha\beta\kappa\lambda}B^n_{\pm\,\alpha\beta}\, .
\ea
The curvature satisfies two Bianchi identities:
\begin{itemize}
\item the differential identity $D_{[\lambda}\mathcal R_{\mu\nu]}=0$, which holds for any connection;
\item as a consequence of torsion-freeness, $d_\omega^2B=0$ gives $\eps^l{}_{km}F^k\wedge B^m=0$. This is equivalent
to the algebraic identity $\mathcal R_{[\mu\nu\,\rho\sigma]}{}^{\kappa\lambda}=0$, the analogue of $R_{[\mu\nu\rho]\sigma}=0$.
\end{itemize}
If $D_\Gamma q=0$, in particular for length metric-induced area metrics, $D^\omega$ is induced by the
Levi-Civita connection. $\mathcal R$ is then the Riemann tensor acting on two-forms,
$\mathcal R_{\mu\nu\,\rho\sigma}{}^{\kappa\lambda}=R_{\mu\nu\rho}{}^{[\kappa}\delta^{\lambda]}_\sigma-R_{\mu\nu\sigma}{}^{[\kappa}\delta^{\lambda]}_\rho$ (with the sign
convention $[\nabla_\mu,\nabla_\nu]v_\rho=-R_{\mu\nu\rho}{}^\kappa v_\kappa$). 

\paragraph{Symmetries and independent components of $\mathcal R$.}
The curvature $\mathcal R_{\mu\nu\,\rho\sigma}{}^{\kappa\lambda}$ of $D^\omega$ has the following symmetries:
\begin{itemize}
\item it is antisymmetric in $\mu\nu$ by definition, and in $\rho\sigma$ and $\kappa\lambda$ (for each
$\mu\nu$, $\mathcal R_{\mu\nu}$ is an endomorphism of $\Lambda^2$);
\item it is $A$-skew, $\mathcal R_{\mu\nu\,\rho\sigma}{}^{\alpha\beta}A_{\alpha\beta\kappa\lambda}=-\mathcal R_{\mu\nu\,\kappa\lambda}{}^{\alpha\beta}A_{\alpha\beta\rho\sigma}$, by condition (i);
\item it commutes with the complex structure, $[\mathcal R_{\mu\nu},\mathcal J]=0$, by condition (ii). Hence
$\mathcal R_{\mu\nu}\in\mathfrak{so}(3,\mathbb C)_+\oplus\mathfrak{so}(3,\mathbb C)_-$; in particular $\mathcal R_{\mu\nu\,\rho\sigma}{}^{\rho\sigma}=0$. The $-$ block is
the complex conjugate of the $+$ block;
\item it satisfies the algebraic Bianchi identity $\mathcal R_{[\mu\nu\,\rho\sigma]}{}^{\kappa\lambda}=0$, by condition (iii), and
the differential Bianchi identity $D_{[\lambda}\mathcal R_{\mu\nu]}=0$.
\end{itemize}
In contrast to the Riemann tensor, $\mathcal R$ has no pair symmetry: the first two indices, which define the directions of the derivatives in the commutator,  and the
endomorphism slots are of different nature, and there is no metric to identify them.

To count the independent components, we use \eqref{Rtwoformomega}: $\mathcal R$ is equivalent to the
curvature $F^k_\pm(\omega)$. Decomposing the two-form index of $F^k_+$ along $\Lambda^2_+\oplus\Lambda^2_-$,
\ba
F^k_+=\Phi^k{}_m\,B^m_++\Psi^k{}_m\,B^m_-\,,\qquad F^k_-=\overline{F^k_+}\,,
\ea
with complex $3\times3$ matrices $\Phi$ and $\Psi$, gives $18$ complex, i.e.\ $36$ real, components. Since
$B_+\wedge B_-=0$, the algebraic Bianchi identity $\eps_{lkm}F^k_+\wedge B^m_+=0$ does not involve $\Psi$ and
reduces to
\ba
\eps_{lkm}\,(\Phi\,\mathbb M_+)^{km}=0\,,
\ea
with $\mathbb M_+^{nm}=\langle B^n_+,B^m_+\rangle\propto\det(e)\,m_+^{nm}$. That is, $\Phi\mathbb M_+$ is symmetric, which removes
$3$ complex components. The curvature therefore has $15$ complex, i.e.\ $30$ real, independent
components:
\begin{itemize}
\item the symmetric matrix $\Phi\mathbb M_+$ ($6$ complex). Its traceless part ($5$ complex) is of Weyl
type, and its trace ($1$ complex) contains a scalar and a pseudoscalar curvature;
\item the matrix $\Psi$ ($9$ complex), which mixes the two chiral sectors.
\end{itemize}
For metric-induced area metrics, and more generally when $D_\Gamma q=0$, $D^\omega$ is induced by the
Levi-Civita connection. In that case only $20$ of these components occur, namely those of the
Riemann tensor.  

\paragraph{Invariants.}
Under internal gauge transformations $O_\pm\in SO(3,\mathbb C)_\pm$ the curvature transforms in the
adjoint representation, $F_\pm\to O_\pm F_\pm$, and the frame index of $B^m_\pm$ rotates with it, $B_\pm\to O_\pm B_\pm$.
The blocks of the decomposition $F^k_+=\Phi^k{}_m\,B^m_++\Psi^k{}_m\,B^m_-$ therefore transform as
\ba
\Phi_\pm\to O_\pm\,\Phi_\pm\,O_\pm^{-1}\,,\qquad \Psi_+\to O_+\,\Psi_+\,O_-^{-1}\,,\qquad \Psi_-=\overline{\Psi_+}\to O_-\,\Psi_-\,O_+^{-1}\,,
\ea
i.e.\ $\Phi_\pm$ are endomorphisms of the internal bundles $V_\pm$, while $\Psi_\pm$ map between them.
Gauge-invariant scalars can thus be formed as follows:
\begin{itemize}
\item from the chiral block alone: $\tr\Phi_+$, $\tr\Phi_+^2$, $\tr\Phi_+^3$, and their complex conjugates.
The real and imaginary parts of $\tr\Phi_+$ are a scalar and a pseudoscalar curvature;
\item from the mixed block: the traces $\tr\big((\Psi_+\Psi_-)^n\big)$, $n=1,2,3$, of the endomorphism $\Psi_+\Psi_-$
of $V_+$;
\item mixed invariants such as $\tr(\Phi_+\Psi_+\Phi_-\Psi_-)$;
\item invariants that also involve the area-metric data $m_\pm=b_\pm b_\pm^T$, which transforms as
$m_\pm\to O_\pm m_\pm O_\pm^T$. Examples are $\tr(\Phi_+m_+)$ and $\tr(\Phi_+^T\Phi_+m_+)$. The algebraic Bianchi identity
states that $\Phi_+m_+$ is symmetric.
\end{itemize}
The invariants entering the action are of the last type. With
$\mathcal X^{ia}_\pm=\tfrac12\Sigma^{a\,\mu\nu}_\pm F^i_{\pm\,\mu\nu}$ and $\tfrac12\Sigma^a_\pm\cdot\Sigma^c_\pm=2\delta^{ac}$, and since
$\Sigma_\pm$ is orthogonal to $B_\mp$, one has
\ba
\mathcal X_\pm=2\sigma_\pm\,\Phi_\pm\,b_\pm\,.
\ea
Hence the BF term is $\tr(b_\pm^T\mathcal X_\pm)=2\sigma_\pm\tr(\Phi_\pm m_\pm)$, and the curvature-squared terms are built from
$\mathcal X_\pm^T\mathcal X_\pm=-4\,b_\pm^T\Phi_\pm^T\Phi_\pm b_\pm$.

Some invariants can be written without reference to the frame. The two-form
\ba
\mathcal K_{\rho\sigma}:=\mathcal R_{\mu\nu\,\rho\sigma}{}^{\mu\nu}
\ea
vanishes in the metric case, where it is the antisymmetric part of the Ricci tensor. It therefore
directly measures the non-metricity. Further invariants follow by contraction with the
inverse area metric. Another invariant is the density $\epst^{\mu\nu\rho\sigma}\tr(\mathcal R_{\mu\nu}\mathcal R_{\rho\sigma})$, which is topological and of Pontryagin type.

\section{Equations of motion for $b$-fields and effective actions for the length metric}\label{Sec:beom}

Let us consider the equations of motion for $b_\pm$. Here we have to take into account the sharp constraints ${\rm Im}\,\tr q_\pm=0$, which is imposed via the Lagrange multiplier terms $\mp\tfrac{\im}{3}\varphi_0 q^\pm_\tr\,\vol$.

Instead of directly varying the effective action \eqref{SBFeff} with respect to $b$, we will use a shortcut. We will vary $S_{BF}$ at \emph{fixed} $\omega$ in the
original BF action \eqref{SBFchiral}, and then re-insert the solution for the connection. 
\ba
\delta S^{\rm eff}_{BF}&=&-\sum_\pm\gamma_\pm
\int\delta B^i_\pm\wedge F^i_\pm(\omega_\pm)  \q \text{with}\q  \delta B^i_\pm=\sigma_\pm\,\delta b_\pm^i{}_a
\Sigma_\pm^a \nn\\
&=:&\;\;\;  \sum_\pm\gamma_\pm\int\vol\;\delta b^i{}_a\,\mathcal X^{ia}_\pm
\ea

 Using \eqref{iddual} we have $\Sigma^a_\pm\wedge F^i_\pm=\pm\tfrac{\im}{2}\,
\Sigma^{a\,\mu\nu}_\pm F^i_{\pm\,\mu\nu}\,\vol$, this cancels out the sector sign $\sigma_\pm$.
We obtain for $\mathcal X_\pm$
\ba\label{Xomega}
\mathcal X^{ia}_\pm&=&\tfrac12\,\Sigma^{a\,\mu\nu}_\pm\,F^i_{\pm\,\mu\nu}(\Gamma_\pm+\rho_\pm)\nn\\
&=&X^{ia}_\pm+\tfrac12\Sigma^{a\,\mu\nu}_\pm (d_\Gamma \rho^i_{\pm})_{\mu\nu}
-\tfrac12\,\eps^i{}_{jk}\,\Sigma^{a\,\mu\nu}_\pm\rho^j_{\pm\,\mu}\rho^k_{\pm\,\nu}\,,
\ea
using $F(\Gamma+\rho)=F(\Gamma)+d_\Gamma\rho
-\tfrac12[\rho,\rho]$.

 With $\Pi_\pm:=\partial L_{\rm pot}/
\partial q_\pm$ from \eqref{Spot}, the equation for $b_+$ reads 
\begin{equation}\label{beom}
\gamma_+\,\mathcal X^{ia}_+
\,=\,\frac{2\im}{3}\,\varphi_0\,b^i{}_a-2\,b^i{}_b\,\Pi_+^{\,ba}\,,
\end{equation}
where we have to insert the solution \eqref{rhoexplicit} for $\rho$ and $\Pi_+$ is given by
\begin{equation}\label{Pidef}
\Pi_+=-\frac{\mu}{q^+_\tr+q^-_\tr}\,\mathring q_+
+\bigg[\frac{\mu\tr\mathring q_+^2+\bar{\mu}\tr\mathring
q_-^2}{2\,(q^+_\tr+q^-_\tr)^{\,2}}-\tfrac13\Lambda\bigg]\delta\,.
\end{equation}


Let us investigate whether $q_\pm=\delta$ can be a solution. Inserting $b_\pm=\delta$ into \eqref{beom} we obtain 
\be
\gamma_+X^{ia}_+=\tfrac{2\im}3\varphi_0\delta^{ia}+\tfrac{2\Lambda}3\delta^{ia}\, ,
\ee
its traceless part gives $\mathring X_+=0$, and its trace, with $\tr X_+=\tfrac R2$
gives 
$(1+\im/\gamma)\tfrac R2=2\im\varphi_0+2\Lambda$. The imaginary part of this equation gives $\varphi_0=R/(4\gamma)$. The real part of this equation corresponds actually to the variation of $b$ in the scaling direction (described in at the end of Section \ref{sec:param}), which is however fixed by the normalisation condition $|\det b_+|=1$. 

 Thus, setting the non-length metric degrees of freedom  to $q_\pm=\delta$ forces\ the Weyl tensor (encoded in $\mathring X_\pm$) to vanish.  Indeed, in the analysis of the linearised theory in \cite{Dittrich26} gravitational waves are dressed with a $b$-field.  We can also illustrate this by assuming $D_{\Gamma_\pm}b=0$ (which is justified if curvature is varying slowly on the scale $\sqrt{|\gamma_+/\mu_+|}$). The traceless part of the $b$ equation of motion, at $q_\tr^\pm=3$ and leading order in $\mathring q_\pm$, is then
\begin{equation}\label{qWeyl}
\mathring q_\pm=\frac{3\gamma_\pm}{\mu_\pm}\,\mathring X_\pm
+O(\mathring q^2)\, .
\end{equation}
Thus the non-length metric degrees of freedom $\mathring q$ are algebraically enslaved to the self-dual Weyl
curvature. 

\paragraph{Expansion in $b=\delta +\tfrac12 \chi$.} To be more systematic we use the second-order action \eqref{Seffchi} of
Section~\ref{sec:intcon}. Its $\mathring\chi$-equation, with $\Box$ the covariant
d'Alembertian built from \eqref{DGammachi},
\begin{equation}\label{chieq}
\frac{\gamma_\pm}{2}\mathring X_\pm-\frac{\hat\mu_\pm}{6}\mathring\chi^\pm
+\frac{\gamma_\pm}{4}\Box\mathring\chi^\pm=O(\mathring X\mathring\chi)\,,
\end{equation}
is solved by
\begin{equation}\label{chisol}
\mathring\chi^\pm=\frac{3\gamma_\pm}{\hat{\mu}_\pm}\Big(1-\frac{\Box}{\mathfrak m_\pm^2}\Big)^{-1}
\mathring X_\pm\,,\qquad
\mathfrak m^2_\pm:=\frac{2\hat{\mu}_\pm}{3\gamma_\pm}
=\frac{2\mu_\pm}{3\gamma_\pm}+\frac{2R}3+\frac{4\Lambda-R}{3\gamma_\pm}\,.
\end{equation}
The curvature shift of the mass has two sources:
\begin{itemize}
\item the trace-sector pieces $\tfrac{4\Lambda}{3\gamma_\pm}$ and $-\tfrac{\gamma_\mp R}
{6\gamma_\pm}$, which  by $\gamma_++\gamma_-=2$
sum to $\tfrac R6+\tfrac{4\Lambda-R}{3\gamma_\pm}$;
\item the kinetic curvature term, which adds the real, chirality-blind $\tfrac R2$.
\end{itemize}

We will see that in the case of a flat background choosing $\mu_\pm=\tfrac32 \gamma_\pm m^2$ with real $m^2$ avoids a ghost term in the effective length propagator. With this value for $\mu_\pm$
the Immirzi-odd part is  proportional to $4\Lambda-R$, the trace of the
background field equations (without matter). It vanishes on backgrounds with $R=4\Lambda$, and off them
it can be removed by a field redefinition at this order.  (For
$R=4\Lambda$, we have  $\mathfrak m^2_\pm=
\tfrac{2\mu_\pm}{3\gamma_\pm}+\tfrac{8\Lambda}3$.) 

Since \eqref{Seffchi} is Gaussian, re-insertion gives half the coupling term,
\begin{equation}\label{SeffWeyl}
S^{\rm eff}[e]=\int\sqrt{-g}\,(R-2\Lambda)
+\sum_\pm\frac{\gamma_\pm}{2}\int\vol\;
\mathring X_\pm\cdot\big(\mathfrak m^2_\pm-\Box\big)^{-1}\mathring X_\pm
+O(\mathring X^3)\,.
\end{equation}
To read this as a curvature action, we use the identity 
\begin{equation}\label{Xsq}
\mathring X_\pm\cdot\mathring X_\pm
=\tfrac12\big(C_{\mu\nu\rho\sigma}C^{\mu\nu\rho\sigma}
\mp\im\,C_{\mu\nu\rho\sigma}\tilde C^{\mu\nu\rho\sigma}\big)\,,\qquad
\tilde C_{\mu\nu\rho\sigma}:=\tfrac12\eps_{\rho\sigma\alpha\beta}C_{\mu\nu}{}^{\alpha\beta}\,,
\end{equation}
so that 
the local     
limit $\Box\ll\mathfrak m^2$ of \eqref{SeffWeyl} is
\begin{equation}\label{SeffLocal}
S^{\rm eff}[e]=\int\sqrt{-g}\,\Big[R-2\Lambda
+\tfrac34\,{\rm Re}\Big(\frac{\gamma_+^2}{\hat{\mu}_+}\Big)\,C_{\mu\nu\rho\sigma}
C^{\mu\nu\rho\sigma}\Big]
+\tfrac34\,{\rm Im}\Big(\frac{\gamma_+^2}{\hat{\mu}_+}\Big)\int\sqrt{-g}\,
C_{\mu\nu\rho\sigma}\tilde C^{\mu\nu\rho\sigma}+O(\Box/\mathfrak m^2)\,,
\end{equation}
an Einstein--Weyl action. $C\tilde C$ is equal to the Pontryagin density $R\tilde R$  in
four dimensions, which is a total derivative.

\paragraph{All orders in $\Box$.}
At quadratic order in the Weyl tensor the form factor can be kept exact. 
Because $\Sigma^a_\pm$ is parallel with respect to the combined Levi-Civita connection
(the Christoffel connection on spacetime indices and the chiral spin connection
$\Gamma_\pm(e)$ on the internal index), the internal $\Box=g^{\mu\nu}D^\Gamma_\mu D^\Gamma_\nu$ acts on
$\mathring X_\pm=\tfrac14\Sigma C\Sigma$ as $\tfrac14\Sigma(\Box_gC)\Sigma$, with $\Box_g$ the Levi-Civita
d'Alembertian on the Weyl tensor.

The polarisation of \eqref{Xsq} therefore holds for any function $f$:\\ \,\,
$\mathring X_\pm\cdot f(\Box)\mathring X_\pm=\tfrac12(C f(\Box)C\mp\im\,Cf(\Box)\tilde C)$,
and \eqref{SeffWeyl} becomes
\ba\label{SeffAll}
S^{\rm eff}[e]&=&\int\sqrt{-g}\,(R-2\Lambda)+\frac12\int\sqrt{-g}\,\Big[C\,F_C(\Box)\,C
+C\,\tilde F_C(\Box)\,\tilde C\Big]+O(C^3)\,, \q \text{with} \nn\\
 && F_C+\im\tilde F_C:=\frac{\gamma_+}{\mathfrak m^2_+-\Box}\,.
\ea
Its first Taylor coefficients reproduce \eqref{SeffLocal}. The next order is
\ba
\tfrac98\int\sqrt{-g}\,[{\rm Re}(\gamma_+^3/\hat\mu_+^2)\,C\Box C+{\rm Im}
(\gamma_+^3/\hat\mu_+^2)\,C\Box\tilde C]\, .
\ea
 There the parity-odd term is no longer
topological, since $C\Box\tilde C=-\nabla C\cdot\nabla\tilde C$ up to a total derivative.

\paragraph{Flat background.}
Around Minkowski space we have $\hat{\mu}=\mu$ and $\mathfrak m^2_\pm=\tfrac23\mu_\pm/\gamma_\pm$.
The $C\tilde F_C\tilde C$ term
vanishes identically at quadratic order in $h$, for any $\tilde F_C$: the momentum-space
kernel of the quadratic Pontryagin density vanishes. On the spin-2 sector the effective
propagator with $P$ the eigenvalue of $\Box$, is
\begin{equation}\label{propflat}
G_{\rm eff}(P)=\frac{|\mathfrak m^2_+-P|^2}{P\,\big(|\mathfrak m^2_+|^2-{\rm Re}(\gamma_+\mathfrak m^2_+)P\big)}
=\frac1P+\frac{r}{P-m^2_{\rm pole}}-\frac{1}{{\rm Re}(\gamma_+\mathfrak m^2_+)}\,,
\end{equation}
where
\begin{equation}\label{flatpole}
m^2_{\rm pole}=\frac{2\,|\mu_+|^2}{3\,|\gamma_+|^2\,{\rm Re}\,\mu_+}\,,\qquad
r=-\frac{|\gamma_+{\rm Re}\,\mu_+-\mu_+|^2}{|\gamma_+|^2\,({\rm Re}\,\mu_+)^2}
\le0\,,
\end{equation}
(Appendix~\ref{app:flatFRW}). The massive mode therefore leads to a ghost unless ${\rm Re}\,\mu_+=\mu_+/\gamma_+$, that is $\mathfrak m^2_\pm=\tfrac23\mu_\pm/\gamma_\pm$ is real. We choose $\mu_\pm=\tfrac32\gamma_\pm m^2$
with real $m^2$ and hence $\mathfrak m^2_\pm=m^2$.
 In that case $G_{\rm eff}=1/P-1/m^2$, the equal-mass, shift-symmetric
case of \cite{Freidel:2008ku, Borissova:2022clg,BDK, Dittrich26}.

\paragraph{Degrees of freedom around flat space.}
The field $\mathring\chi^+$ is complex, with $\mathring\chi^-=\overline{\mathring\chi^+}$,
so there are ten real non-length components, each with a second-order kinetic term in
\eqref{Seffchi}. Naively one would therefore expect two massive spin-2 fields.
However, the curvature coupling $\tfrac{\gamma_\pm}{2}\mathring X_\pm(h)\cdot\mathring\chi^\pm$
contains two derivatives and mixes $\mathring\chi$ kinetically with the metric. On flat
space $\mathring X\simeq\tfrac12\Box h$ on each component of the off-shell spin-2
projection of $h$. Per component, with amplitudes $c_\pm$ of
$\mathring\chi^\pm$ and $P$ the eigenvalue of $\Box$, the quadratic Lagrangian is
\begin{equation}\label{Lspin2}
L=\tfrac14Ph^2+\sum_\pm\Big[\tfrac{\gamma_\pm}{4}P\,h\,c_\pm+\tfrac{\gamma_\pm}{8}P\,c_\pm^2
-\tfrac{\mu_\pm}{12}c_\pm^2\Big]\,.
\end{equation}
The matrix of $P$-coefficients, i.e.\ the kinetic matrix of $(h,c_+,c_-)$, has
determinant
\begin{equation}\label{detK}
\det K=\frac{\gamma_+\gamma_-}{64}\Big[\frac14-\frac{\gamma_++\gamma_-}{8}\Big]=0\,.
\end{equation}
It vanishes identically as $\gamma_++\gamma_-=2$, for every $\gamma$ and every $\mu_\pm$;
this is the shift symmetry of \cite{Borissova:2022clg,BDK,Dittrich26}.
One
combination of $(h,c_+,c_-)$ is therefore auxiliary. Per component only a massless and
a single massive field propagate, and gauge invariance reduces the massless one to the
two graviton polarisations. The spectrum thus consists of the graviton and one massive
spin-2 field, $2+5=7$ degrees of freedom. The trace of $b$ adds no scalar, since it is
fixed at second order by \eqref{tauFix}. This agrees with the canonical analysis of
\cite{BDK}, with the single massive pole of \eqref{propflat}, and with the flat
determinant of the linearised theory.
 
The limit $\gamma\to\infty$ ($\gamma_\pm=1$) with real $\mu$ makes the structure
explicit. Write $c_+=u+\im v$. The Lagrangian becomes
$P[\tfrac14(h+u)^2-\tfrac14v^2]-\tfrac{\mu}{6}(u^2-v^2)=P[\tfrac14 H^2-\tfrac14v^2]-\tfrac{\mu}{6}(u^2-v^2)$, with $H:=h+u$. Integrating out $u$ gives $u=0$ and leaves
the massless field $H$ and the massive field $v={\rm Im}\,c_+$, with Lagrangian
$-\tfrac14(P-\mathfrak m^2)v^2$ and $\mathfrak m^2=\tfrac23\mu$. The massive mode
thus has a negative kinetic term, i.e.\ it is the negative-energy massive branch of
\cite{BDK,Dittrich26}, and in this limit it does not couple to $h$. At finite $\gamma$ the
massive direction is rotated. It decouples from the metric channel exactly for
$\mu_\pm=\tfrac32\gamma_\pm m^2$, and otherwise contributes the residue $r\le0$ of
\eqref{flatpole}. The degeneracy \eqref{detK} that removes the second spin-2 field is
therefore also what allows the remaining one to decouple.

\paragraph{FRW background.}
Take matter coupled to the tetrad only, e.g.\ a scalar field with stress tensor
$T_{\mu\nu}$; the tetrad is $e[B]$ in the gauge $|\det b|=1$. On a conformally flat FRW
background:
\begin{itemize}
\item $q_\pm=\delta$ remains an exact solution of the full theory, with multiplier 
$\varphi_0=R/(4\gamma)$;
\item the background obeys general relativity exactly, $G_{\mu\nu}+\Lambda g_{\mu\nu}=
\tfrac12T_{\mu\nu}$, so $R=4\Lambda-\tfrac12T$;
\item with $\mu_\pm=\tfrac32\gamma_\pm m^2$, \eqref{chisol} gives $\mathfrak
m^2_\pm=m^2+\tfrac{2R}3+\tfrac{4\Lambda-R}{3\gamma_\pm}$, hence ${\rm Im}\,\mathfrak m^2_+
=-\gamma T/(6(1+\gamma^2))$.
\end{itemize}
For non-conformal matter ($T\neq0$) the massive mode thus leaks into the metric
channel. In the adiabatic regime, where we have for the Hubble rate $H\ll m$ and where the rate of change of $T$ is  also small as compared to $m$, the flat-space analysis applies with frozen coefficients. Thus the residue of the massive mode is
\begin{equation}
r\simeq-\frac{T^2}{36(1+\gamma^2)\,m^4}
\end{equation}
(Appendix~\ref{app:flatFRW}). The consequences are a ghost-sign Yukawa term in the
potential of a static source and ghost emission by sources with $\omega>m$, both
negligible for Planckian $m$. Still, ghost freedom (with minimal matter coupling to the length metric) holds only on backgrounds with
$T=0$, i.e.\ in vacuum with $\Lambda$ and for conformal matter.

\emph{Restoring ghost freedom by coupling matter to $\chi$.} The background equation of motion $R=4\Lambda-\tfrac12T$ lead to the mass parameter
\begin{equation}
\hat\mu_\pm=\mu_\pm+\gamma_\pm\Lambda+\tfrac34\gamma_\pm R+\tfrac18\gamma_\mp T\,,
\end{equation}
whose last term is not proportional to $\gamma_\pm$. This last term can be cancelled by a coupling of
matter to the non-metric sector that is quadratic in $\mathring\chi$ and proportional to $T$,
\begin{equation}\label{SintT}
S_{\rm int}=\sum_\pm\int\vol\;\frac{\gamma_\mp}{96}\,T\,\mathring\chi^\pm\cdot\mathring\chi^\pm
=\int\vol\;\frac{T}{48}\,{\rm Re}\big(\gamma_-\,\mathring\chi^+\cdot\mathring\chi^+\big)\, .
\end{equation}
For $\mu_\pm=\tfrac32\gamma_\pm m^2$ one then has
$\hat\mu_\pm=\gamma_\pm\big(\tfrac32m^2+\Lambda+\tfrac34R\big)$, so that $\mathfrak m^2_\pm=m^2+\tfrac23\Lambda+\tfrac12R$
is real. The  residue (computed with frozen coefficients) vanishes for any matter content. But $\mathfrak m^2_\pm$ becomes time-dependent through $T$ and terms with $\dot T$ need also to be checked.

A covariant realisation of this coupling is to
couple matter to the rescaled metric $\Omega^2g$ with
\begin{equation}
\Omega^2=1+\tfrac1{24}\,{\rm Re}\big(\gamma_-\tr\mathring q_+^{\,2}\big)+O(\mathring q^3)\,,
\end{equation}
since $\delta S_m=\tfrac12\int\sqrt{-g}\,T\,(\Omega^2-1)$ to this order and $\mathring q_\pm=\mathring\chi^\pm+O(\mathring\chi^2)$. The combination
${\rm Re}(\gamma_-\tr\mathring q_+^2)={\rm Re}\tr\mathring q_+^2+\gamma^{-1}{\rm Im}\tr\mathring q_+^2$ is a quadratic invariant of the
non-metric part of the area metric, with a parity-odd part weighted by the Barbero--Immirzi
parameter. Such a coupling is non-minimal: matter feels the non-metric field at order
$\mathring q^2$, which violates the standard equivalence principle at that order and adds $\mathring q^2$-terms to the
matter stress tensor.

\paragraph{Higher-order potential.}
Taking higher orders in $\mathring \chi_\pm$ into account will generate higher order invariant contractions of $\mathring X_\pm$, also at lowest order in the derivative expansion. Since $\mathring X_\pm$ is a symmetric traceless
$3\times3$ matrix its independent invariants are the second order $I_\pm=\tr\mathring X_\pm^2$ and the third order 
$J_\pm=\tr\mathring X_\pm^3$, higher order invariants can be build from polynomials in $I_\pm$ and $J_\pm$. Note that mixtures between the $+$ and $-$ sectors can occur, as we have already seen indications for such a mixture at quadratic order.

 $I_\pm$ is given in (\ref{Xsq}) and $J_\pm$ by
\begin{equation}\label{Xcube}
\tr\mathring X_\pm^3
=\tfrac12\big(C_{\mu\nu}{}^{\rho\sigma}C_{\rho\sigma}{}^{\alpha\beta}C_{\alpha\beta}{}^{\mu\nu}
\mp\im\,C_{\mu\nu}{}^{\rho\sigma}C_{\rho\sigma}{}^{\alpha\beta}\tilde C_{\alpha\beta}{}^{\mu\nu}\big)\, .
\end{equation}
 Thus, the real and imaginary parts of $J_\pm=\tr\mathring X_\pm^3$ are the
parity-even and parity-odd cubic Weyl invariants (in four dimensions the other
index contraction of three Weyl tensors are not independent).   

Keeping a quadratic potential, the couplings for the higher order terms are determined by $\mu_\pm$ and $\Lambda$ and $\gamma$.  Introducing a higher order potential as discussed in Section \ref{higherorderpot} introduces additional couplings for these terms, independent of $\mu_\pm$.

\section{Poincar\'e gauge form of the action}\label{sec:poincare}

We have seen that the area metric action can be interpreted in terms of curvature and torsion. We also have, apart from the massless graviton, an extra propagating massive spin 2 degree of freedom, that can be linked to the torsion. 

Poincar\'e gauge theories \cite{Blagojevic:2003cg,Blagojevic:2013xpa} with curvature-squared terms can propagate torsion.  The variables are a tetrad and an independent Lorentz connection $\omega^{IJ}$ with Riemann--Cartan curvature $F^{IJ}(\omega)=d\omega^{IJ}+\omega^I{}_K\wedge\omega^{KJ}$.

Our theory can be brought into the same Poincar\'e-gauge form. Before integrating out the connection $\omega$ the
field $b$ enters the action algebraically, but non-polynomially: $B\wedge F(\omega)$ is linear in $b$ and the
potential depends on $q$ non-polynomially. One may therefore integrate out $b$
instead of $\omega$.

After integrating out the Lagrange multipliers the modified Plebanski action (with quadratic potential) is given by
\begin{equation}\label{SPGT}
S[e,\omega,b,\psi]=\int d^4x\,|e|\,\Big[\sum_\pm\gamma_\pm\tr\big(b_\pm\mathcal X_\pm\big)
-\frac{\mu\tr\mathring q_+^{\,2}+\bar\mu\tr\mathring q_-^{\,2}}{4\,q^+_\tr}-\tfrac23\Lambda\,q^+_\tr\Big]\,,
\end{equation}
and is subject to $|\det b_\pm|=1$, ${\rm Im}\,q^+_\tr=0$ and $b_-=\overline{b_+}$. Here we made use of the chiral $\Sigma$-frame components of the curvature
\begin{equation}
\mathcal X^{ia}_\pm(e,\omega):=\tfrac12\,\Sigma^{a\,\mu\nu}_\pm(e)\,P_{\pm i\,IJ}\,R^{IJ}{}_{\mu\nu}(\omega)\,,
\end{equation}
with $\mathcal X_-=\overline{\mathcal X_+}$. 

The equations of motion are the same as derived in Section (\ref{Sec:beom}) (but now with an independent connection). Tat is
\ba\label{beomPGT}
\gamma_\pm\,\mathcal X_\pm &=&-2\,b_\pm (\Pi_\pm\mp\tfrac{\im}{3}\varphi_0\,\delta)   \, 
\q \text{with}\nn\\
\Pi_\pm&:=&
-\frac{\mu_\pm}{q^+_\tr+q^-_\tr}\,\mathring q_\pm
+\Big[\frac{\mu\tr\mathring q_+^{\,2}+\bar\mu\tr\mathring q_-^{\,2}}{2\,(q^+_\tr+q^-_\tr)^2}-\frac\Lambda3\Big]\delta\, .
\ea
Here $\varphi_0$ is the Lagrange multiplier implementing ${\rm Im}\,q^+_\tr=0$. The real part of the trace of
\eqref{beomPGT} is the scaling direction $(e,b)\to(\lambda e,\lambda^{-2}b)$, fixed by
$|\det b_\pm|=1$.

Since $\hat\Pi_\pm$ is a function of $q_\pm$ and hence commutes with it,
\eqref{beomPGT} can be rewritten into the following equation system for $b, q=b^Tb$ and $\varphi_0$:
\begin{equation}\label{bsolPGT}
b_\pm=-\frac{\gamma_\pm}{2}\,\mathcal X_\pm\,\hat\Pi_\pm^{-1}\,,\qquad
\mathcal X_\pm^T\mathcal X_\pm=\frac{4}{\gamma_\pm^2}\,q_\pm\,\hat\Pi_\pm^{\,2}\, , \q \text{with} \q \hat\Pi_\pm:=\Pi_\pm\mp\tfrac{\im}{3}\varphi_0\,\delta \, .
\end{equation}

To determine the Lagrange multiplier we multiply \eqref{beomPGT} from the left
by $b_+^T$ and take the trace, which gives, with $q^+_\tr$ real,
\begin{equation}
\gamma_+\tr\big(b_+^T\mathcal X_+\big)+2\tr\big(q_+\Pi_+\big)=\tfrac{2\im}{3}\,\varphi_0\,q^+_\tr\, .
\end{equation}
The real part is the scaling
direction, which is fixed by $|\det b_\pm|=1$.
The imaginary part determines $\varphi_0$:
\begin{equation}\label{phi0sol}
\varphi_0=\frac{3}{2q^+_\tr}\Big[{\rm Im}\big(\gamma_+\tr(b_+^T\mathcal X_+)\big)-\frac{{\rm Im}(\mu\,\tr\mathring q_+^{\,2})}{q^+_\tr}\Big]\,.
\end{equation}

The system of equations of motion for $b$ is non-linear, but we can solve it in a perturbative expansion $b_\pm=\delta$ with $b_\pm=\delta+\tfrac12\mathring\chi^\pm+\tfrac16\chi^\pm_\tr
\,\delta$  (with $\mathring\chi^\pm$ symmetric and traceless).  The linearised solution is $\mathring\chi^\pm=3\gamma_\pm\mathring{\mathcal X}^s_\pm/\tilde\mu_\pm$. In order to express the action evaluated on this solution we decompose  $\mathcal X_\pm=\tfrac13(\tr\mathcal X_\pm)\delta+\mathring{\mathcal X}^{s}_\pm+\mathcal X^{a}_\pm$
(trace, symmetric traceless and antisymmetric parts).
We obtain to quadratic order in
$\mathring{\mathcal X}^s$
\begin{equation}\label{SPGTquad}
S_{\rm PGT}[e,\omega,\psi]=\int d^4x\,|e|\,\Big[\sum_\pm\gamma_\pm\tr\mathcal X_\pm-2\Lambda
+\sum_\pm\frac{3\gamma_\pm^2}{4\tilde\mu_\pm}\,\tr\big(\mathring{\mathcal X}^{s}_\pm\big)^2\Big]
+O(\mathring{\mathcal X}^3)\,,
\end{equation}
\begin{equation}
\tilde\mu_\pm=\mu_\pm+2\Lambda-\tfrac12\gamma_\mp\tr\mathcal X_\mp\,,
\end{equation}
where $\mu_+=\mu$, $\mu_-=\bar\mu$.

The first term is the Einstein--Cartan--Holst action; the last
contains the parity-even and parity-odd squares of the Weyl-type part of the Riemann--Cartan
curvature, with coefficients ${\rm Re}$ and ${\rm Im}$ of $3\gamma_+^2/(4\tilde\mu_+)$. 

Within Poincare gauge theories, torsion
bigravity \cite{Nair:2008yh,Nikiforova:2009qr,
Damour:2019oru,Nikiforova:2020sac} has also a spectrum consisting of
a massless and a massive spin-2 field, as in our case.   Its action has however significant differences to (\ref{SPGTquad}). 
Torsion bigravity squares the symmetric Ricci part of
 the connection determined curvature and includes a separate Ricci scalar $R(e)$ built from the tetrad. In contrast, equation \eqref{SPGTquad} squares the chiral
Weyl part and has no separate $R(e)$ term.  Additionally torsion bigravity is parity even, whereas our action includes parity even and parity odd terms. 

It would be interesting to understand whether torsion bigravity and other Poincare gauge theories can be treated
within the area-metric framework.

\section{Discussion}\label{sec:disc}

We have constructed non-linear area-metric dynamics from modified Plebanski actions \cite{Krasnov:2006du,Krasnov:2008fm,Krasnov:2009iy}. Instead of a tetrad one form $e^I_\mu$ the Plebanski framework \cite{Plebanski:1977zz} works with a $\mathfrak{so}(3,1)$ valued two-form field $B^{IJ}_{\mu\nu}$, on which simplicity constraints are imposed, so that $B=\pm \star (e\wedge e)$.  The modified Plebanski actions replace the simplicity constraints with a potential, allowing fluctuations of $B$ away from simple $B=\pm \star (e\wedge e)$ configurations. In this work, as in \cite{Borissova:2022clg}, we split the simplicity constraints into two sets. One set is sharply imposed, the second set is replaced by a potential. The resulting configuration space is the space of
cyclic area metrics, so that the theory provides a non-linear completion of the area-metric
dynamics found in the continuum limit of effective spin foams \cite{Dittrich:2022yoo} and in  \cite{Borissova:2022clg}.

A main outcome of this work is a set of geometric tools for area metrics that do not rely
on perturbation theory. We have given an algorithm that extracts a length metric from a
given area metric, and shown that the corresponding map is locally invertible under a mild
non-degeneracy condition. The $B$-fields of the Plebanski
formulation provide the soldering form that area-metric geometry otherwise lacks. This
allowed us to define a connection on two-forms that is uniquely determined by compatibility
with the area metric, preservation of the split into self-dual and anti-self-dual parts,
and torsion-freeness. The complex structure underlying this split can itself be read off from
the area metric. The curvature of this connection contains the Riemann tensor together with
additional components generated by the non-metricity, and its invariants can be used to build
area-metric actions well beyond the modified Plebanski class considered here. Length metric,
connection, curvature and invariants together provide what is needed to construct and
analyse area-metric actions systematically at the non-linear level.

Applied to the modified Plebanski action, these tools give the dynamics in several
equivalent forms. Integrating out the Lagrange multipliers we obtain a first order version of the area metric actions. Integrating out the connection leads to an action for the length metric
and the non-metric fields. Around flat space it describes the graviton together with one
massive spin-2 field. For a particular family of masses in the potential the massive field does not lead to extra poles or ghosts in the propagator for the effective length dynamics, but only to a contact term.  Keeping the connection and
eliminating the non-metric fields instead gives a Poincar\'e gauge theory with propagating
torsion, which places the model within the broader family of torsion theories of gravity. We
also showed that Friedmann--Robertson--Walker geometries are exact solutions. With matter
coupled only to the length metric, the ghost freeness holds only for
conformal matter.  But it can be restored by a coupling of matter to the non-length metric degrees of freedom.

Finally, the sharply imposed constraints include a trace constraint, which removes the totally antisymmetric part of
the area metric. Relaxing this constraints adds a healthy pseudoscalar field with an axion-like coupling.

With these tools in place, applications to cosmology are now within reach. The non-linear
action allows one to study the early universe and the evolution of perturbations on
cosmological backgrounds, and to follow the parity-violating terms, which are parametrised by the Barbero--Immirzi parameter, beyond the linear regime.
If the electromagnetic field couples to the area metric rather than to the length metric,
the non-metric modes and the axion-like trace mode lead to birefringence. This links the
framework to observational tests with gravitational waves and with the polarisation of the
cosmic microwave background. This would strengthen the link between spin foams and the derivation of parity violating terms in effective continuum actions \cite{Bianchi:2024mrt}.  Strong-field configurations such as black holes \cite{Borissova:2025nvj},
where the non-metricity can become large, are a second natural field of application.

Several conceptual questions remain open. A complete canonical analysis of the non-linear
theory is needed to settle the number of degrees of freedom beyond perturbation theory and
the well-posedness of the evolution, in particular in the presence of matter.  As we have seen in the example of FRW cosmology, the type of matter coupling can influence these questions. It will be interesting to see whether matter couplings to area metrics can be determined by demanding a stable dynamics. The
curvature invariants constructed here open the way to more general area-metric actions and
to a systematic comparison with other Poincar\'e gauge theories with propagating torsion,
such as torsion bigravity. Here one can also compare how matter is coupled in the two formulations.

The massive spin-2 field
carries negative energy \cite{BDK, Dittrich26}. It is therefore important to understand whether its decoupling
from the metric persists at the non-linear level, whether it is stable under quantum
corrections, and how the couplings behave under the renormalisation group
\cite{Borissova:2025frj,Krasnov:2006du,Krasnov:2008fm,Krasnov:2009iy}. One will also need to understand which classes of area metrics to admit. One possibility is to demand certain regularity conditions with regard to causality and matter field evolution as in \cite{Schuller:2009hn}. We note however, that this question is not even fully settled for quantum gravity approaches working with (discrete) length metrics \cite{Dittrich:2026rxr}. 

On the side of loop quantum gravity, the parameters of the
continuum action should be related to those of spin foams, and the non-linear
dynamics compared with their (perturbative) continuum limit beyond the linearised level \cite{Dittrich:2021kzs,Dittrich:2022yoo}. The area metric action captures the effects resulting from the extension of the configuration space in spin foams from length to area metrics. Of course, there could also be quantum effects that the (classical) action misses, and it is important to understand the regime in which the continuum action does apply. To this end it will be helpful to improve our understanding of the non-perturbative continuum limit of spin foams \cite{Dittrich:2012jq,Dittrich:2014ala,Delcamp:2016dqo,Bahr:2016hwc,Asante:2022dnj,Bruno:2026xzh}.

The geometric tools here might also be helpful to improve spin foam dynamics on the discrete level. The construction of a unique connection determined by the area metric can be translated to the discrete set-up in \cite{Dittrich:2023ava} and will provide a better geometric understanding of the discrete geometry, possibly in terms of a higher category set-up \cite{Asante:2019lki,Girelli:2021zmt}.  Since this geometry can be interpreted in terms of area metrics, constructing matter couplings in spin foams \cite{Oriti:2002bn, Kisielowski:2018oiv,Ali:2022vhn} will also profit from a better understanding of how to couple matter to area metrics in the continuum. 

In summary, area metrics are a feature that appears across a number of quantum gravity approaches and might help to identify common properties \cite{deBoer:2022zka}. Having a systematic way to construct area metric actions will help to study the viability of models and extract phenomenological predictions. 

\vspace{2cm}

\section*{Acknowledgement}	
	
Research at Perimeter Institute is supported in part by the Government of Canada through the Department of Innovation, Science and Economic Development Canada and by the Province of Ontario through the Ministry of Colleges and Universities.

\newpage

\begin{appendix}

\section{The chiral Cartan equation and $\Gamma_\pm(\Sigma)$}\label{app:Gamma}

\paragraph{Chiral projection.}
To obtain the Cartan equation for $\Sigma_\pm$ 
we apply the Cartan equation twice in 
\begin{equation}\label{dee}
d(e^I\wedge e^J)=de^I\wedge e^J-e^I\wedge de^J
=-\Gamma^I{}_K\wedge e^K\wedge e^J-\Gamma^J{}_K\wedge e^I\wedge e^K\, \, .
\end{equation}
 Now apply $P_{\pm a\,IJ}$. The left-hand side gives $d\Sigma^a_\pm$, with
$\Sigma^a_\pm=P_{\pm a\,IJ}e^I\wedge e^J$. The right-hand side is the adjoint action
$-[\Gamma,e\wedge e]^{IJ}$. For it we use the intertwiner property of the chiral projectors,
$P_{\pm a}[\theta,\,\cdot\,]=\eps_a{}^{bc}(P_\pm\theta)_b(\,\cdot\,)_c$ for $\theta\in
\mathfrak{so}(3,1)$, which expresses that the two chiral subalgebras commute.
The right-hand side thus becomes $\eps^a{}_{bc}\Gamma^b_\pm\wedge\Sigma^c_\pm$, with
$\Gamma^a_\pm=P_{\pm a\,IJ}\Gamma^{IJ}$, which is \eqref{cartanchi}. Conversely, for a
non-degenerate tetrad in four dimensions, $d_\Gamma(e\wedge e)=0$ implies vanishing
torsion. Both sectors of \eqref{cartanchi} together are therefore equivalent to the
Cartan equation, and each sector separately determines $\Gamma_\pm$ uniquely, as shown
below.

\paragraph{Two lemmas.}
First, for any one-form $\alpha$,
\begin{equation}\label{dualemma}
\frac{1}{3!\det e}\epst^{\sigma\mu\nu\rho}(\alpha\wedge\Sigma^a_\pm)_{\mu\nu\rho}
=\pm\im\,(\Sigma^a_\pm)^{\sigma\nu}\alpha_\nu\,,
\end{equation}
which follows from \eqref{iddual}. 

Second, the operator $\mathbb S_\pm$ ( defined in \eqref{Sop}: $(\mathbb{S}_\pm t)^p{}_\mu:=\eps^{pkl}\,(\Sigma^l_\pm)^\nu{}_\mu\,t^k{}_\nu$)
satisfies $\mathbb S_\pm^2+\mathbb S_\pm-2=0$, by \eqref{id75}. Its inverse is therefore $\mathbb S^{-1}=\tfrac12(\mathbb1+\mathbb S)$. This also follows from representation theory: the internal index $k$ of the tensors $t^k{}_\nu$, on which $\mathbb S_\pm$ acts indicates a $(1,0)$ (for the $+$ sector) or $(0,1)$ (for the $-$ sector) representation space, $\nu$ is the vector representation $(\tfrac12,\tfrac 12)$ of $SL(2,\mathbb C)_+\times SL(2,\mathbb C)_-$. For the $+$ sector the 12-dimensional tensor product of the representations decomposes into $(1,0) \otimes (\tfrac12,\tfrac 12)=(\tfrac32,\tfrac12)\oplus(\tfrac12,\tfrac12)$, that is a $8$- and a $4-$dimensional representation (exchange entries for the $-$ sector). The spectrum of $\mathbb S_\pm$ is $1$ with multiplicity $8$ and $-2$ with multiplicity $4$, and this explains why $\mathbb S_\pm^2+\mathbb S_\pm-2=0$ holds.

\paragraph{Solving for $\Gamma$.}
Dualising \eqref{cartanchi} with \eqref{dualemma} gives $G^a_\pm=\mp\im\,(\mathbb S_\pm
\Gamma_\pm)^a$, with $G$ of \eqref{Gdual}. Inverting,
\begin{equation}
\Gamma_\pm=\pm\im\,\mathbb S_\pm^{-1}G_\pm=\pm\tfrac\im2(\mathbb1+\mathbb S_\pm)G_\pm\,,
\end{equation}
which is \eqref{Gammasol}.

\section{Solution for $\rho$}\label{app:rho}
We solve \eqref{rhoeq} in three steps.
\begin{enumerate}
\item \emph{Dualise.} Dualise both sides with \eqref{dualemma}, applied to each
one-form separately. On the right-hand side this produces the second line of
\eqref{Tdual}, multiplied by $\det b\,(b^T)^{-1}$.
\item \emph{Pull back to the $\Sigma$-frame.} Write $\rho^j=b^j{}_p\tilde \rho ^p$ and use the
cofactor identity $\eps_{ijk}b^i{}_lb^j{}_pb^k{}_a=\det b\,\eps_{lpa}$. After multiplying
by $b^T/\det b$ the left-hand side becomes $\mp\im\,(\mathbb S_\pm\tilde \rho_\pm)$, so that
\begin{equation}
\mp\im\,\mathbb S_\pm\tilde \rho_\pm=\hat T_\pm\,.
\end{equation}
\item \emph{Invert.} This gives $\tilde \rho_\pm=\pm\im\,\mathbb S_\pm^{-1}\hat T_\pm=\pm\tfrac\im2
(\mathbb1+\mathbb S_\pm)\hat T_\pm$ and $\rho=b\,\tilde \rho$, which is \eqref{rhoexplicit}.
\end{enumerate}
The solution is unique because $\mathbb S_\pm$ is invertible. 

\section{The contorsion term}\label{app:contorsion}

\paragraph{Derivation of \eqref{cubicrho}.}
With $\tilde \rho:=b^{-1}\rho$ we proceed in four steps.
\begin{enumerate}
\item $\mathbb S_\pm$ is symmetric for the pairing $s\cdot t=g^{\mu\nu} {s^k}_\mu t_{k_\nu}$.  Indeed
$t\cdot\mathbb Ss=\eps^{pkl}t^{p\mu}s^{k\nu}\Sigma^l_{\nu\mu}$ is invariant under
$(p,\mu)\leftrightarrow(k,\nu)$, since $\eps^{pkl}$ and $\Sigma^l_{\nu\mu}$ both change sign.
\item \eqref{rhoexplicit} reads $(\mathbb1+\mathbb S)\hat T=\mp2\im\tilde \rho$,   with $(\mathbb1+\mathbb S)^{-1}=\tfrac12\mathbb S$ we therefore have
$\hat T=\mp\im\,\mathbb S\tilde \rho$. Using step 1,
\begin{equation}
\hat T\cdot(\mathbb1+\mathbb S)\hat T=-2\,\tilde \rho\cdot\mathbb S\tilde \rho\,,
\end{equation}
which carries no sector sign.
\item Insert $\tilde \rho=b^{-1}\rho$ and use the cofactor identity $\eps^{pkl}(b^{-1})^p{}_i
(b^{-1})^k{}_j=\det(b)^{-1}\eps_{ijm}b^m{}_l$. This gives
\begin{equation}
\tfrac{\det b}{4}\hat T \cdot(\mathbb1+\mathbb S)\hat T=\tfrac12\eps_{ijm}\rho^{i\mu}\rho^{j\nu}
(b\Sigma)^m_{\mu\nu}\,.
\end{equation}
\item Convert to forms with \eqref{iddual} and $b\Sigma=\sigma_\pm^{-1}B$. This gives $\tfrac12\eps_{ijk}B^i\wedge\rho^j\wedge\rho^k=-\tfrac{\det b}4
\hat T(\mathbb1+\mathbb S)\hat T\,\vol$, which is \eqref{cubicrho}.
\end{enumerate}

Making use of the second line of \eqref{Tdual} the terms $\hat T\cdot\hat T$ and
$\hat T\cdot\mathbb S\hat T$ can be expressed as
\begin{align}
(\det b)^2\,\hat T\cdot\hat T&=-\,m_{jj'}\Big[D^j{}_{a\lambda}D^{j'a\lambda}+\eps^{aa'c}\Sigma^{c\,\lambda\lambda'}
D^j{}_{a\lambda}D^{j'}{}_{a'\lambda'}\Big]\,,\qquad m=bb^T\,,\label{UUred}\\
(\det b)^2\,\hat T\cdot\mathbb S\hat T&=-\,N^l{}_{jj'}D^j{}_{a\lambda}D^{j'}{}_{a'\lambda'}\Big[\delta^{la'}\Sigma^{a\,\lambda'\lambda}
+\eps^{la'a}g^{\lambda\lambda'}+\delta^{aa'}\Sigma^{l\,\lambda\lambda'}-\delta^{la}
\Sigma^{a'\,\lambda\lambda'}\Big]\,,\label{UEUred}
\end{align}
with $D^j{}_{a\lambda}:=(D_\Gamma b)^j{}_{a\lambda}$ and 
$N^l{}_{jj'}:=\eps^{pkl}(b^T)^p{}_j(b^T)^k{}_{j'}=\det b\,\eps_{jj'n}(b^{-1})^l{}_n$.

\section{Second-order expansion}\label{app:expansion}

\paragraph{Trace sector.}
For $b_\pm=\delta+\tfrac12\mathring\chi^\pm+\tfrac16\chi^\pm_\tr\,\delta$ the trace is fixed by \eqref{tauFix},
$\chi^\pm_\tr=\tfrac14\tr\mathring\chi_\mp^{\,2}+O(\mathring\chi^3)$. This gives $q^+_\tr=3+\tfrac12{\rm Re}\,\tr\mathring\chi_+^2$, so
that the term $-\tfrac23\Lambda q^+_\tr$ in \eqref{Spot} yields the $2\Lambda$ in $\hat\mu_\pm$ of \eqref{Ahat}. The
curvature term $\gamma_\pm\tr(X_\pm b^T_\pm)$ contains the product of the trace of $X_\pm$ with the trace
of $b_\pm$, $\tfrac{\gamma_\pm R}{12}\chi^\pm_\tr=\tfrac{\gamma_\pm R}{48}\tr\mathring\chi_\mp^2$; this is the cross-chiral $-\tfrac14\gamma_\mp R$ in
\eqref{Ahat}.

\paragraph{Kinetic term.}
\eqref{UUred}--\eqref{UEUred} are quadratic in $D_\Gamma b=\tfrac12D\mathring\chi$.  For the second order expansion we can therefore set 
 $m=\delta$ and $N^l{}_{jj'}=\eps_{jj'l}$ in these formulas.
This gives:
\begin{itemize}
\item the $\eps^{la'a}g$-structure of \eqref{UEUred} reproduces the $g$-term of
\eqref{UUred};
\item the other $\Sigma$-structures either reproduce $\mathcal K:=\eps^{abc}\Sigma^{c\,\mu\nu}D_\mu
\mathring\chi^{ja}D_\nu\mathring\chi^{jb}$ or vanish for symmetric $\mathring\chi$.
\end{itemize}
This results in \eqref{Qdef}. Since $\Sigma$ is parallel with respect to the combined
Levi-Civita connection (cf.\ Section~\ref{Sec:beom}),
\begin{equation}
\mathcal K=\nabla_\mu(\eps^{abc}\Sigma^{c\,\mu\nu}\mathring\chi^{ja}D_\nu\mathring\chi^{jb})
-\tfrac12\eps^{abc}\Sigma^{c\,\mu\nu}\mathring\chi^{ja}[D_\mu,D_\nu]\mathring\chi^{jb}\,.
\end{equation}
On each internal index $[D_\mu,D_\nu]$ acts as $-\eps^i{}_{nk}F^n_{\mu\nu}$, and
$\Sigma^{c\,\mu\nu}F^n_{\mu\nu}=2X^{nc}$. The second term therefore equals
$\eps^{abc}\mathring\chi^{ja}(\eps^{jnk}X^{nc}\mathring\chi^{kb}+\eps^{bnk}X^{nc}\mathring
\chi^{jk})$. For traceless symmetric $\mathring\chi$ its two pieces reduce as follows:
\begin{itemize}
\item the second piece gives $(\tr X)\,\mathring\chi\cdot\mathring\chi-\tr(X\mathring\chi^2)$,
by $\eps\eps=\delta\delta-\delta\delta$;
\item the first piece gives $(\tr X)\,\mathring\chi\cdot\mathring\chi-2\tr(X\mathring\chi^2)$,
by the mixed-discriminant identity for $3\times3$ matrices.
\end{itemize}
Hence $\mathcal K\simeq2\tr X\,\mathring\chi\cdot\mathring\chi-3\tr(X\mathring\chi^2)$, and with
$X=\tfrac R6\delta+\mathring X$ the term $-\tfrac18\mathcal K$ is \eqref{twistcurv}.

\section{Flat and FRW backgrounds}\label{app:flatFRW}

For a real plane wave $h_{\mu\nu}=\epsilon_{\mu\nu}\cos(k\cdot x)$ with $\epsilon$
transverse and traceless with respect to the four-momentum $k$ (not necessarily on shell),
let $\langle\cdot\rangle$ denote the average over one period, which removes total
derivatives. We find symbolically
\begin{equation}
\langle\sqrt{-g}R\rangle^{(2)}=\tfrac14\langle h\Box h\rangle\,,\qquad
\langle C^2\rangle=\tfrac12\langle h\Box^2h\rangle\,,\qquad
\langle C\tilde C\rangle=0\,,
\end{equation}
the last for linear and circular polarisations.  The parity-odd term vanishes for any
form factor. The quadratic Pontryagin density is a total derivative, so its kernel,
bilinear in $h(k)$ and $h(-k)$, vanishes pointwise, and a form factor only multiplies it
by $\tilde F(-k^2)$.
 
On spin~2, $S^{(2)}=\tfrac14\int h\,\Box[1+\Box F_C(\Box)]h$. With $F_C={\rm Re}[\gamma_+/
(\mathfrak m^2_+-\Box)]$ and ${\rm Re}\,\gamma_+=1$,
\begin{equation}
1+\Box F_C=\frac{|\mathfrak m^2_+|^2-{\rm Re}(\gamma_+\mathfrak m^2_+)\Box}{|\mathfrak m^2_+-\Box|^2}\,.
\end{equation}
Its inverse, times $1/\Box$, is \eqref{propflat}. The partial fractions are:
\begin{itemize}
\item residue $1$ at $P=0$;
\item residue $r$ of \eqref{flatpole} at $m^2_{\rm pole}$;
\item the contact term $-1/{\rm Re}(\gamma_+\mathfrak m^2_+)$.
\end{itemize}
 For real $\mu$, $r=-1/(1+\gamma^2)$, and
$m^2_{\rm pole}=\tfrac23|\mu|^2/[(1+\gamma^{-2}){\rm Re}\,\mu]$ is the
physical mass of the linearised theory. For real $\mathfrak m^2_+=m^2$ the numerator
factorises as $m^2(m^2-\Box)$ and the massive pole cancels.
 
\paragraph{FRW.}
The background is conformally flat. Every term of \eqref{SeffAll} beyond
Einstein--Hilbert is quadratic in $C$, so it does not contribute to the background
equations. In the full theory, $q=\delta$ solves the traceless $b$-equations because
$\mathring X=0$, $D_\Gamma\delta=0$ and $\Pi\propto\delta$. The imaginary part of the trace
equation fixes $\varphi_0=R/(4\gamma)$. Its real part is the gauge direction of the
scaling redundancy once matter couples to $e[B]$ in the gauge $|\det b|=1$.
 
For $\mu_\pm=\tfrac32\gamma_\pm m^2$, write $\mathfrak m^2_+=M_R+\im M_I$ with
$M_I=-\gamma(4\Lambda-R)/(3(1+\gamma^2))=-\gamma T/(6(1+\gamma^2))$. To leading order in $M_I$,
\begin{equation}
m^2_{\rm pole}\simeq M_R+\gamma^{-1}M_I\,,\qquad
r\simeq-(1+\gamma^{-2})\frac{M_I^2}{M_R^2}=-\frac{T^2}{36(1+\gamma^2)M_R^2}\, .
\end{equation}
 With frozen coefficients, the FRW propagator is
\eqref{propflat} with these time-dependent parameters, up to $O(H^2/m^2)$ corrections.
During inflation $T\simeq-24H^2$, so $|r|\simeq16H^4/[(1+\gamma^2)m^4]$. With the matter coupling \eqref{SintT}, $M_I=0$ and the residue vanishes.

\section{The trace constraint and the acyclic part of the area metric}\label{app:acyclic}

In the main text the multiplier $\varphi_0$ imposes the sharp constraint ${\rm Im}\,\tr q_+=0$,
\eqref{trconstr}, and the area metric \eqref{Adef} is cyclic by construction. Here we show that
the one parameter removed by this constraint is precisely the totally antisymmetric (acyclic)
part of the area metric built from the $B$-fields.

\paragraph{The acyclic part.} Consider the area metric without the subtraction in
\eqref{Adef},
\begin{equation}
A^{\rm full}_{\mu\nu\rho\sigma}=\sum_\pm\frac{1}{2\sigma_\pm^2}\,\delta_{ij}\,B^i_{\pm\,\mu\nu}B^j_{\pm\,\rho\sigma}\,,
\qquad A=A^{\rm full}-\frac{1}{4!}\big(\epst^{\alpha\beta\gamma\delta}A^{\rm full}_{\alpha\beta\gamma\delta}\big)\,\epst_{\mu\nu\rho\sigma}\,.
\end{equation}
Contracting \eqref{qext} with $\delta_{ij}$ and using $\sigma_\pm^2=-1$ gives
$\delta_{ij}B^i_\pm B^j_\pm\epst=\mp8\im\det(e_\pm)\,\tr q_\pm$, and therefore
\begin{equation}\label{acycpart}
\epst^{\mu\nu\rho\sigma}A^{\rm full}_{\mu\nu\rho\sigma}
=\sum_\pm\big(\pm4\im\det(e_\pm)\,\tr q_\pm\big)
=-8\,{\rm Im}\big(\det(e_+)\,\tr q_+\big)\,,
\end{equation}
where we used $\det e_-=\overline{\det e_+}$ and $\tr q_-=\overline{\tr q_+}$. Under the
normalisation \eqref{phaseCondnew} $\det e_+$ is real, so the acyclic part is
\begin{equation}
A^{\rm full}_{\mu\nu\rho\sigma}-A_{\mu\nu\rho\sigma}=-\tfrac13\,\det e\;{\rm Im}(\tr q_+)\;\epst_{\mu\nu\rho\sigma}\,.
\end{equation}
Three observations follow:
\begin{itemize}
\item \emph{The $\varphi_0$-constraint is the vanishing of the acyclic part.} By \eqref{acycpart} the
$\varphi_0$-equation in \eqref{sharp}, $\delta_{ij}(B^i_+\wedge B^j_++B^i_-\wedge B^j_-)=0$, is equivalent to
$\epst^{\mu\nu\rho\sigma}A^{\rm full}_{\mu\nu\rho\sigma}=0$. On the constraint surface the subtraction in
\eqref{Adef} is therefore trivial, and the area metric built from the $B$-fields is automatically
cyclic.
\item \emph{Counting.} A complex symmetric $q_+$ with $|\det q_+|=1$ has 11 real parameters,
which together with the 10 parameters of the tetrad gives 21. This is the number of components
of a general area metric, 20 cyclic ones plus the acyclic one. Imposing ${\rm Im}\,\tr q_+=0$
leaves the 20 components of a cyclic area metric. The remaining trace direction ${\rm Re}\,\tr q_+$
is the scaling redundancy $(e,b)\to(\lambda e,\lambda^{-2}b)$ and is removed by $|\det b_\pm|=1$.
\item \emph{Eigenvalues.} For the endomorphism \eqref{endo} built from $A^{\rm full}$, the
eigenvalues on $\Lambda^2_+$ are $\lambda_i=2\im\det e\,\mu_i$, with $\mu_i$ the eigenvalues of
$m_+$. Hence $\sum_i{\rm Re}\,\lambda_i=-2\det e\,{\rm Im}\,\tr q_+$. The acyclic part shifts all
eigenvalues by a common real amount; for a cyclic area metric the real parts sum to zero. 
\end{itemize}

\paragraph{Relaxing the constraint.} If the $\varphi_0$-constraint is dropped, ${\rm Im}\,\tr q_+$
becomes an additional field. We note the following, without having carried out the
analysis in detail:
\begin{itemize}
\item \emph{Parity.} The acyclic part is a pseudoscalar density, i.e.\ parity odd.
\item \emph{Kinetic term.} Write $b_\pm=(1+\tfrac16\chi^\pm_\tr)\delta$ with $\chi^\pm_\tr=\pm2\im\,t$; ${\rm Re}\,\chi_\tr$ is fixed by $|\det b|=1$. Then $D_\Gamma b=\tfrac16\,d\chi_\tr\,\delta$.
Solving the $\rho$-equation \eqref{rhoeq} for this source and evaluating the contorsion term of
\eqref{SBFrhostar} gives
\begin{equation}\label{Lcontr}
\mathcal L_{\rm contorsion}=-\tfrac16\,(\partial t)^2=\tfrac16\,\dot t^2-\tfrac16\,(\nabla t)^2\,.
\end{equation}
This is Lorentz covariant, has the healthy sign and is independent of $\gamma$, since only
$\gamma_++\gamma_-=2$ enters. The same computation shows that the contorsion term does not mix $\chi_\tr$ with
$\mathring\chi$: the mixed coefficient vanishes for all momenta, while the $\mathring\chi\mathring\chi$ coefficient reproduces
$-\tfrac{\gamma_\pm}{8}(\partial\mathring\chi^\pm)^2$ of \eqref{Seffchi}. In addition, the curvature coupling $\tfrac{R}{12}\chi^\pm_\tr$ in $\tr(X_\pm b_\pm^T)$ gives the
parity-odd coupling
\begin{equation}
\sum_\pm\gamma_\pm\,\tfrac R{12}\,\chi^\pm_\tr=-\tfrac{1}{3\gamma}\,t\,R\,.
\end{equation}
In the Einstein frame this adds $-\tfrac{1}{6\gamma^2}(\partial t)^2$, so that the total kinetic term is
$-\tfrac16|\gamma_+|^2(\partial t)^2$, with $|\gamma_+|^2=1+\gamma^{-2}$.
\item \emph{Mass.} Replacing the sharp trace constraint by a potential term
$-\tfrac\mu2\,({\rm Im}\,\tr q_+)^2\,\vol$, where ${\rm Im}\,\tr q_+=2t$, gives the mass
\begin{equation}\label{mtrace}
m_t^2=\frac{12\,\mu}{|\gamma_+|^2}\qquad(16\pi G=1)\,.
\end{equation}
\item \emph{Electromagnetism.} If the Maxwell field couples to the area metric through the
constitutive tensor $A^{\rm full}$, the acyclic part enters as an axion-like coupling
$\theta\,F\wedge F$ with $\theta\propto\det e\,{\rm Im}\,\tr q_+$. A dynamical ${\rm Im}\,\tr q_+$ would
then rotate the plane of polarisation (circular birefringence), which cannot occur for the
cyclic area metrics considered in the main text.
\end{itemize}

\end{appendix}

\bibliographystyle{jhep}
\bibliography{references}

  \end{document}